\documentclass[journal=jctcce,manuscript=article]{achemso}
\usepackage{graphicx}
\graphicspath{{images/}}
\usepackage{epstopdf}
\usepackage{dsfont}
\usepackage{chngcntr}
\usepackage{amsmath,amssymb}
\usepackage{amsthm}

\usepackage{graphicx}
\usepackage{color}
\usepackage{dcolumn}
\usepackage{bm}
\usepackage{changepage}
\definecolor{background-color}{gray}{0.98}
\usepackage[algoruled,vlined]{algorithm2e}

\usepackage{xcolor}
\usepackage{xr}

\newtheorem{remark}{Remark}

\usepackage{longtable}

\usepackage[utf8]{inputenc} 
\usepackage[T1]{fontenc}    
\usepackage{hyperref}       
\usepackage{url}            
\usepackage{booktabs}       
\usepackage{amsfonts}       
\usepackage{nicefrac}       
\usepackage{microtype}      
\usepackage{verbatim}
\usepackage{amsmath}

\DeclareMathOperator*{\argmin}{arg\!\min}
\usepackage{caption}
\usepackage{subcaption}

\newcommand{\labelnotempty}[1]{
\def\temp{#1}\ifx\temp\empty
\else
    \label{#1}
\fi
}

\author{Zineb~Belkacemi}
\email{zineb.belkacemi@sanofi.com}
\affiliation{CERMICS, Ecole des Ponts, Marne-la-Vallée, France}
\alsoaffiliation{ Structure Design and Informatics, Sanofi 1371 R\&D, 91385 Chilly-Mazarin}
\author{Paraskevi~Gkeka}
\affiliation{ Structure Design and Informatics, Sanofi 1371 R\&D, 91385 Chilly-Mazarin}
\author{Tony~Lelièvre}
\affiliation{CERMICS, Ecole des Ponts, Marne-la-Vallée, France}
\alsoaffiliation{MATHERIALS team-project, Inria Paris, France}

\author{Gabriel~Stoltz}
\affiliation{CERMICS, Ecole des Ponts, Marne-la-Vallée, France}
\alsoaffiliation{MATHERIALS team-project, Inria Paris, France}

\title{Chasing collective variables using autoencoders and biased trajectories}

\begin{document}
\begin{tocentry}
    \centering
    \includegraphics[width=0.8\textwidth]{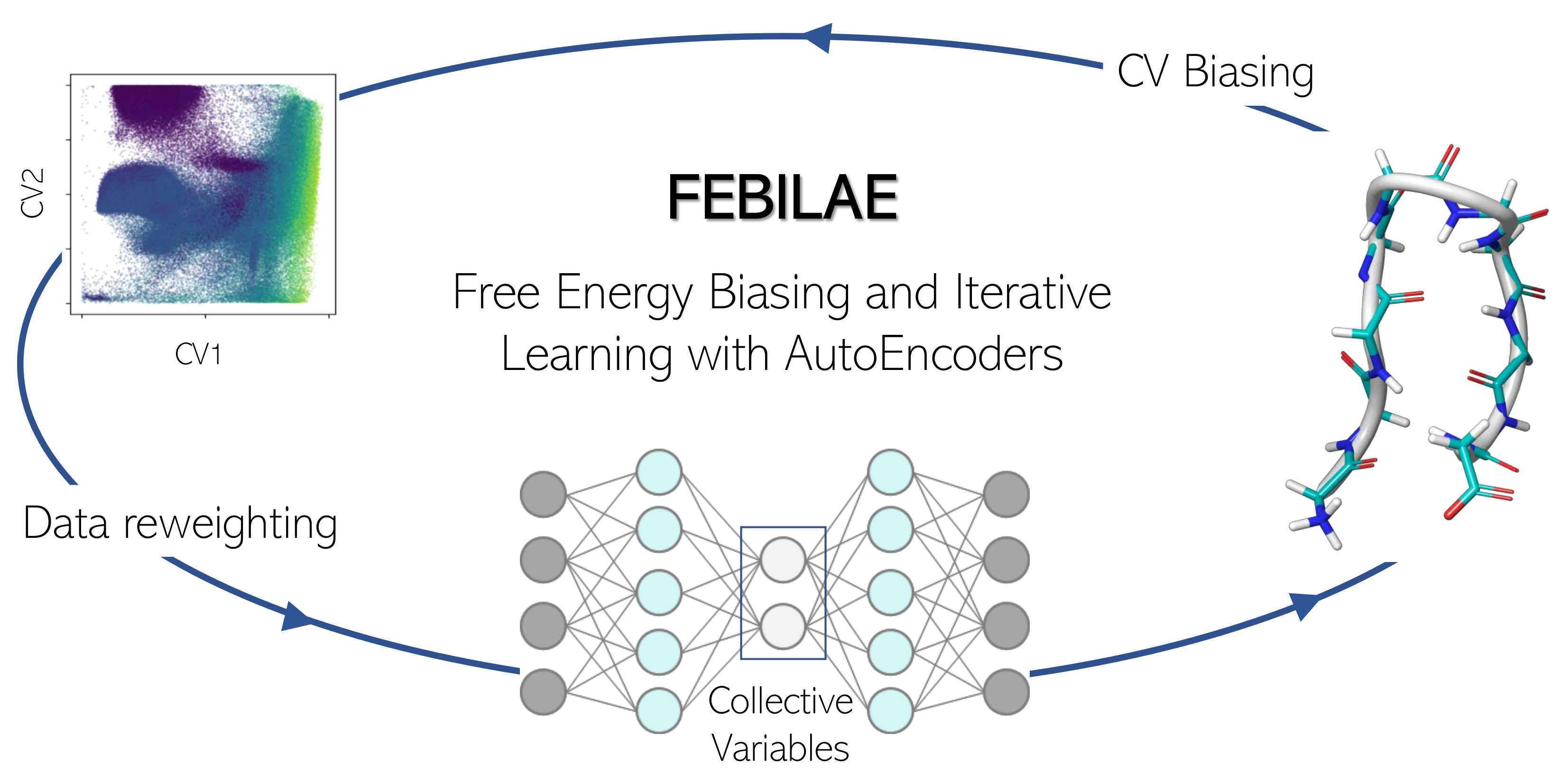}
\end{tocentry}
\clearpage
\begin{abstract}

Free energy biasing methods have proven to be powerful tools to accelerate the simulation of important conformational changes of molecules by modifying the sampling measure. However, most of these methods rely on the prior knowledge of low-dimensional slow degrees of freedom, i.e. Collective Variables (CV). Alternatively, such CVs can be identified using machine learning (ML) and dimensionality reduction algorithms. In this context, approaches where the CVs are learned in an iterative way using adaptive biasing have been proposed: at each iteration, the learned CV is used to perform free energy adaptive biasing to generate new data and learn a new CV.  In this paper, we introduce a new iterative method involving CV learning with autoencoders: Free Energy Biasing and Iterative Learning with AutoEncoders (FEBILAE). Our method includes a reweighting scheme to ensure that the learning model optimizes the same loss at each iteration, and achieves CV convergence. Using the alanine dipeptide system and the solvated chignolin mini-protein system as examples, we present results of our algorithm using the extended adaptive biasing force as the free energy adaptive biasing method. 

\end{abstract}

\section{Introduction}
\label{sec:intro}

In the last decades, molecular dynamics (MD) simulations have helped gain insight into the microscopic and macroscopic properties of biomolecular processes. However, the time scales accessible to MD simulations are often significantly smaller than the times needed for the observation of slow conformational changes of the systems under study~\cite{MD1, MD3}. This is due to the presence of energy or entropy traps in the energy landscape, which causes the system to be stuck within metastable states and thus hinders the full exploration of the configurational space. As a consequence, thermodynamic quantities (obtained from trajectorial averages) can not be accurately estimated. 

To cope with this issue, several methods for enhanced sampling have been designed to mitigate the sampling difficulties associated with metastability~\cite{ES1, ES2}.
Most of these methods can be broadly divided into two categories according to whether or not they use collective variables~(CV), also known as reaction coordinates, which are low dimensional or coarse-grained representations of the system:
\begin{itemize}
\item Collective variable free methods alter the canonical distribution by e.g. modifying the system temperature or the system Hamiltonian in order to accelerate crossing energetic or entropic barriers. This category includes, for example, simulated tempering~\cite{temp2}, parallel tempering~\cite{paratemp}, replica exchange MD~\cite{temp1}, multicanonical simulation~\cite{temp3}, temperature-accelerated dynamics~\cite{temp4} or the Wang-Landau algorithm~\cite{wanglandau}.
\item Collective variable based methods modify the system's dynamics by adding a bias in order to accelerate the dynamics by flattening the energy barriers along a chosen CV. Most of these methods simultaneously calculate the free energy associated to these CVs. Notable examples include metadynamics~\cite{metadynamics, metadynamics2}, which biases the system's evolution using a potential constructed as a sum of Gaussian variables centered along the trajectory of the CV; likewise, umbrella sampling~\cite{umbrella, umbrella2} adds a biasing potential along the CVs, often of harmonic form, to force the system to visit intermediate regions between metastable states. Adaptive biasing force (ABF) methods~\cite{abf, abfHC, abfDP} add a biasing force to the system dynamics so as to eliminate any average force acting along the CVs, rendering the dynamics more diffusive along these directions. The efficiency of these methods crucially relies on the prior knowledge of a proper low dimensional CV which contains most of the metastable functions of the dynamics, and thus in particular clearly distinguishes between the metastable states. 
\end{itemize}
In simple cases, the CV can be chosen based on intuition and prior knowledge of the system at hand. This is for example the case for alanine dipeptide~\cite{dialaMA}, for which two known CVs are the $\phi$ and $\psi$ dihedral angles. For more complex systems, adequate CVs are often difficult to guess. There have been attempts in the last years to extract CVs using dimensionality reduction and machine learning~(ML) methods~\cite{dinnermlmd}. 

Collective variable discovery methods range from simple linear projections such as principle component analysis (PCA) or factor analysis, to more elaborate algorithms involving non linear and/or dynamically relevant projections of the configurational space.  Below, we recall some notable works on automatic CV discovery and design, but we do not aim to be exhaustive nor do we give a detailed comparison of these algorithms. For a more complete overview on optimization and learning methods for CV discovery and in molecular dynamics in general, we refer the reader to Refs.\citenum{cecam, glielmoreview, sidkyreview}. We choose here to distinguish between three broad categories of methods: Operator-based methods, which aim at building a coarse grained description of the dynamics;  metastable state separation methods, which use supervised machine learning; and unsupervised learning methods for dimensionality reduction. 

Operator-based methods account for the dynamical properties of the system under study by using the approximation of a transfer operator or generator associated with the dynamics. Notable examples are the variational approach to conformational dynamics~\cite{vac, ticadmd} (VAC) and its linear version, time lagged independent component analysis (tICA), the extended dynamic mode decomposition~\cite{ticadmd, ticadmd2} (EDMD), Diffusion Maps~\cite{Dmap, Dmap2} or Markov State Models~\cite{msm1, msm2}, which can also be considered a special case of VAC. Methods like VAC can often be optimized by using a well defined dictionary of functions to represent the system (instead of using the system coordinates in the case of tICA~\cite{tica}). For instance, VAMPNets~\cite{Vampnets} and state-free reversible VAMPNets~\cite{srv} learn this dictionary using neural networks. 

Methods for metastable state separation use supervised learning for feature selection and feature engineering~\cite{SVM-LR, randomFor, CVscreen1}. Supervised learning models are trained on a labeled dataset and learn a mapping from the datapoints to their labels. In CV discovery, the dataset is the set of sampled configurations and the labels are usually their corresponding states (which are thus assumed to be known). In particular, feature selection based CV learning uses a set of candidate CVs as input features. Each feature is then given an importance score according to its contribution to the learned model, and the most relevant features are selected as final CVs. Feature engineering methods on the other hand compute a new CV from the input features. Notable examples include Ref~\citenum{SVM-LR} where decision functions learned by support vector machines are used as collective variables; Ref~\citenum{spvpathCV} where a 1-dimensional path CV is constructed as a nonlinear combination of classifier CVs obtained from a supervised neural network model; or Ref~\citenum{deepLDA} where a neural network is also used in combination with an LDA objective function. 

Unsupervised learning models for CV identification, on the other hand, are trained on unlabeled datapoints (namely the sampled configurations without any additional information). These learning models aim at recovering patterns and intrinsic properties of the visited configurational space. Currently, the most used non linear ML dimensionality reduction method for CV discovery is auto-associative neural networks, also known as autoencoders (AE)~\cite{AANN}. Autoencoders are a type of neural networks that aim at learning low dimensional representations of the data. The network is composed of two parts. The first part is the encoder that learns the representation or encoding of the input. The second part is the decoder, it simultaneously learns to reconstruct the input from the encoding. Then, the obtained encoding map can be used as a CV.  Many state-of-the-art methods for CV unsupervised learning include autoencoders in their framework. For example, time lagged autoencoders~\cite{tae} train the decoder to predict the time-lagged configuration instead of the input configuration, thus factoring time evolution into the learned CV. Some works~\cite{DeepBayes} use variational autoencoders~\cite{VAE}, a class of models  which combine autoencoder structure with Bayesian inference to learn a latent variable distribution through a probabilistic (instead of deterministic) encoder and decoder, resulting in a generative decoder in addition to the dimensionality reducing encoder.Variational dynamics encoders~\cite{VDE} employ the time-lag refinement on variational autoencoders. To learn CVs which resolve the different metastable states, the authors in Ref~\citenum{GMVAE} use Gaussian mixture variational AEs. Another example is Ref~\citenum{extendedAE} where extended autencoders are used to predict the committor function. Of course, many methods for unsupervised CV learning do not include autoencoders, instead using, e.g. Bayesian models~\cite{nodeepBayes} or information theory based methods~\cite{amino}.

Naturally, any learning or dimensionality reduction model requires a certain amount of good quality data to learn relevant information. In the case of molecular dynamics, the lack of data (incomplete sampling) is the initial problem at hand. A natural solution to overcome this difficulty is to use learning models in some iterative fashion. This can be done by training models on available data and using the learned CVs for enhanced sampling to generate additional learning data, and so on. Autoencoders have been used in iterative methods for CV learning in previous literature. For example, molecular enhanced sampling with autoencoders (MESA)~\cite{mesa1, mesa2} is a framework that alternates between autoencoder learning of CVs and free-energy biasing (more specifically umbrella sampling) along those CVs. Note first that this kind of iterative procedure can in theory be used for any dimensionality reduction model and any CV biasing method. However, it is important to note that biased sampling makes the distribution of the sampled configurations drift from the Boltzmann--Gibbs density. Changing the distribution of the data results in changing the loss function optimized by any model. 
In the case of iterative algorithms for CV learning and enhanced sampling, this implies that at each iteration, the training data coming from a biased simulation has a different distribution, and the model optimizes a different loss. This means that iterative models such as MESA are not guaranteed to converge to a certain CV, regardless of whether they end up obtaining a good sampling of the configurational space. Even in cases when the search for a collective variable of the system is not the goal, this still poses the issue of how to design a stopping rule for the iterative procedure. 
    
Here, we present a new iterative algorithm for CV learning with autoencoders, named FEBILAE for "Free Energy Biasing and Iterative Learning with AutoEncoders". Our algorithm is inspired from methods such as MESA, but adds a simple yet crucial reweighting step of the data sampled from free energy biased simulations, so as to ensure the consistency of the optimized loss, and thus the convergence of the learned CVs. Note that a reweighting protocol is also used in other iterative methods, notably the reweighted autoencoded variational Bayes for enhanced sampling (RAVE)~\cite{rave, rave2019}. RAVE uses a variational autoencoder that takes as input one or more pre-selected variables and iteratively learns a CV and its distribution. This CV is then used for biasing. In RAVE, the learned CV is always one dimensional in practice and is learned as a linear combination of the preselected variables (through a linear encoder). This is the key difference of the specific method compared to FEBILAE: our CV is learned as a nonlinear function of the input coordinates, rather than a linear combination of a preselected set of order parameters or candidate collective variables. Consequently, the FEBILAE CV is representative of the whole system, and does not require prior knowledge of the system to select the input, although it lacks interpretability. 

The FEBILAE algorithm can be used with any collective variable based biasing technique and any dimensionality reduction method that provides a differentiable mapping from the configurational space to the CV space. Moreover, in order to accelerate convergence, we propose an iterative procedure where information from each step, namely the trajectory data, the estimated free energy, or the learned model, are judiciously used in the following steps of the iterative procedure. We finally present the results of our implementation of the algorithm, which we coin AE-ABF, as it uses ABF to perform enhanced sampling using autoencoder learned CVs. 
    
This article is organised as follows. We first provide in Section~\ref{sec:Theory} an introduction to autoencoders and their use for dimensionality reduction and then move on to a theoretical analysis highlighting that the learned model is always dependent on the distribution of the training data. We continue by demonstrating how a simple reweighting procedure of the loss function can be applied to target a specific density distribution. We then introduce in Section~\ref{sec:ae-abf} our algorithm for iteratively learning a CV, and present some refinements that can be incorporated to the method for more efficiency, and possibly for a faster convergence. For an immediate illustration of the theoretical points made in Sections~\ref{sec:Theory} and~\ref{sec:ae-abf}, we intertwine the presentation of these theoretical concepts with numerical results obtained on a 2-dimensional toy example. Section~\ref{subsec:matmet} then summarizes the practical details and parameters of the implementation of AE-ABF for molecular systems. Finally, Section~\ref{sec:results} presents the results obtained when applying AE-ABF to two different systems: alanine dipeptide in vacuum, and solvated chignolin. Section~\ref{sec:conclusion} contains our conclusions and some variants of the method which could be useful for applications to other systems. Finally, additional results and details are given in the supplementary material.

\section{Learning autoencoder collective variables from (un)biased samples}
\label{sec:Theory}

We briefly introduce in Section~\ref{subsec:autoenc} autoencoders and their use for dimensionality reduction. Section~\ref{subsec:qtox} then recalls the usual processing steps required to convert trajectory data to autoencoder inputs. We then describe in Section~\ref{subsec:learning} how autoencoders are trained. We can then make precise in Section~\ref{subsec:diffdist} how the training is impacted by the distribution of the data, and present a reweighting procedure to correct a bias in the distribution of data points.  Finally, Section~\ref{subsec:2D} introduces a 2-dimensional toy example, which we use to illustrate the points made in Sections~\ref{subsec:learning} and \ref{subsec:diffdist} in particular.

\subsection{Autoencoders}
\label{subsec:autoenc}

Autoencoders (AE)\cite{AANN} are a type of neural network designed for unsupervised learning tasks.  The aim is usually to learn a new representation of the data, called an encoding. The AE is composed of two parts: the \emph{encoder} learns the new representation and the \emph{decoder} simultaneously learns to reconstruct the original data from this representation. The AE thus seeks to approximate the identity function. When the encoder is composed of one fully connected layer which reduces the dimension, together with a \emph{linear} activation function, its learned representation is essentially the same as that of a PCA projection of the same dimensionality~\cite{encispca1, encispca2}: more precisely, the two models project on the same bottleneck space, but not using the same vectors. In general, however, AEs are used with \emph{nonlinear} activation functions. This allows for nonlinear encoding functions, and thus potentially better encoders than those restricted to stay within the smaller class of linear functions. 

AEs can have different topologies depending on the learning task, the data size and dimensionality, etc. Here, we describe the general autoencoder topology used in this work. We denote by $\mathcal{X} \subseteq \mathbb{R}^D$ the data space, and by $\mathcal{A} \subseteq \mathbb{R}^d$ a lower dimensional space ($d < D$). Figure~\ref{fig:autoenc} presents the typical topology of the AEs used in our work. The autoencoder can be represented by a mapping $f = f_{\text{dec}} \circ f_{\text{enc}}$ where $f_{\text{enc}}: \mathcal{X} \xrightarrow{} \mathcal{A}$, $f_{\text{dec}}: \mathcal{A} \xrightarrow{} \mathcal{X}$ and $\circ$ is the function composition operator i.e.~$f_{\text{dec}} \circ f_{\text{enc}}(x) = f_{\text{dec}}\left(f_{\text{enc}}(x)\right)$.  It is symmetrical in structure, fully connected, and contains $2L$ layers. Each hidden layer is of dimension $d_{\ell}=d_{2L-\ell}$ for $\ell=1 \dots L$, and the output layer is of dimension $d_{2L} = D$ (Note that, by convention, the input layer does not count as a layer of the network). Each layer $\ell \in {1,\dots,2L}$ has an activation function $g_{\ell}$ and is connected to the previous layer by a projection matrix $W_{\ell} \in \mathbb{R}^{d_{\ell}\times d_{\ell-1}}$, and a bias vector $b_{\ell} \in \mathbb{R}^{d_{\ell}}$. There are thus $K= \displaystyle \sum_{\ell=1}^L d_{\ell} d_{\ell-1}+ d_{\ell}$ learnable real parameters denoted by $(p_1, \dots, p_K) \in \mathbb{R}^K$. As the activation functions are predefined and do not change during learning, the autoencoder function is fully described by its parameters $\textbf{p} = (p_k)_{k=1,\dots,K}$. We indicate this dependence as~$f_{\textbf{p}}$. The general formula for~$f_\textbf{p}$ is:    
$$ 
\forall x \in \mathcal{X}, \qquad f_{\textbf{p}}(x) = g_{2L} \left[b_{2L} + W_{2L}\,g_{2L-1}\left(b_{2L-1}+W_{2L-1} \dots g_1(b_1 +  W_1 x) \right) \right] \text{,}  
$$
where the activation functions are by convention applied element-wise: for $z=(z_1,\dots,z_{d_{\ell}})$, it holds that  $g_{\ell}(z) = (g_{\ell}(z_1),\dots,g_{\ell}(z_{d_{\ell}}))$. Note that $f_\textbf{p}$, as well as $f_{\text{enc}}$, are differentiable functions when all the activation functions ~$(g_{\ell})_{1\leq \ell \leq 2L}$ are.

\begin{figure}[ht!]
        \centering
        \includegraphics[width=0.9\textwidth]{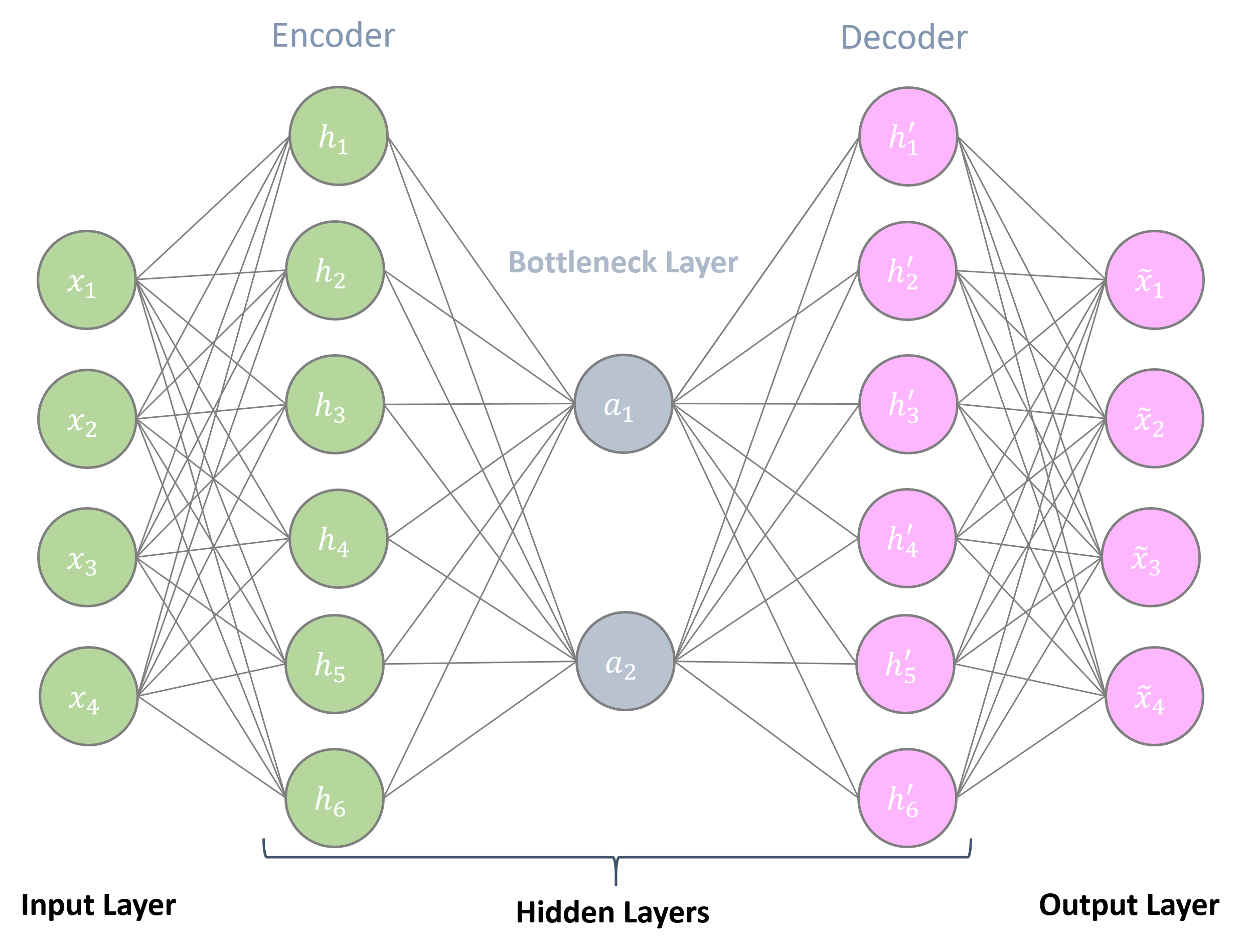}
        \caption{Example of an autoencoder topology. Here, the dimension of the data is $D=4$ and there are $2L=4$ layers: 3 hidden layers (including the bottleneck) and the output layer, of respective dimensions $d_1=6$, $d_2=2$, $d_3=6$ and $d_4=4$. The parameters can be represented by four matrices $W_1$, $W_2$ (for the encoder) and $W_3$, $W_4$ (for the decoder), such that:
        $\textbf{a} = (a_1, a_2) = f_{\text{enc}}(\textbf{x}) = g_2(b_2+ W_2 g_1(b_1+W_1\textbf{x}))$ and $\tilde{\textbf{x}} = f_{\text{dec}}(\textbf{a}) = g_4(b_4+W_4 g_3(b_3+W_3\textbf{a}))$ where  $g_1, \dots, g_4$  are activation functions, and $b_1,\dots,b_4 $ are biases (not shown in the topology).
        }
        \labelnotempty{fig:autoenc}
\end{figure}

\subsection{From trajectory to training data}
\label{subsec:qtox}
There is a distinction to be made between the configurational space of the simulation, and the data space over which the learned model is optimized. Indeed, a preprocessing step is usually necessary to obtain a usable dataset from the sampled molecular trajectory. A notable example is the elimination of rotational and translational invariances through centering and structural alignment of the configurations to a reference structure, or by using internal coordinates (distances, angles, etc). Another example is using only a subset of the coordinates to reduce the data dimensionality, for example by taking out solvent molecules and hydrogen atoms, whose motions are often considered irrelevant. For the remainder of the paper, we denote by $q$ the configurations of the system, and by~$x$ the corresponding post-processed data points. We however allow for an abuse of notation by keeping the same notation for various objects whose argument is either~$q$ or~$x$, depending on the context. For example, the probability measures~$\mu(dq)$ and~$\mu(dx)$ are denoted by the same symbol, even though $\mu(dx)$ is actually the image of $\mu(dq)$ by the application $q\mapsto x$. 
\subsection{Learning from unbiased samples}
\label{subsec:learning}
Consider a probability distribution $\mu$ on $\mathcal{X} \subset \mathbb{R}^D$.  Typically in MD,  $\mu$ is the Boltzmann-Gibbs distribution and $\mathcal{X}$ is the configurational space of the system under study.
We seek to encode configurations of $\mathcal{X}$ sampled from $\mu$ in a smaller dimensional space $\mathcal{A} \subset \mathbb{R}^{d}$, where $d < D$, using an autoencoder (AE). 
Theoretically, the optimal parameters $\textbf{p}_{\mu}$ minimize the expected loss 
\begin{equation}
\label{eq:reference_loss}
\mathcal{L}(\mu, \textbf{p}) = \mathbb{E}_{\mu}(\|X-f_{\textbf{p}}(X)\|^2) = \int_{\mathcal{X}}~\|x-f_{\textbf{p}}(x)\|^2 ~\mu(dx) ~\text{,}
\end{equation}
where the subscript $\mu$ in $\mathbb{E}_{\mu}$ indicates that $X$ is a random variable distributed according to~$\mu$. We denote by $\textbf{p}_{\mu}$ a solution to the minimization of $\mathcal{L}(\mu, \textbf{p})$, provided it exists (which is always assumed here): 
\begin{equation}
    \textbf{p}_{\mu} \in \argmin_{\textbf{p} \in \mathbb{R}^K} \mathcal{L}(\mu,\textbf{p}) \text{.}
\label{pmu}
\end{equation}  
 Note that $\textbf{p}_{\mu}$ is a priori not unique, and depends of course on the distribution~$\mu$. 

\subsubsection*{Empirical distribution}

In practice, the optimization problem~\eqref{pmu} is not easily solved, in particular because $\mathcal{L}$ is unknown and must be approximated. For this, the distribution~$\mu$ is replaced by an empirical distribution $\hat{\mu}$ corresponding to an available dataset of $N \in \mathbb{N}$ points $x^1,\dots, x^N$ drawn from the distribution~$\mu$. In MD, these datapoints are typically the configurations sampled during a simulation of the system, represented e.g by atomic positions or internal coordinates. The autoencoder is thus optimized in practice using the empirical distribution:  
$$\hat{\mu} = \frac{1}{N} \sum_{i=1}^{N} \delta_{x^i} \text{ .}$$
The AE parameters are thus optimized in order to minimize the empirical loss: 
\begin{equation}
     \mathcal{L}(\hat{\mu}, \textbf{p}) = \frac{1}{N} \sum_{i=1}^N \|x^i-f_{\textbf{p}}(x^i)\|^2 \text{ .} 
\label{EmpLoss}
\end{equation}
Using the law of large numbers (ergodic theorem), it holds that:
\[
\mathcal{L}(\hat{\mu}, \textbf{p}) \xrightarrow[N \to \infty]{} \mathcal{L}(\mu, \textbf{p}) \qquad \text{almost surely},
\]
which motivates minimizing with respect to $\textbf{p}$ the empirical loss $\mathcal{L}(\hat{\mu}, \textbf{p})$ instead of $\mathcal{L}(\mu, \textbf{p})$. The optimization problem thus becomes: 
\begin{equation}
\text{Find  ~~}
    \textbf{p}_{\hat{\mu}} \in \argmin_{\textbf{p} \in \mathbb{R}^K} \mathcal{L}(\hat{\mu},\textbf{p}) \text{.}
\label{pmuhat}
\end{equation}

\subsection{Learning from biased samples}
\label{subsec:diffdist}
To accelerate the exploration of the phase space of the system under study, MD simulations can be biased to target another distribution than the Boltzmann-Gibbs reference measure. 
The resulting dataset is thus not drawn from the original distribution of interest $\mu$, but from a biased distribution $\widetilde{\mu}$. Since we want to optimize the loss~\eqref{eq:reference_loss}, we need to reweight the configurations~$x$ sampled from~$\widetilde{\mu}$ by a factor
\[
w(x) = \displaystyle\frac{ \mu(x)}{\widetilde{\mu}(x)}.
\]
Note that to ensure that $w(x)$ is finite, we always assume that $\mu$ is absolutely continuous with respect to $\widetilde{\mu}$ (i.e. $\mu(A)=0 $ for all measurable sets $A \subset \mathcal{X}$ such that $\widetilde{\mu}(A)=0$). 

The reweighting corresponds to considering the loss function
\begin{equation}
\label{newLoss}
    \mathcal{L}(\mu, \textbf{p}) =  \mathcal{L}\left(\frac{\mu}{\widetilde{\mu}} \widetilde{\mu}, \textbf{p}\right) =  \int_{\mathcal{X}} \left\|x-f_{\textbf{p}}(x)\right\|^2  w(x) \widetilde{\mu}(dx).
\end{equation}
Reweighting ensures that the same optimization problem as~\eqref{pmu} is solved, even when the data points are not distributed according to~$\mu$. 

To approximate the expectation~\eqref{newLoss} with respect to~$\widetilde{\mu}$, we again rely on an empirical weighted distribution:
\begin{align}
\label{hatw}
	\widehat{\mu}_{\text{wght}} = \sum_{i=1}^N \widehat{w}_i \delta_{x^i} 
	\qquad 
 	\widehat{w}_i = \displaystyle\frac{ \mu(x^i)/\widetilde{\mu}(x^i)}{ \displaystyle\sum_{j=1}^N 		 \mu(x^j)/\widetilde{\mu}(x^j)}.
\end{align}
The new discrete loss is thus a weighted average loss over the~$\widetilde{\mu}$ sampled data: 
\begin{equation}
\label{lossw}
    \mathcal{L}(\widehat{\mu}_{\text{wght}}, \textbf{p}) = \sum_{i=1}^N \widehat{w}_i \|x^i-f_{\textbf{p}}(x^i)\|^2 \text{ ,}
\end{equation}  
which converges by the ergodic theorem almost surely to $\mathcal{L}(\mu,\textbf{p})$ when $N \xrightarrow{} \infty$. 

Note that computing $(\widehat{w}_i)_{1\leq i\leq N}$ only requires the knowledge of $\mu$ and $\widetilde{\mu}$ up to a multiplicative constant.

\subsection{Learning from free energy biased simulations: 2D toy example}
\label{subsec:2D}
This section provides a 2D example for free energy biasing. This system is then used to illustrate the necessity of reweighting when training models over biased simulations. We use throughout the section some common definitions of free energy and Boltzmann--Gibbs density measure,recalled in Supp.~Mat.~Section~\ref{subsec:FECV} for completeness.

Let us introduce the following three well potential, previously considered in other works~\cite{threewellpot}:

\begin{equation} 
\label{2Dpot}
\begin{aligned}
V(x_1,x_2) &= 3\mathrm{e}^{-x_1^2} \left(\mathrm{e}^{-(x_2-1/3)^2} - \mathrm{e}^{-(x_2-5/3)^2} \right) - 5\mathrm{e}^{-x_2^2} \left(\mathrm{e}^{-(x_1-1)^2} + \mathrm{e}^{-(x_1+1)^2} \right)\\
&\ \ +0.2 x_1^4 +0.2(x_2-1/3)^4 \text{ .}
\end{aligned}
\end{equation}
In this example, the configurations $q$ are represented by the $2$ dimensional coordinates of the particle: $q = (x_1,x_2)$. The potential $V$ has two deep wells centered at $q_{\mathrm{L}} = (-1.113, -0.03685)$, $q_{\mathrm{R}} = (1.113, -0.03685)$, and a shallow well around $q_{\mathrm{C}} = (0, 1.7566)$.  
We consider the overdamped Langevin dynamics: 
\begin{equation}
\label{Langevin}
    dq_t = -\nabla V(q_t)~dt + \sqrt{\frac{2}{\beta}}~dB_t \text{,}
\end{equation}
where $B_t$ is a $2$-dimensional Brownian motion. The dynamics are discretized using the Euler--Maruyama scheme: 
\begin{equation}
 q^{j+1} = q^j -\nabla V(q^j) \Delta t + \sqrt{\frac{2\Delta t}{\beta}}G^j \text{ ,} 
 \label{eulermaru}
 \end{equation}
where $\{G^j\}_{j \geq 0}$ is a sequence of independent standard normal random vectors and $q^j \approx q_{j\Delta t}$.  
We simulate a sample trajectory of this system, with inverse temperature $\beta=4$ and time-step $\Delta t = 10^{-3}$, starting from initial condition~$q^0=(-1,0)$. Figure \ref{fig:scattertraj} gives a scatter plot of this trajectory as well as the time evolution of the coordinate~$x_1$ .
\begin{figure}[ht!]
    \centering
    \begin{subfigure}{0.5\columnwidth}
        \centering
        \includegraphics[width=\textwidth]{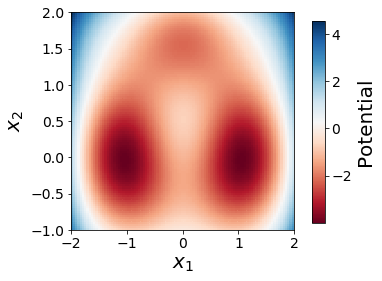}
        \caption{}
        \labelnotempty{subfig:traj}
    \end{subfigure}
    ~
    \begin{subfigure}{0.35\columnwidth}
        \centering
            \includegraphics[width=\textwidth]{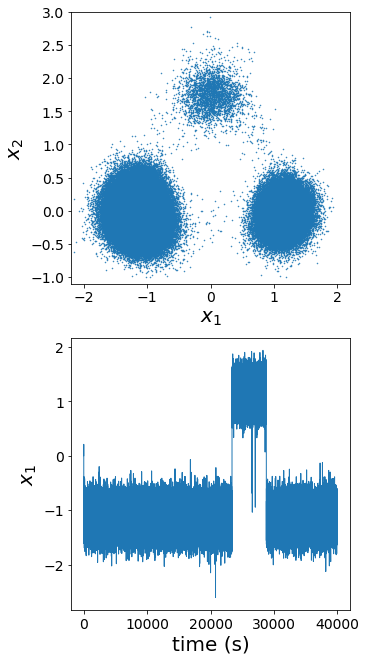}
        \caption{}
        \labelnotempty{subfig:xevol}
    \end{subfigure}
    \caption{Simulations with the 2D three well potential~\eqref{2Dpot} at $\beta=4$. Left: Potential function, two deep wells and one shallow well can be seen. Right: Overdamped Langevin trajectory. Top. Scatter plot of the sampled points.  Bottom: Time evolution of the coordinate $x_1$ through the example trajectory. Metastability is observed.}
\labelnotempty{fig:scattertraj}
\end{figure}

The metastability of the dynamics is quite well characterized by the direction~$x_1$, whereas~$x_2$ is not enough to distinguish between the two deep wells. In the sequel, we consider $\xi_1(q) = x_1$  and $\xi_2(q) = x_2$ as two different choices of CV. 

The free energies corresponding to the CVs $\xi_1$ and $\xi_2$ are respectively:
\begin{align}
\label{F(x)}
 F_1(x_1)  = -\beta^{-1}\ln\left(\int_{\mathbb{R}} \mathrm{e}^{-\beta V(x_1,x_2)} dx_2\right) \text{, ~~~}
 F_2(x_2)  = -\beta^{-1}\ln\left(\int_{\mathbb{R}} \mathrm{e}^{-\beta V(x_1,x_2)}  dx_1\right)  \text{ .}
\end{align}
The quantities $F_1(x_1)$ and $F_2(x_2)$ can be easily approximated in this low dimensional example by a numerical quadrature.
We can therefore easily sample from the three following probability measures: the unbiased Boltzmann-Gibbs probability distribution~$\nu$ associated with the potential~$V$, and the biased Boltzmann-Gibbs probability distributions $\nu_{F_i}$ associated with the biased potentials $V-F_i \circ \xi_i$ for $i=1,2$.



We now perform experiments to compare the results obtained from biased trajectories using different collective variables, with those from an unbiased trajectory. For this, we first simulate three trajectories. All three are long enough to sample the three potential wells of our system. The first trajectory is unbiased, following~\eqref{eulermaru} with a timestep $\Delta t = 10^{-3}$ for $N = 4\times 10^7$ steps. Every $40$th configuration is kept as a datapoint. The second trajectory is free energy biased using $\xi_1(x_1,x_2) = x_1$ as CV, i.e., it follows~\eqref{eulermaru} with potential $V-F_1\circ \xi_1$ instead of $V$, and a timestep $\Delta t=10^{-3}$ for $T = 2\times 10^6$ steps. Configurations are saved every 2 timesteps. The third trajectory is free energy biased using $\xi_2(x_1,x_2)=x_2$ as a collective variable, and using the same simulation settings as the second trajectory. Note that the second and third trajectories are shorter: the duration of these trajectories were chosen so that they visit the full configurational space. These three trajectories serve as our training datasets. The datapoints are thus sampled respectively from the unbiased Boltzmann--Gibbs measure $\mu=\nu$,  the free energy biased Boltzmann--Gibbs measure $\widetilde{\mu}_1=\nu_{F_1}$, and the free energy biased Boltzmann--Gibbs measure $\widetilde{\mu}_2=\nu_{F_2}$. 

We next build five small autoencoders with the following topology: $D=2$ (input data dimensionality), $2L=2$ layers containing respectively $1$ and $2$ nodes, and $g_1(z)=\tanh(z)$ and $g_2(z)=z$, the identity function. The number of parameters is thus $K=7 $ (including $3$ bias parameters). We use $\tanh$ as the bottleneck activation function in order to obtain a bounded learned collective variable with known bounds $[-1,1]$. 
All five autoencoders are initialized with the same random network parameters.
We then separately train these autoencoders for a maximum of $200$ epochs each as follows:
\begin{itemize}
    \item The first AE is trained on the unbiased trajectory;
    \item The second  AE is trained on the $\xi_1$-biased trajectory;
    \item The third AE is trained on the $\xi_1$-biased trajectory, but with reweighting, meaning that each data point $x^i = (x_1^i,x_2^i)$ contributes to the learning loss with  
$$\widehat{w}_i = N \frac{\mu(x^i)/\widetilde{\mu}_1(x^i)}{\displaystyle \sum_{j=1}^N \mu(x^j)/\widetilde{\mu}_1(x^j)} = N \displaystyle \frac{\mathrm{e}^{-\beta F_1(x^i)}}{\displaystyle \sum_{j=1}^N \mathrm{e}^{-\beta F_1(x^j)}} \text{ .}$$
    \item The fourth  AE is trained on the $\xi_2$-biased trajectory;
    \item The fifth AE is trained on the $\xi_2$-biased trajectory, but reweighted using
$$\widehat{w}_i = N \frac{\mu(x^i)/\widetilde{\mu}_2(x^i)}{\displaystyle \sum_{j=1}^N \mu(x^j)/\widetilde{\mu}_2(x^j)} = N \displaystyle \frac{\mathrm{e}^{-\beta F_2(x^i)}}{\displaystyle \sum_{j=1}^N \mathrm{e}^{-\beta F_2(x^j)}} \text{ .}$$
\end{itemize}
As discussed in Supp.~Mat.~Section~\ref{wnormal}, the multiplication by the number of samples $N$ allows to comply with the default normalization of the weights of the ML package we use. A subset of the data is kept as validation set (here, $10\%$), and early stopping is applied when the validation loss no longer improves for $20$ consecutive epochs.

We compare the obtained encoders using heat maps. The values of the encoded bottleneck functions are computed over the (discretized) space $(x_1,x_2) \in [-2,2]\times[-1,2.5]$. The results are given in Figure~\ref{fig:Exp2}. Note that the aim here is to compare encodings obtained from the different models, and possibly to have insight on how these encodings depend on the coordinates $x_1$ and $x_2$. Let us emphasize in particular that it is for example possible to obtain a function of $x_1$ instead of $x_1$ itself as encoder variable. For this reason, and to simplify the comparison between the obtained CVs, they are renormalized to have a range that is exactly between $0$ and $1$: 
\begin{equation}
\label{eq:renorm_xi}
    \xi_{\text{AE}}^{\text{norm}}(x) = \frac{\xi_{\text{AE}}(x)-\xi_{\text{AE}}^{\min}}{\xi_{\text{AE}}^{\max}-\xi_{\text{AE}}^{\min}} \text{,} 
\end{equation}
where $\xi_{\text{AE}}^{\min}$ and $\xi_{\text{AE}}^{\max}$ are respectively the minimum and maximum values taken by the encoder function~$\xi_{\text{AE}}$ over the $2$-dimensional space $[-2,2]\times[-1,2.5]$.
We draw two important conclusions upon interpreting the obtained results:
\begin{itemize}
    \item First, the encoding learned through unbiased training on the reference data obtained from a long unbiased trajectory (top panel) is a function of $x_1$, which we recall is considered a good CV as it distinguishes the three potential wells. The direction~$x_1$ therefore corresponds to the target unbiased representation that we want to find using the remaining models (in particular with the reweighted learning). The encodings obtained from the reweighted training over the two biased trajectories (middle and bottom right panels) also represent bijective functions of~$x_1$. The encodings obtained through biased unweighted learning, however, encode different variables (middle and bottom left panels). This is therefore an example where reweighting proves necessary for obtaining results that are consistent with unbiased learning. 
    \item It is interesting to note the difference between the CV obtained from biased learning over the $\xi_1$-trajectory (middle left panel) and the one obtained from the $\xi_2$-trajectory (bottom left panel). It is quite clear that the latter is closer to a monotonic function of the target learning result, $x_1$, than the former. These results can be intuitively explained as follows: Free energy biasing changes the distribution along the CV (here $\xi_1$ or $\xi_2$) to a uniform law, making the corresponding direction no longer relevant. This prompts the model to learn other directions that are still relevant in the biased simulation. In the case of the trajectory obtained by biasing with $F_1\circ \xi_1$, the $x_1$ direction is no longer a relevant feature of the data space and the encoder does not recover it. On the other hand,  biasing using $F_2\circ \xi_2 $ does not completely annihilate the relevance of the variable $x_1$ in the sampled data. This explains why the CV learned from biasing with $\xi_2$ is closer to~$x_1$. 
\end{itemize}
This last point is very important as it means that if an iterative learning run is performed without reweighting, then whenever a good CV is learned at iteration~$n$, this CV may be cancelled out in the next iteration, making convergence difficult to achieve. 

\begin{figure}[ht!]
    \centering
    \begin{subfigure}{\linewidth}
        \centering
        \includegraphics[width=0.4\textwidth]{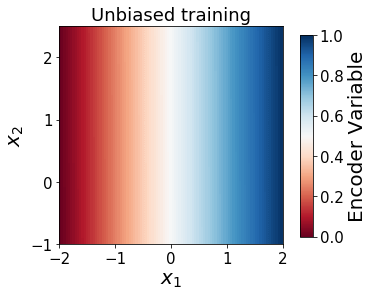}
        \labelnotempty{subfig:unbiased}
        \caption{No biasing}
    \end{subfigure}
    \begin{subfigure}{\linewidth}
        \centering
            \includegraphics[width=0.7\textwidth]{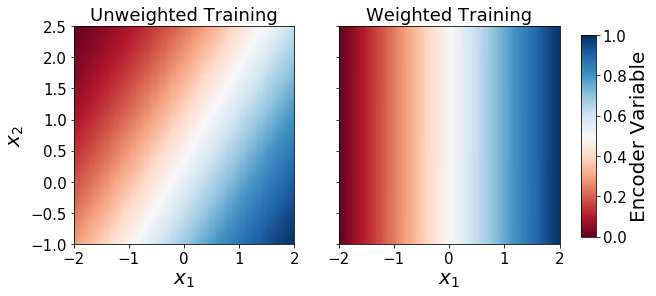}
        \labelnotempty{subfig:biased}
        \caption{Biasing with CV $\xi_1 = x_1$}
    \end{subfigure}
        \begin{subfigure}{\linewidth}
        \centering
            \includegraphics[width=0.7\textwidth]{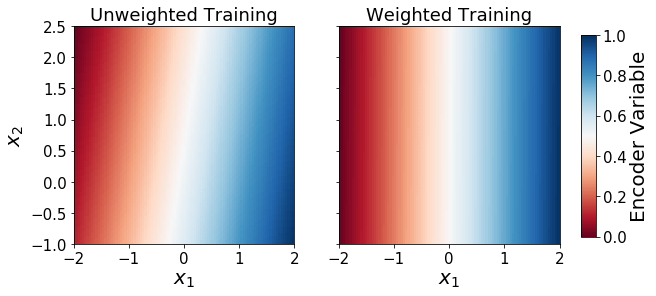}
        \labelnotempty{subfig:biased2}
         \caption{Biasing with CV $\xi_2 = x_2$}
    \end{subfigure}
    \caption{Heatmaps of the values of the encoder variables over the discretized 2D space $[-2,2]\times[-1,2.5]$: (a) Unbiased learning outputs a CV close to $\xi_1(x) = x_1$. (b) and (c) Left: Training on biased simulations leads to learned CVs that depend on the free energy biasing. Right: Reweighting enables learning a CV that is independent of the sampling procedure.}
    \labelnotempty{fig:Exp2}
\end{figure}


\section{Iterative reweighted learning of CVs with autoencoders}
\label{sec:ae-abf}
In this section, we introduce our algorithm for the iterative learning of a CV from biased trajectory data. As mentioned in the introduction, our algorithm is in part inspired by the work done in Refs.~\citenum{mesa1} and~\citenum{mesa2}, where MESA, an iterative method for learning CVs on the fly while performing enhanced sampling was devised. MESA alternates between learning CVs with autoencoders and using those CVs for extrapolation with Umbrella Sampling, until convergence. 

The major difference between our algorithm and MESA is that we use reweighting of the configurations sampled from free energy biasing to target the unbiased loss corresponding to the Boltzmann-Gibbs distribution. Indeed, as observed in the experiments of Section~\ref{subsec:2D}, training on biased data does not yield the same results (same encoded CVs) as with unbiased data. Additionally, we saw that performing free energy biasing using a certain choice of CV may make this direction irrelevant within the sampled data, which prevents the model from learning it. This means first that a naive iterative learning algorithm without reweighting is not theoretically guaranteed to converge, as each new CV (learned with a new AE) is different from the previous one; and second, that this can result in iterations of the algorithm where the learned CV is not necessarily a relevant choice for the next free energy biasing. Reweighting the loss is expected to solve this issue in the iterative method the same way it did in the experiments shown in Section~\ref{subsec:2D}.  At each step, we train the autoencoder on the reweighted loss, using the reweighting introduced in \eqref{lossw}, so as to remove bias from the data sampled in the free energy biasing simulation. 
Another major difference in our approach compared to MESA is that we rely on adaptive techniques to sample configurations and compute the free energy for reweighting them. Let us however emphasize that our approach could also be used with other methods to compute free energy profiles.

We first give in Section~\ref{pipeline} a detailed description of our algorithm. We then introduce in Section~\ref{AB} adaptive biasing methods in general  and the adaptive biasing force method in particular, as it constitutes our choice of biasing procedure. Then, some additions and refinements of the algorithm are  discussed in Section~\ref{transfer} to transfer information between consecutive iterations of the algorithm. Finally,  Section~\ref{2Dres} provides a first application of the algorithm to the 2-dimensional toy example introduced in Section~\ref{subsec:2D}.

\subsection{Iterative algorithm for CV learning: General description}
\label{pipeline}
This section provides a description of the steps of our iterative algorithm, which we call FEBILAE, for "Free Energy Biasing and Iterative Learning with AutoEncoders". The algorithm is also summarized in the pseudo-code of Algorithm~\ref{algo:max}. 
\subsubsection*{Initialization}
The first step of the algorithm is to produce an initial unbiased trajectory to start from, traj$_0$. This requires providing the algorithm with a simulation setup $S$ (dynamics, physical conditions, etc), and an initial configuration $q^0$. 
The trajectory traj$_0$ is preprocessed as necessary (see Section~\ref{subsec:qtox}) to obtain an initial training dataset $(x^1, \dots, x^N)$  where $N$ is the number of samples, provided as an input to the method. As traj$_0$ is unbiased, the sample weights associated to this first training are uniform: $\widehat{w}_j=1$ for all $j\in \{1,\dots,N\}$. A first autoencoder, AE$_0$ is then initialized (with a given topology, hyper-parameters and random parameters $A_{\text{init}}$) and trained on this dataset.
A first collective variable, $\xi_0$, is thus computed from AE$_0$. 
\subsubsection*{Iterative procedure and stopping rule}
At each iteration $i \geq 1$ of the algorithm, the following steps are performed: 
\begin{itemize}
    \item \textbf{New trajectory.} The previous CV  $\xi_{i-1}$ is used to perform adaptive free energy biasing under a given simulation method and  setup~$S_{\text{AB}}$, and starting from the same initial configuration $q^0$ as the one used in the initialization step, or from a new initial configuration (e.g. the last sampled configuration of the previous trajectory). The biased trajectory traj$_i$ is saved, along with the estimation of the free energy $F_i$ corresponding to the CV $\xi_{i-1}$. 
    \item \textbf{Preprocessing and sample weight computation.} The trajectory is preprocessed to obtain the new training dataset $(x^1,\dots, x^N)$, and the weight of each sampled datapoint is computed as defined in~\eqref{hatw} and~\eqref{lossw} (with the weights multiplied by $N$ in order to comply with the default setting of the ML package used for implementation , see Supp.~Mat.~Section~\ref{wnormal}).  
    
    \item \textbf{Autoencoder training and new CV.} A new autoencoder, with the same topology and initial parameters as in the previous iterations ($A_{\text{init}}$), is trained on the new dataset traj$_i$ to optimize the reweighted loss~\eqref{lossw}. As mentioned in Section~\ref{subsec:autoenc}, the autoencoder has the advantage of outputting a  mapping $\xi_i$ from the input configuration to the collective variable, as well as its gradient, which is also required for most CV biasing methods. Additionally, the interval(s) of the CV, over which biasing is applied, must be determined. These intervals can be estimated using the extreme values which the CV takes over the training data traj$_i$. 
    \item \textbf{Stopping rule.} The algorithm stops when one of the two following conditions is met: either the algorithm has reached a given maximum number of iterations $I_{\max}$, or the CV has converged. The last step of each iteration thus requires to assess CV convergence. In this paper, we consider that the learned CV has converged when the new CV $\xi_i$ can be mapped (using e.g. a simple regression model) to the previous CV $\xi_{i-1}$, with a high enough precision score, higher than a given threshold $s_{\min}$. We use linear regression as a mapping model to check whether $\xi_i \approx \Phi(\xi_{i-1})$. We refer to Supp.~Mat.~Section~\ref{appendix:regscore} for a more detailed discussion on how to compute the regression score. The value of $s_{\min}$ depends on the system under study, the complexity and dimension of the computed CVs, and the regression model used to compute the regression score. It is determined separately for each of the systems studied in this paper, using a bootstrap procedure made precise in Supp.~Mat.~Section~\ref{appendix:regscore}. 
    When the algorithm stops, the last learned CV is kept as the final CV. Its free energy is estimated and used to fully sample the configurational space of the system. 
    
\end{itemize}

\begin{remark}[CV dimensionality]
\label{cvdimknee}
When the optimal dimensionality of the CVs is unknown, one can use the following method proposed in Ref.\citenum{mesa1} to determine it: several autoencoders $\Phi_1,\dots,\Phi_M$, with different values for the dimensionality~$d$ of the bottleneck layer ($d \in \{1,\dots,M\}$), are all trained at each iteration. The optimal value of~$d$ is then determined by plotting the FVE (fraction of variance explained) $  1 - \frac{\sigma_{\mathrm{res}}(d)}{\sigma_{\mathrm{tot}}}$ as a function of $d$, where 
\[
\sigma_{\mathrm{res}}(d) = \sum_{i=1}^N \|x^i - \Phi_d(x^i)\|^2,
\]
is the residual sum of squares and
\[
\sigma_{\mathrm{tot}} =  \displaystyle \sum_{i=1}^N \|x^i - \bar{x}\|^2, 
\quad 
\bar{x} = \frac{1}{N}  \displaystyle \sum_{i=1}^N x^i \text{,}
\]
is the total sum of squares. The optimal value of $d$ corresponds to a plateau or "knee" in the FVE curve, meaning that no considerable improvement in the optimized loss is obtained by adding another dimension to the bottleneck space. This is similar to what is done in PCA, where spectral gaps in the distribution of eigenvalues are used to determine the optimal number of principal components to keep. 
\end{remark}

\begin{algorithm}
\DontPrintSemicolon 
\KwIn{Initial condition $q^0$, autoencoder topology and initialization parameters $A_{\text{init}}$, number of samples  $N$, simulation procedure $S$ and adaptive biasing procedure $S_{\text{AB}}$, maximum number of iterations $I_{\text{max}}$, minimum convergence score $s_{\text{min}}$.}

\KwOut{CV $\xi_{\text{final}}$ and corresponding PMF $F_{\text{final}}$. }

  Sample traj$_0 \gets S(q^0, N)$. 
Initialize autoencoder AE$_0 \gets A_{\text{init}}$.

Train AE$_0$ on traj$_0$ with weights $(\widehat{w}_0,\dots,\widehat{w}_N) = (1, \dots 1)$.

Extract the encoder function $\xi_0: x \mapsto \xi_0(x)$.

Set $i \gets 0$,  $s \gets 0$.

\While {$i < I_{\rm{max}}$ \& $s < s_{\rm{min}}$} {
   Set $i \gets i+1$.
   
   Sample (traj$_i$, $F_i) \gets S_{\text{AB}}(q^0, N, \xi_{i-1})$. 
   
   Compute weights $(\widehat{w}_j)_{1\leq j\leq N}$ as $\widehat{w}_j = N \displaystyle \frac{ \text{e}^{-\beta F_i(\xi_{i-1}(x^j))}}{\displaystyle \sum_{n = 1}^N \text{e}^{-\beta F_i(\xi_{i-1}(x^n))}} $.
   
   Initialize autoencoder AE$_i \gets A_{\text{init}}$.
   
   Train AE$_i$ on traj$_i$ with sample weights $(\widehat{w}_j)_{1\leq j\leq N}$.
   
   Extract the encoder function $\xi_i: x \mapsto \xi_i(x)$.
   
   Set $s \gets$ regscore$(\xi_{i-1}, \xi_i)$.
}
\vspace{4pt}
Set $\xi_{\text{final}} \gets \xi_{i}$.

Sample traj$_\text{final}$, $F_{\text{final}} \gets S_{\text{AB}}(q^0, N_{\text{final}}, \xi_{\text{final}})$ with $N_{\text{final}}$ large enough to ensure PMF convergence.

\caption{FEBILAE}
\label{algo:max}
\end{algorithm}

\subsection{Sampling with the (extended) Adaptive Biasing Force method}
\label{AB}
With the exception of low dimensional systems ($D \leq 3$) such as the example used in Section~\ref{subsec:2D}, the free energy of a system associated with a given CV cannot be easily estimated. Adaptive sampling algorithms replace the actual free energy $F$ by an estimated function $F_t$ in the biased dynamics at time $t$, meaning that the potential becomes $ V-F_t\circ \xi$. The estimate $F_t$  is updated on-the-fly so that it converges to $F$ as the molecular dynamics simulation proceeds. 
The two categories of adaptive biasing techniques are~\cite{lelievre2007computation} Adaptive Biasing Potential (ABP) methods, where the free energy $F_t$ is estimated and its gradient, the so-called mean force, is then used in the dynamics; and Adaptive Biasing Force (ABF) methods where the mean force is estimated directly, and the free energy is subsequently obtained by numerical integration through a Helmholtz projection~\cite{helmholtz}.

In this work, we choose to work with ABF, and more precisely extended system ABF (eABF)~\cite{tony} for free energy biasing. We call this version of the algorithm AE-ABF.  Section~\ref{ABF} gives a brief description of ABF. We then recall its main limitation, namely that it requires the computation of the second order derivatives of the used CVs. This motivates using eABF, which is described in Section~\ref{eABF}.
 
\subsubsection{Adaptive biasing force}
\label{ABF}

The adaptive biasing force method estimates the mean force associated with the collective variable $\xi$. For ease of notation, we assume in this section and the following one that $\xi$ is one dimensional, with values in $\mathcal{A} \subseteq \mathbb{R}$, but the generalization to a higher dimensional CV is straightforward.
By differentiating both sides of Equation~\eqref{F(z)}, one obtains:
\begin{equation} 
\label{eq:MF} 
F'(z) = -\frac{1}{\beta} \int_{\Sigma(z)} f(q) \frac{\mathrm{e}^{-\beta V(q)} \delta_{\xi(q)-z} (dq)}{\mathrm{e}^{-\beta F(z)}}  \text{ ,}
\end{equation}
where $f$ is called the local mean force: 
\begin{equation}
\label{locmeanforce}                                                             
f =\nabla V\cdot \frac{\nabla\xi}{|\nabla\xi|^2} -\frac{1}{\beta} \text{div} \left(\frac{\nabla\xi}{|\nabla\xi|^2}\right) \text{ .}
\end{equation}
Note that the equation above contains second order derivatives of $\xi$. 
Equation~\eqref{eq:MF} shows that the derivative of the free energy $F'(z)$ is related to the conditional average of the local mean force as 
\begin{equation}
\label{meanforce}
F'(z) = \mathbb{E}_{\nu} \left(f(q) | \xi(q)=z \right) \text{ ,}
\end{equation}
with~$\nu$ defined in~\eqref{BGPos}. This suggests, for instance, that the biasing force at time $t$ can be estimated as follows when considering a single long trajectory:
$$ \Gamma_t(z) = \frac{\int_0^t f(q_s)~\delta^{\varepsilon}\left(\xi(q_s)-z\right)ds}{\int_0^t \delta^{\varepsilon}\left(\xi(q_s)-z\right)ds} \text{ ,} $$
where $\varepsilon>0$ and $\delta^{\varepsilon}$ is an approximation of the Dirac mass. 

The estimated mean force is then used to bias the dynamics with the force $\Gamma_t(\xi(q_t))$. For example, the dynamics of ABF used in conjunction with Langevin dynamics is:
\[
\left\{
\begin{aligned}
    dq_t &= M^{-1}p_t~dt, \\
    dp_t &= -\nabla V(q_t)dt +\Gamma_t(\xi(q_t))\nabla \xi(q_t) dt - \gamma p_t dt +\sqrt{2\gamma M \beta^{-1}} dB_t \text{,}
\end{aligned}
\right.
\]
where $M$ is the mass matrix and $\gamma>0$ is the friction coefficient.

\subsubsection{Extended system Adaptive biasing force}
\label{eABF}
Equation~\eqref{locmeanforce} shows that regular ABF requires the knowledge of second order derivatives of the CV $\xi$ to compute the local mean force $f$. The analytical expression of this quantity is quite cumbersome for most choices of CVs, especially when $\xi$ is vector valued. In particular, as our CVs are encoder based mappings and can involve complex non-linear activation functions, extracting the second order derivatives is not a viable option. 

To overcome the limitations in computing the second term of~\eqref{locmeanforce}, a method coined extended system ABF (eABF)\cite{tony} was devised. A fictitious degree of freedom $\lambda$ is added to the configurational space and the new potential is:
\begin{equation}
\label{eq:Vext}
V^{\text{ext}}(q,\lambda) = V(q) + \frac{\kappa}{2} |\xi(q)-\lambda|^2 \text{ ,}
\end{equation}
 where $\kappa$ is a force constant. The collective variable $\xi^{\text{ext}}(x, \lambda) = \lambda$ is used instead of the original CV $\xi$. Using Equation~\eqref{F(z)}, the resulting free energy~$F^{\text{ext}}$ is a convolution of a Gaussian kernel and the free energy associated with $\xi$: 
 $$\mathrm{e}^{-\beta F^{\text{ext}}(\lambda)} = \int_{\mathcal{A}} \chi_{\kappa}(\lambda-z)\mathrm{e}^{-\beta F(z)}dz \text{ ,}$$ 
where $\chi_{\kappa}$ is a Gaussian kernel with variance $1/(\kappa\beta)$. When~$\kappa \xrightarrow{} \infty$,~$F^{\text{ext}}(\lambda)$ converges to~$F(\lambda)$. Thus, a simple estimator of the real free energy $F$ is to use $F = F^{\text{ext}}$ directly. This naive estimator is biased of course, given that in practice $\kappa < \infty$. 
The new extended mean force does not depend on second order derivatives of the CV~$\xi$: Only the gradient of~$\xi$ is needed for computing the gradient of~$V^{\text{ext}}$. 
ABF can therefore easily be applied to the new extended system.

Denoting by $\rho$ the momentum of $\lambda$, and by $M^{\text{ext}}$ the extended mass matrix (which includes~$m_{\lambda}$, the fictitious mass of $\lambda$), the Langevin dynamics of the eABF trajectory are: 
\begin{equation}
\label{eabfdyn}
\left\{
 \begin{aligned}
    dq_t^{\rm{ext}} &= (M^{\text{ext}})^{-1}p_t^{\rm{ext}}~dt \text{,} \\
    dp_t^{\rm{ext}} &= \left(-\nabla V^{\rm{ext}}(q_t, \lambda_t)+\Gamma_t^{\rm{ext}}(\lambda_t)\mathbf{u}\right)dt - \gamma p_t^{\rm{ext}} dt +\sqrt{2\gamma \beta^{-1}M^{\text{ext}}} dB_t \text{,}
\end{aligned}
\right.
\end{equation}
where $q_t^{\text{ext}} = (q_t,\lambda_t)$, $p_t^{\text{ext}} = (p_t, \rho_t)$, $\mathbf{u}^T = (0,\dots,0,1)$ and $\Gamma_t^{\text{ext}}$ is the estimate at time $t$ of the mean force associated with $\lambda$ in the extended system:
\begin{align*}
 \Gamma_t^{\text{ext}}(\lambda) = \frac{\int_0^t f^{\text{ext}}(q_s)~\delta^{\varepsilon}\left(\lambda_s-\lambda\right)ds}{\int_0^t \delta^{\varepsilon}\left(\lambda_s-\lambda\right)ds} \text{ ,}
 \qquad 
 f^{\text{ext}}(q) = \frac{\partial V}{\partial \lambda}(q) = \kappa (\lambda-\xi(q)) \text{.}
\end{align*}
In practice, the above equation is discretized in time using a timestep $\Delta t$. Denoting by $q^{\text{ext},j},p^{\text{ext},j}$ the approximations of $q_{j\Delta t}^{\text{ext}},p_{j\Delta t}^{\text{ext}}$ at iteration $j$, the estimated mean force $\Gamma_t^{\text{ext}}$ is considered to be constant in discrete bins of $\xi^{\text{ext}}=\lambda$ centered around points ${z_1 \dots, z_k}$ (uniformly spaced, for simplicity of presentation): for all $\ell \in \{1 \dots, k\}$, at time $j\Delta t$,
\begin{equation}
\label{dicreteMF}
\forall z \in [z_\ell-\varepsilon,z_\ell+\varepsilon[ , \qquad  \Gamma_j^{\text{ext}}(z) = \frac{\sum_{i=0}^j f^{\text{ext}}(q^{\text{ext},i})\mathds{1}_{\{z_\ell-\varepsilon\leq \lambda^i < z_\ell+\varepsilon\}}}{\sum_{i=0}^j \mathds{1}_{\{z_\ell-\varepsilon\leq \lambda^i < z_\ell+\varepsilon\}}}  \text{,}
\end{equation}
where $2\varepsilon = \frac{z_k-z_1}{k-1}$.

\subsection{Transferring information between algorithmic iterations}
\label{transfer}
This section discusses two ways of using part of the information learned in previous iterations to possibly improve or accelerate the learning or sampling in the next iteration. Following the notation in Algorithm~\ref{algo:max}, the iteration index is denoted by $i$.

\subsubsection{Using previous trajectories}
\label{previoustraj}
At each iteration, the training dataset can be a combination of a fixed number~$n_T \geq 1$ of previously sampled trajectories. Standard FEBILAE corresponds to $n_T=1$. Using more than one trajectories provides the autoencoder with a larger, and therefore possibly more complete dataset. This refinement of the algorithm is very straightforward, and is thus directly included in the implementation used to obtain the results presented in Section~\ref{dialares} with $n_T=2$. The results of this simple addition on the 2-dimensional example as compared to the basic algorithm are presented in Supp.~Mat.~Section~\ref{appendix:prevtraj}.

\subsubsection{Free energy initialization}
\label{previousFE}
The algorithm AE-ABF as presented above starts each new eABF with a free energy profile initialized to $0$.
However, as the CV progressively converges to an optimal value throughout the algorithm, the successive CVs from one iteration to the other are somewhat similar. Therefore, the free energy profile at the end of the previous iteration of eABF could be used to suggest a better initialization of the free energy or mean force for the new eABF run.

In order to make this discussion more precise, let us assume here again that the CV is one dimensional for the simplicity of exposition, although our approach can be extended to CVs of dimension larger or equal to~$2$. At iteration~$i$, we call $\xi_i: \mathcal{X} \xrightarrow{} [a_i,b_i]$ the CV used in~\eqref{eq:Vext} to perform the eABF simulation for this round, which we call $\text{eABF}_i$. The corresponding estimated free energy and mean force are respectively denoted by $F_i$ and $F'_i$. 
Before starting $\text{eABF}_{i+1}$, we want to determine a function  $\displaystyle \widetilde{F}_{i+1}$ (respectively $\widetilde{F}_{i+1}'$) as a good initialization of the free energy (respectively the mean force) using the previous free energy $F_i$ and the CVs~$\xi_i$ and~$\xi_{i+1}$. We first assume that the CVs have converged, namely that:
$$ \xi_{i+1} = \Phi \circ \xi_i \text{,}$$
for some strictly monotonic function $\Phi \in C^1([a_i,b_i])$.
We can then easily calculate $F_{i+1}$ from $F_i$ under this assumption. Indeed, for $Z = \Phi(z)$,
$$\mathrm{e}^{-\beta F_{i+1}(Z)} = \int_{\Sigma_{i+1}(\Phi(z))} \mathrm{e}^{-\beta V(q)} \delta_{\xi_{i+1}(q)-\Phi(z)}(dq) \text{,} $$
where
$$
\Sigma_{i+1}(\Phi(z)) = \{q \in \mathbb{R}^D | \xi_{i+1}(q) = \Phi(z) \} 
					  = \{q \in \mathbb{R}^D | \Phi(\xi_i(q)) = \Phi(z) \} 
					  = \Sigma_i(z) \text{.}
$$
By the co-area formula (Equation (3.14) in Ref.~\citenum{LRS10})
$$
\delta_{\xi_{i+1}(q)-\Phi(z)}(dq) = \frac{|\nabla\xi_i(q)|}{|\nabla\xi_{i+1}(q)|} \delta_{\xi_i(q)-z}(dq) = \frac{1}{|\Phi'(z)|} \delta_{\xi_i(q)-z}(dq) \text{.}                       
$$
Therefore, 
\begin{equation}
\mathrm{e}^{-\beta F_{i+1}(\Phi(z))} = \int_{\Sigma_i(z)} \mathrm{e}^{-\beta V(q)} \frac{1}{|\Phi'(z)|} \delta_{\xi_i(q)-z}(dq) = \frac{1}{|\Phi'(z)|} \mathrm{e}^{-\beta F_i(z)} \text{.}
\label{extrapolation}
\end{equation}
The above equation provides an expression of $F_{i+1}$ from $F_i$ when $\Phi$ is known. In practice, we approximate $\Phi$ as an affine function using a linear regression model optimized using the datapoints sampled from the trajectory of $\text{ABF}_i$ (and possibly also from previous trajectories as suggested in Section~\ref{previoustraj}). This provides an approximate mapping: 
$$
\xi_{i+1}(x) = \omega_1\xi_i(x)+ \omega_2 \text{,}
$$
for which the following equality holds: 
$$
 \forall Z \in [a_{i+1},b_{i+1}], \qquad   \mathrm{e}^{-\beta F_{i+1}(Z)} = \frac{1}{|\omega_1|}\mathrm{e}^{-\beta F_i\left(\frac{Z-\omega_2}{\omega_1}\right)} \text{,}
$$
and thus 
\begin{equation}
 \forall Z \in [a_{i+1},b_{i+1}] , \qquad   F_{i+1}(Z) =F_i \left(\frac{Z-\omega_2}{\omega_1}\right) +\frac{\ln(\omega_1)}{\beta} \text{.}
    \label{cvmappreg}
\end{equation}
Note that the additive constant $\ln(\omega_1)/\beta$ on the right hand side of the previous equality is irrelevant, since free energies are defined up to an additive constant.

\paragraph{Discretization.} In practice, the ranges of $\xi_i$ and $\xi_{i+1}$ are discretized into $k$ bins centered at~$\{z_1, \cdots, z_k\}$ and~$\{Z_1, \cdots, Z_k\}$. The centers of the bins are matched as follows: if $\omega_1>0$, then $z_\ell$ is associated with $ Z_\ell$; while if $\omega_1<0$, then $z_\ell$  is associated with $Z_{k-\ell+1} $. The initialization of the free energy $F_{i+1}$ is thus computed for each bin $\ell'$ of $\xi_{i+1}$ using its final value in the  corresponding bin $\ell$ of $\xi_i$, and Equation~\eqref{cvmappreg}. In practice, because ABF uses the mean force rather than the free energy, it is the mean force at each bin, $Z_{\ell'}$ that is initiated using the final estimate of the mean force in bin $z_\ell$, i.e the value of the numerator of Equation~\eqref{dicreteMF} for time $N\Delta t$.
Additionally, the number of samples for each bin of $\xi_{i+1}$ is initialized with the values of the previous ABF run in the same manner (i.e. according to the sign of $\omega_1$, using the final values (at time $N\Delta t$) of the denominator in Equation~\eqref{dicreteMF}). The new ABF simulation, $\text{ABF}_{i+1}$ is then actually equivalent to continuing an ABF run with the CV $\xi_{i+1}$, instead of starting a new simulation. 

Some results obtained with the free energy initialization scheme described in this section are given in Supp.~Mat.~Section~\ref{appendix:feinit}.

\subsection{Toy two-dimensional example}
\label{2Dres}
In order to illustrate the methodology developed in this section, we come back to the 2 dimensional example introduced in Section~\ref{subsec:2D}, with the same parameters as in this section unless otherwise specified. For this toy example, the routines for unbiased simulations as well as eABF simulations were implemented in python using the \texttt{numpy} module.

We demonstrate on this simple 2D example that reweighting the sampled configurations leads to a fast convergence to the CV $\xi_1(x) = x_1$ (which is the CV obtained by training on a long unbiased trajectory), whereas not reweighting can lead the algorithm to be trapped, continually learning different CVs between consecutive iterations. 
We start with the initial unbiased simulation, which is purposedly stopped before it crosses the first energy barrier: The trajectory is simulated starting from the initial condition $q^0=(-1,0)$, for $2 \times 10^7$ timesteps, with $\Delta t= 10^{-3}$, keeping $1$ out of $50$ datapoints. The time horizon of the simulation is thus $T = 2 \times 10^4$, and the  dataset contains $N = 4\times 10^5$ samples belonging to only one metastable state.

We then use this dataset to start two separate learning frameworks. The first one uses reweighting at each iteration as discussed in Section~\ref{pipeline}. The second one does not use reweighting before achieving a new round of training (i.e. the weights $\widehat{w}_j$ in~\eqref{lossw} are all set to $1$). All autoencoders are initialized with the same parameters at each iteration, and are trained using $80\%$ of the data for training (and $20\%$ for validation), with batch size $b=400$, and $N/b = 8 \times 10^2$ steps per epoch for a maximum of $100$ epochs with early stopping when the validation loss does not improve for $20$ consecutive epochs. 
At each new iteration, for both schemes, eABF is performed for $1.2 \times 10^6$ timesteps with $\Delta t = 10^{-3}$, and $1$ in every $3$ configurations are kept.  The time horizon of the simulation is thus $T = 1.2 \times 10^3$ and the datasets all contain $N = 4\times 10^5$ samples. eABF is run with a force constant $\kappa=50$. Note again that the biased simulations have a smaller time horizon than the unbiased simulation, since biasing accelerates the exploration of all states of the configurational space. The CV intervals are determined using the range $[z_{\min}, z_{\max}]$ of the values the CV takes over the training data. 
Regardless of the chosen CV interval, we use a grid of $200$ bins to discretize the CV. Each simulation is started at the same point $(x_1,x_2) = (-1,0)$. The estimated mean force over a certain bin is only applied after 100 samples are collected inside this bin. Finally, the CV gradients needed for ABF are extracted from the autoencoders using \texttt{keras} functions.

Figure~\ref{MESA2D} illustrates heat maps of the encodings obtained at each iteration for both schemes. All CVs were renormalized using Equation~\eqref{eq:renorm_xi} to have a range within~$[0,1]$. As shown in Figure~\ref{MESA2D} left, after only two iterations, the reweighted model finds the CV $x_1$ and stabilizes. 
The biased model, on the other hand, learns a different direction at each new iteration, so that the learned CV does not converge, but rather seems to oscillate between two different functions. 

\begin{figure}[h!]
    \centering
    \begin{subfigure}{0.37\linewidth}
        \centering
        \includegraphics[width=\textwidth]{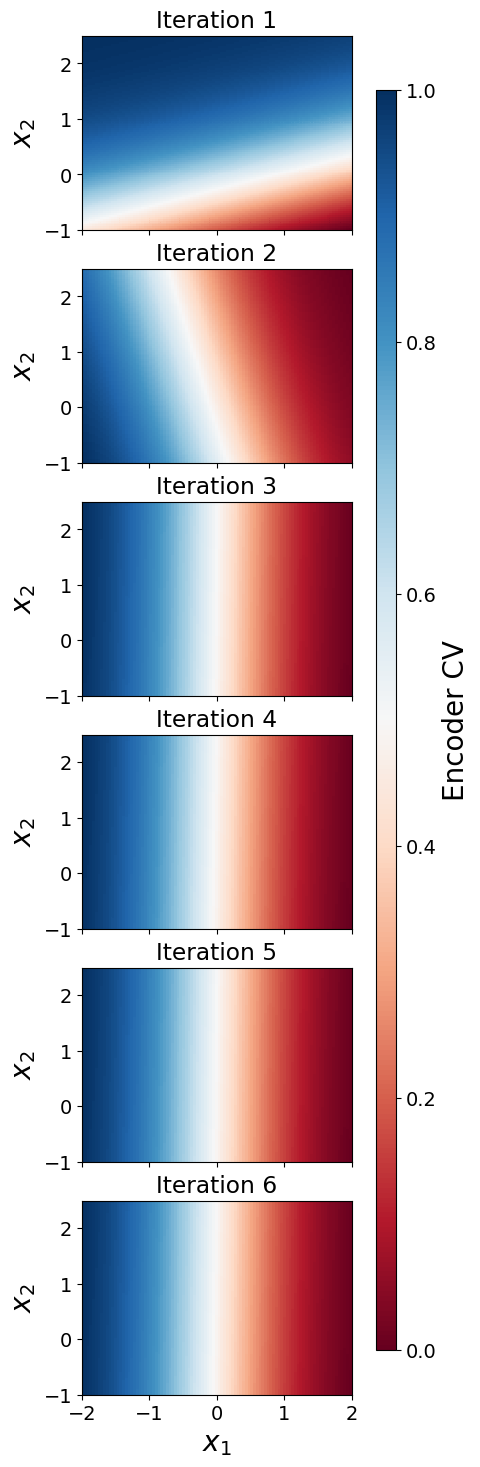}
        \labelnotempty{mesaw}
    \end{subfigure}
    ~~~~~~~~~
    \begin{subfigure}{0.37\linewidth}
        \centering
            \includegraphics[width=\textwidth]{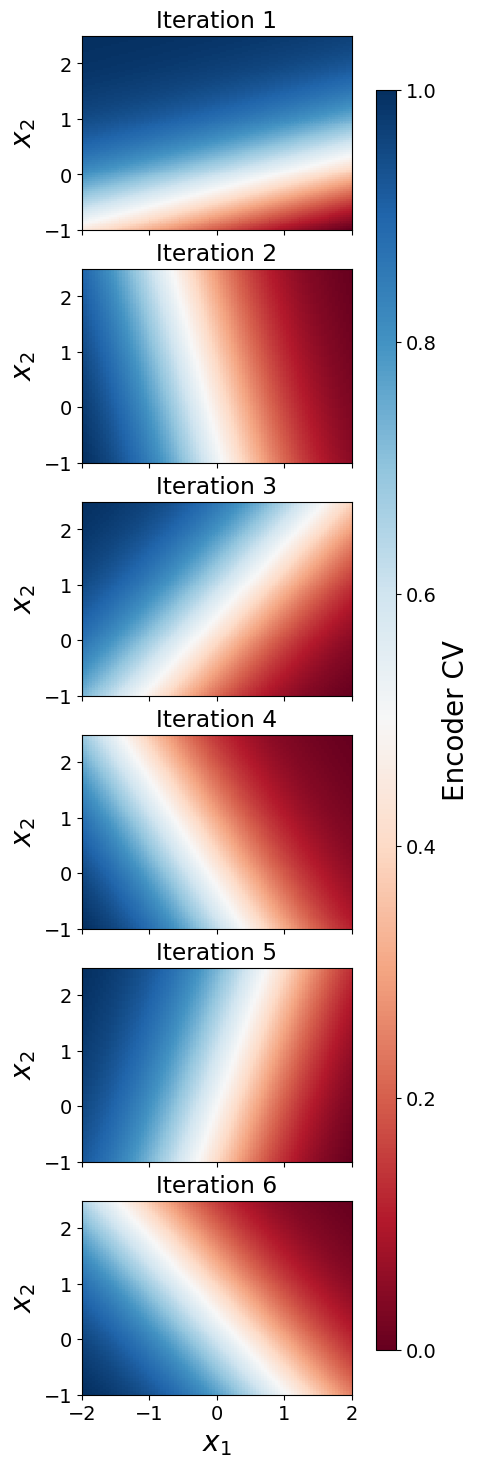}
        \labelnotempty{mesanw}
    \end{subfigure}
    \caption{Encodings obtained at each iteration of AE-ABF with (left) and without (right) reweighting. Left: Reweighting guides learning to the same CV at each iteration, and the algorithm quickly converges to the CV $\xi_1(x) = x_1$. Right: The simulations are biased using a different free energy at each iteration. The CV does not stabilize. }
    \labelnotempty{MESA2D}
\end{figure}
\section{Computational details}
\label{subsec:matmet}
\subsection{Software packages and libraries}
\label{subsec:tools}
We use the \texttt{keras}~\cite{keras} library to wrap the \texttt{tensorflow}~\cite{tensorflow} neural network module in \texttt{python}. 

In order to construct an entirely automated code, all molecular simulations were performed with the \texttt{openmm} software~\cite{openmm} within its \texttt{python} API. The adaptive biasing and collective variable analysis were mainly performed using  \texttt{plumed}~\cite{plumed1,plumed2,plumedeabf}. To link these two modules, we used the \texttt{openmmplumed} plugin module~\cite{ommplumed}. Additionally, the \texttt{colvar}~\cite{colvar} \texttt{abf integrate} utility was used to recover the potential of mean force from the gradients computed by eABF. 

Our goal is to have an implementation that is automated and entirely runnable from \texttt{python}. Consequently, our implementation is practical and easy to use, but not necessarily computationally optimized. 

\subsection{Alanine dipeptide in vacuum}
\label{subsec:implem}

\subsubsection{Parameters}
\label{setup}
The algorithm was implemented with the following ML and MD parameters.
\begin{itemize}
\item \textbf{Machine Learning.} The input features are chosen as the Cartesian coordinates of only the backbone atoms of alanine dipeptide, instead of the complete peptide, making the input data $x$ of dimension $D=3 \times 8 = 24$. Structural rotational alignment (using the Kabsch algorithm~\cite{Kabsch}) to a reference configuration and re-centering are included in preprocessing to respectively eliminate rotational and translational invariances. Note that the same reference structure, namely a configuration which falls within the C$5$ state, is used for all our experiments. 

All autoencoders used for AE-ABF runs have the same symmetrical topology: In addition to the input layer of dimension $D=24$, the encoder contains two fully connected layers, of respective dimensions $G=40$ and $d=2$ (computed using the FVE curve as explained in Remark~\ref{cvdimknee}), and all activation functions are $\tanh$.
This topology is the same as the one used by the authors of MESA\cite{mesa1, mesa2} on the same system. Note however that additional experiments were done with $G=8$ instead. The obtained results (namely the learned CV) were the same.

We use the \texttt{adam}~\cite{Adam} optimization algorithm. The model is trained for a maximum of $2000$ epochs, and early stopping~\cite{earlystopping} is used to stop training when the validation loss stops decreasing for $50$ consecutive epochs, making the actual number of epochs in practice approximately $300$.  The learning rate used is always $\eta = 10^{-3}$.

At each new iteration of the algorithm, both the new trajectory and the previous one are combined to construct the training set. As shown in Supp.~Mat.~Section~\ref{appendix:prevtraj} (for the $2$D example), this can help the algorithm converge with smaller simulation times, as well as possibly speed up the convergence. 
 
\item \textbf{Molecular Dynamics.} All simulations (biased and unbiased) are performed using the Amber ff99SB forcefield~\cite{amber99sb}, under Langevin dynamics at a temperature $T=300$~K,  non periodic conditions with cutoff $r_{\rm{c}}=1$nm, a friction coefficient $\gamma=1~\text{ps}^{-1}$, no bond constraints and a timestep $\delta t = 1$~fs. For consistency, the same peptide configuration is used as input in all runs, this configuration is obtained after $500$ steps of energy minimization.

For eABF simulations, collective variable biasing intervals are selected at each iteration of the AE-ABF algorithm as the ranges (along the two CV directions) of the CV values obtained from the encoders on the training data. A grid of a fixed value of $50$ bins for each of the two dimensions of the CV is used. The force constant $\kappa$ is kept to its default \texttt{plumed} value: $\kappa\beta= \displaystyle \frac{1}{ (\delta z)^2} $, where $\delta z$ is the grid spacing. The fictitious mass $m_{\lambda}$ is set to $m_{\lambda} =\kappa \displaystyle \left(\frac{\tau }{2\pi}\right)^2$, where $\tau=0.5$ is the default value for the relaxation time in \texttt{plumed} (also called extended time constant in \texttt{colvars}). The estimated biasing force is applied to a sampled point after a minimum of~$500$ samples are collected in the corresponding bin. The gradients of the CVs are numerically estimated in \texttt{plumed} instead of being extracted from the autoencoders. Additionally, \texttt{plumed} ensures a correct propagation of all calculated forces with respect to any translational and rotational alignments applied to the system. 

\end{itemize}

\subsubsection{Simulation speed}
\label{speed}
Under the setup described in Sections~\ref{subsec:tools}~and~\ref{setup}, unbiased simulations ran with a speed of $\sim 1500$ns/day, while biased eABF simulations using encoder CVs  had a speed of $\sim 200$ ns/day. As a comparison, biased simulations using regular CVs of the system (namely the dihedral angles $\Phi$ and $\Psi$) ran at a speed of $\sim 1000$ ns/day. The computation of the encoder CVs and their approximate derivatives in \texttt{plumed} is thus computationally quite expensive compared to the other steps of the algorithm for this low dimensional system in vacuum. Note that for larger systems, in particular solvated systems, the overhead will be much smaller in proportion. Let us also emphasize again that no specific effort was made to obtain a more efficient computational chain, as the focus of this work is primarily methodological. 

\subsection{Chignolin in explicit solvent}
\label{setupchigno}
Chignolin, a 10-residue miniprotein with a beta-hairpin structure, is another well-studied system, particularly for its distinct folding states captured using MD at accessible timescales~\cite{howffpf, chignostate}. Lacking an X-ray for the wild-type form, herein, we use the CLN025 mutant (PDB ID: 5AWL) in explicit solvent as a more realistic and challenging system of our approach. After the addition of hydrogen atoms, the system contains $166$ atoms.
\subsubsection{Parameters}
\label{chignoparam}
For chignolin, the AE-ABF algorithm was implemented using the following parameters: 
\begin{itemize}
\item \textbf{Machine Learning.} The selected input features are the Cartesian coordinates of the $C{\alpha}$ atoms of the chignolin miniprotein. The input data is thus of dimension $D=3\times 10=30$. Similarly to alanine dipeptide, structural rotational alignment to a reference structure, and centering are used to eliminate rotational and translational invariances. The reference structure used is a folded state conformation. 

The autoencoders used for AE-ABF runs have an encoder made of two fully connected layers, of respective dimensions $G=10$ and $d=2$. All activation functions are tanh. The \texttt{adam} optimization scheme is used for a maximum of $2000$ epochs with early stopping after no improvement over $50$ consecutive epochs. The learning rate used is $\eta = 5\times 10^{-4}$. 

As for alanine dipeptide, each new autoencoder training uses both the new and the previous trajectory as a combined dataset.

\item \textbf{Molecular Dynamics.} Chignolin is solvated using a water box with a padding of at least $1.2$ nm. Na$^{+}$ and Cl$^{-}$ ions are added to render the system neutral. All simulations are performed using the Amber ff99SB-ILDN forcefield~\cite{amber99sbildn} and the TIP3P water model~\cite{tip3Pwater}, under  Langevin dynamics at temperature $T = 340 $K and friction coeficient $\gamma=1$ ps$^{-1}$, with a timestep $\delta t=2$ fs. Nonbonded interactions are computed with the particle mesh Ewald with cutoff $r_{\text{c}}=1$ nm, and constrained hydrogen bonds (using the LINCS algorithm~\cite{lincs}). All simulations are performed in the NVT ensemble. The initial configuration for the simulations is obtained after a performing $1000$ steps of energy minimization. 
For eABF simulations, collective variable biasing intervals are selected at each iteration of AE-ABF as the ranges (along the two CV directions) of the CV values obtained from the encoders on the training data. The variables are then rescaled to have a range of $[-1,1]$. A grid of $50$ bins with equal sizes is used for each of the two dimensions of the CV. All other parameters are set to the same values as for alanine dipeptide, with the exception of the force constant which was manually set to $\kappa\beta=353.74$. This change to a lower value of $\kappa$ was made to ensure the stability of the biased simulations. the other \texttt{plumed} parameters are the same as the ones used on alanine dipeptide.
\end{itemize}

\subsubsection{Simulation speed}
In the case of solvated chignolin, and under the computational framework described in Sections~\ref{subsec:tools}~and~\ref{chignoparam}, unbiased simulations ran at a speed of $500-600$ns/day, while biased simulations (using autoencoder CVs) ran at $50-70$ns/day. 

\section{Results}
\label{sec:results}
This section shows results of the AE-ABF algorithm applied to the systems of alanine dipeptide in vacuum and solvated chignolin. 
\subsection{Alanine dipeptide in vacuum}
\label{dialares}
Conventionally the 2D intrinsic manifold of alanine dipeptide in vacuum is described by the $\Phi$ and $\Psi$ backbone dihedrals~\cite{dialaCVs, dialadihedre}. As mentioned in Remark~\ref{cvdimknee}, we have used the plateau method to check that the bottleneck optimal dimensionality is indeed $d=2$. See Supp.~Mat.~Section~\ref{appendix:diala} where we provide the FVE curve computed using a long unbiased trajectory of alanine dipeptide.

We first present in Section~\ref{groundtruth} the ground truth CV, which is the CV obtained from training an autoencoder over a long unbiased simulation. The quality of this CV is assessed by measuring its ability to recover the free energy landscape of the dihedral angles $\Phi,\Psi$. Section~\ref{aeabfondiala} contains the results of applying AE-ABF to alanine dipeptide. For a concise presentation of this toy system, see Supp.~Mat.~Section~\ref{appendix:diala}.

\subsubsection{Autoencoder ground truth collective variable}
\label{groundtruth}
The goal of AE-ABF is to obtain the same CV as the one we would obtain from training an autoencoder on a long unbiased simulation, and to use this CV to bias the dynamics for a better sampling. In order to check that the first goal is attained, we construct a reference CV which will serve as a ground truth to compare our results to. Additionally, this ground truth CV's ability to efficiently bias the dynamics is assessed.
\paragraph{Constructing the ground truth CV.} We first sample a $1.5~\mu$s trajectory, saving frames every $1.5$~ps. 
The Ramachandran scatter plot of this trajectory is given in Figure~\ref{GTsim}. We then train an autoencoder of the same topology and the same initial parameters and hyper-parameters (learning rate, activation functions, etc) as for the model described in Section~\ref{setup}. 
The resulting CV's projection on the $1.5~\mu$s trajectory is plotted in Figures~\ref{GTphi} and~\ref{GTpsi}. It can be observed that this ground truth CV first clearly separates the  C$7$ax state from the others, and also distinguishes between the C$5$ and C$7$eq states as well as the dihedral angles $(\Phi,\Psi)$ do. Moreover, a strong correlation between the CVs and the dihedral angles can be observed (visually illustrated by color plots of the $\Phi$ and $\Psi$ with respect to the CV space in Figures~\ref{GTphi} and~\ref{GTpsi}). This correlation is further confirmed with a high value of the regression score between the ground truth CV and $(\Phi,\Psi)$: $R^2=0.967$. 

\begin{figure}[ht!]
        \centering
    \begin{subfigure}{0.4\linewidth}
        \centering
            \includegraphics[width=\textwidth]{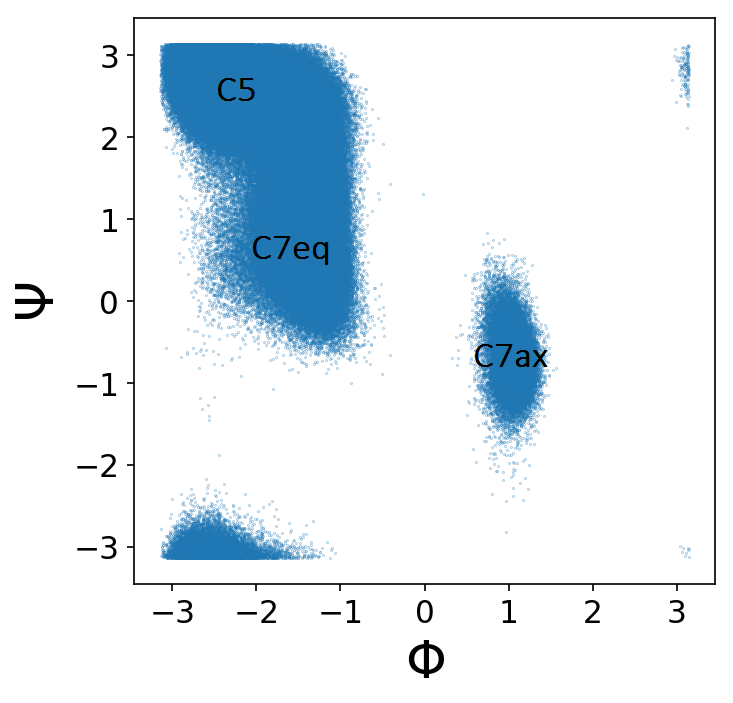}
        \caption{}
        \labelnotempty{GTsim}
    \end{subfigure}
    \\
    \begin{subfigure}{0.45\linewidth}
        \centering
            \includegraphics[width=\textwidth]{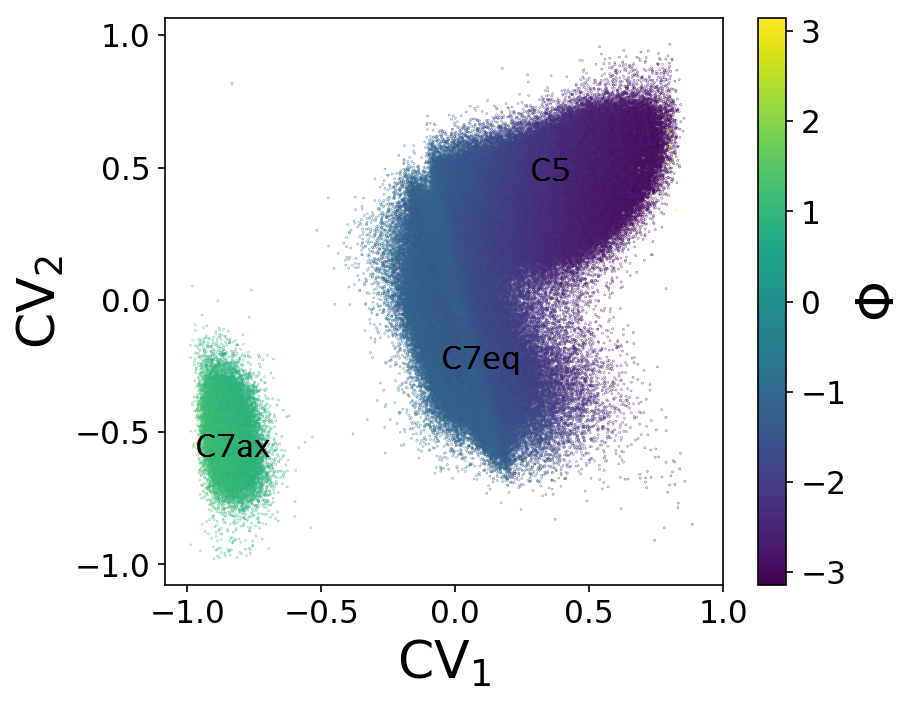}
        \caption{}
        \labelnotempty{GTphi}
    \end{subfigure}
    ~~~
    \begin{subfigure}{0.45\linewidth}
        \centering
            \includegraphics[width=\textwidth]{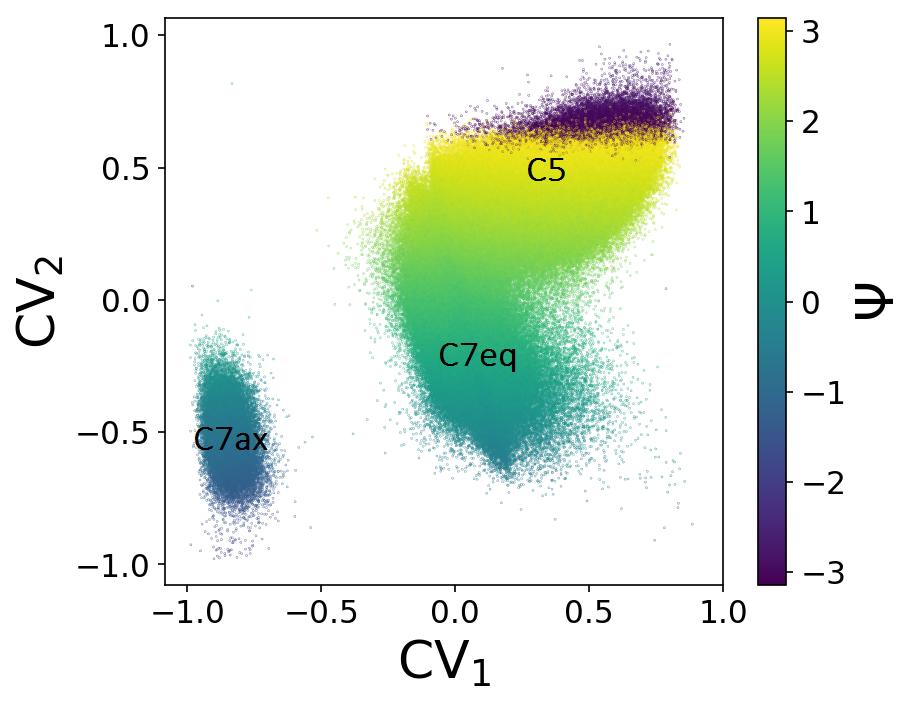}
        \caption{}
        \labelnotempty{GTpsi}
    \end{subfigure}
        \caption{(a) Ramachandran scatter plot of a $1.5~\mu$s unbiased trajectory of alanine dipeptide in vacuum which visits the three metastable states of the molecule. (b) Scatter plot of the autoencoder CVs learned from the unbiased trajectory, using $\Phi$-based coloring. (c) Scatter plot of the autoencoder CVs learned from the unbiased trajectory, using $\Psi$-based coloring. The CVs are projected on the same $1.5~\mu$s unbiased trajectory. The regions corresponding to C$5$, C$7$eq and C$7$ax are approximately defined using $\Phi$ and $\Psi$ values.}
        \labelnotempty{GT}
\end{figure} 

\paragraph{Ground truth CV biasing efficiency.} In order to assess the ground CV's efficiency for biasing the dynamics, we measure its ability to estimate the system's free energy, i.e the $(\Phi,\Psi)$ free energy surface. For this, we first run a long $500$~ns eABF simulation using $(\Phi,\Psi)$ as collective variables in order to obtain a reference free energy landscape $F$. We then run a $500$~ns eABF simulation using the ground truth CV. At any time $t$, the estimate $G_t$ of the CV's free energy is then used to compute an approximation $\tilde{F_t}$ of $F$ by reweighting histograms: using $k$ bins to discretize $(\Phi,\Psi)$ in each direction, the estimate $\tilde{F_t}^{j,l}$ in the bin $[z_{1,j}-\varepsilon,z_{1,j}+\varepsilon)\times [z_{2,l}-\varepsilon,z_{2,l}+\varepsilon)$ for $1\leq j,l\leq k$  is given by:
\begin{equation}
\label{eq:free_energy_estimation_histogram}
  \exp(-\beta\tilde{F_t}^{j,l} )= \sum_{i=1}^{N_t} \exp(-\beta G_t \circ \xi(x^i))  \mathds{1}_{\{z_{1,j}-\varepsilon\leq \Phi(x^i) < z_{1,j}+\varepsilon\}}  \mathds{1}_{\{z_{2,l}-\varepsilon\leq \Psi(x^i) < z_{2,l}+\varepsilon\}}   \text{~,}
\end{equation}
where $\xi$ is the ground truth CV, $N_t$ is the number of samples collected at time $t$ and  $2\varepsilon = \frac{2\pi}{k}$. When a certain bin~$(j,l)$ of the $(\Phi, \Psi)$ space is not visited, the corresponding value of $\exp(-\beta\tilde{F_t}^{j,l} )$ is of course $0$, making the free energy in these bins infinite. Instead, the free energy $\tilde{F_t}^{j,l}$ in these unvisited bins is made equal to the maximum of the free energy values encountered in the visited bins. 

\begin{figure}[ht!]
\centering
\includegraphics[width=\linewidth]{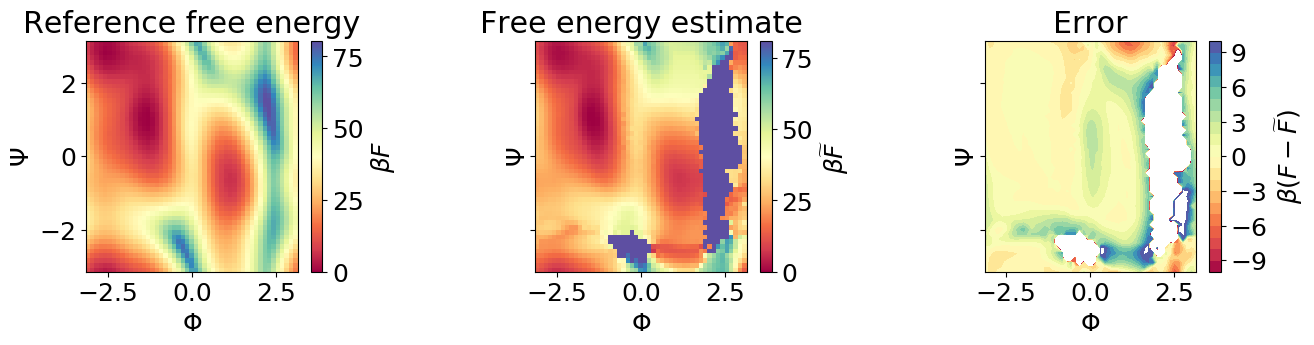} 
\caption{Left. Reference free energy computed by eABF over $\Phi,\Psi$. Middle. Estimate of the free energy computed by reweighting histograms. Transition states located on the right of the C$7$ax basin are not visited. Right. Filled contour plot of the difference $\beta(F-\tilde{F})$ in each bin. The free energies are optimally aligned.}
     \label{fediff}
\end{figure}

The free energies $F$ and $\tilde{F_t}$ are of course defined up to additive constants, which are chosen in order minimize the error between these two quantities (see Remark~\ref{l2distance} in Supp.~Mat.~Section~\ref{sec:gt_CV_vs_ub_CV}).
Figure~\ref{fediff} shows the reference free energy, the free energy estimate at the final time $\tilde{F_{500~\text{ns}}}$ and the error per bin between these two values. The C$7$ax state is less precisely estimated in shape than the C$5$ and C$7$eq states, particularly at extreme values of $\Psi$, where an incorrect local minimum is identified ($\Phi \approx 1$ and $\Psi \approx -2.5$). In addition, some transition regions are not visited during the simulation (dark purple regions in the middle plot of Figure~\ref{fediff}). These regions include transition states located on the right and on the bottom left of the C$7$ax basin, indicating that transitions between C$7$ax and C$5$/C$7$eq through those paths are very rare. Similar errors in the free energy landscape, both in magnitude and localization, were observed in similar works~\cite{mesa1}.

In order to further assess the sampling efficiency of dynamics biased by the free energy associated with the ground truth CV, we also compute the number of transitions between metastable states per nanosecond encountered in a typical eABF simulation. For this, we run three eABF simulations of $50$~ns each, using the ground truth CV as reaction coordinate, and obtain an average number of transitions of about $\sim 63$ per ns. This number is quite close to the value of about $\sim 80$ transitions per ns obtained for eABF simulations using $(\Phi,\Psi)$ as collective variables. However, biasing with the ground truth CV does not enable the sampling of the transition states on the right of C$7$ax in the Ramachandran plot, contrarily to biasing with $(\Phi,\Psi)$. For more analysis on the sampling efficiency of the ground truth CV, see Supp.~Mat.~\ref{biasvsnonbias}, where this CV is compared against $(\Phi,\Psi)$, as well as a CV obtained by training an autoencoder on configurations produced by a biased simulation without reweighting.  

\subsubsection{Results of AE-ABF}
\label{aeabfondiala}
We now apply AE-ABF to this system, and compare our learned CVs to the ground truth shown in Figure~\ref{GT}.  
Instead of stopping the algorithm at CV convergence, we run AE-ABF for a fixed number of iterations $I_{\max} = 9$ and later check at which iteration convergence has occured. This is to check whether the algorithm really stabilizes at the converged CV. We consider that convergence is reached at iteration $i$ when the linear regression score $s(\xi_i, \xi_{i-1}) \geq s_{\min} = 0.996$.

To determine the simulation time per iteration of AE-ABF, a compromise should be made between the duration of the simulation (which we want to be as small as possible) and the time needed for the free energy estimate to be stable (indicating it has converged to the free energy) so that it can be used for the reweighting of sampled data. The results given below correspond to a simulation time of $10$ ns per iteration. Other runs of AE-ABF using larger values of simulation times were also performed, and gave similar results. 

For each simulation, atomic positions are recorded every $100$~timesteps. This corresponds to $N = 10^5$ datapoints at each iteration. Figure~\ref{AECVRamach} illustrates the sampled trajectories (Ramachandran plots) through 7 iterations of AE-ABF on alanine dipeptide in vacuum. 

To compare the consecutive learned CVs, we project the encoders obtained from each iteration on the $1.5 \, \mu$s unbiased trajectory. Figure~\ref{AEs} shows scatter plots of the autoencoder CVs in $\Phi$ and $\Psi$-based coloring. Regression scores between consecutive CVs, and between the CVs from each iteration and the ground truth CV (illustrated in~Figure \ref{GT}) are also given in Table~\ref{tab:scoretable}. These regression scores are computed using data from the $1.5 \, \mu$s unbiased trajectory, and serve only for the analysis of our results. Note that during the AE-ABF algorithm itself, the regression scores between consecutive CVs are computed using the AE-ABF trajectories (precisely the two last sampled trajectories at each iteration). This is the score mentioned in Algorithm~\ref{algo:max}, which is in practice used to monitor convergence during the algorithm and determine a stopping rule. These scores, which have values similar to the ones obtained by comparing with the $1.5 \, \mu$s unbiased trajectory, are reported in Supp.~Mat.~Section~\ref{appendix:diala}.


\begin{figure}
    \centering
     \begin{subfigure}{0.4\linewidth}
        \centering
        \hspace{-3.3em}
        \includegraphics[width=0.75\textwidth]{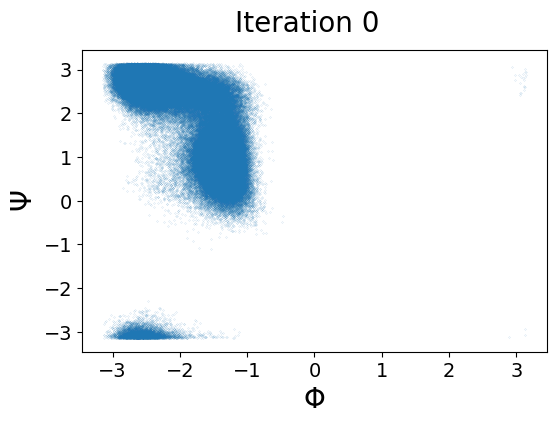}
        \labelnotempty{}
    \end{subfigure}           \begin{subfigure}{0.8\linewidth}
        \centering
        \includegraphics[width=\textwidth]{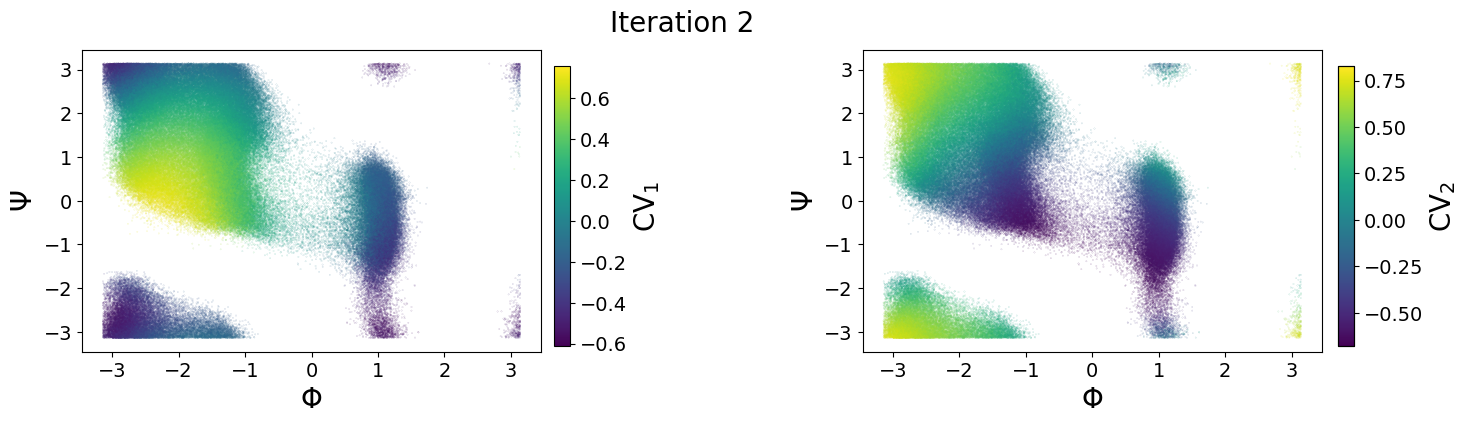}
        \labelnotempty{}
        \vspace{-1em}
    \end{subfigure}            \begin{subfigure}{0.8\linewidth}
        \centering
        \includegraphics[width=\textwidth]{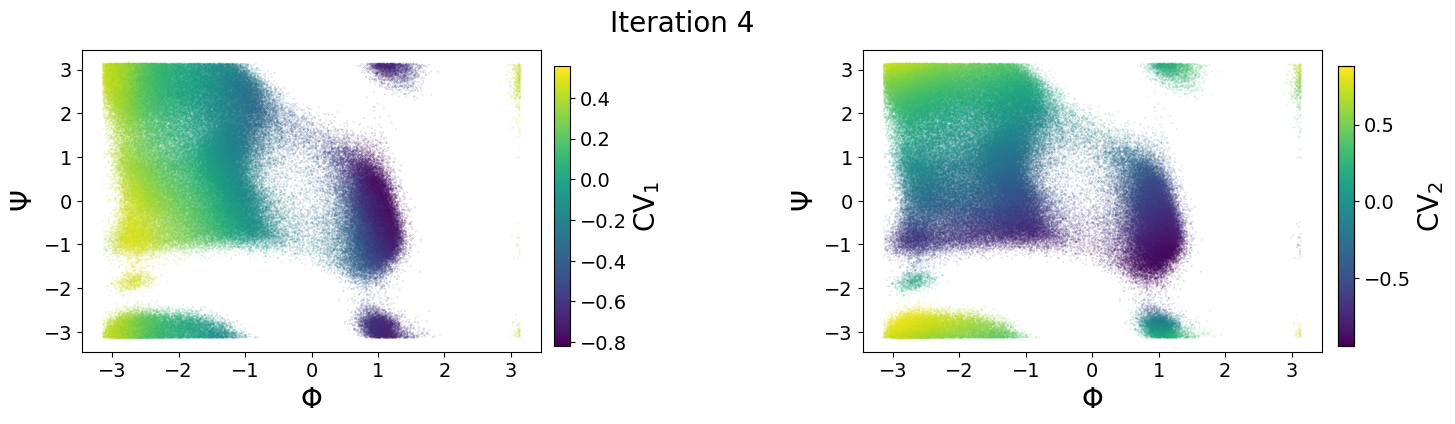}
        \labelnotempty{}
        \vspace{-1em}
    \end{subfigure}            \begin{subfigure}{0.8\linewidth}
        \centering
        \includegraphics[width=\textwidth]{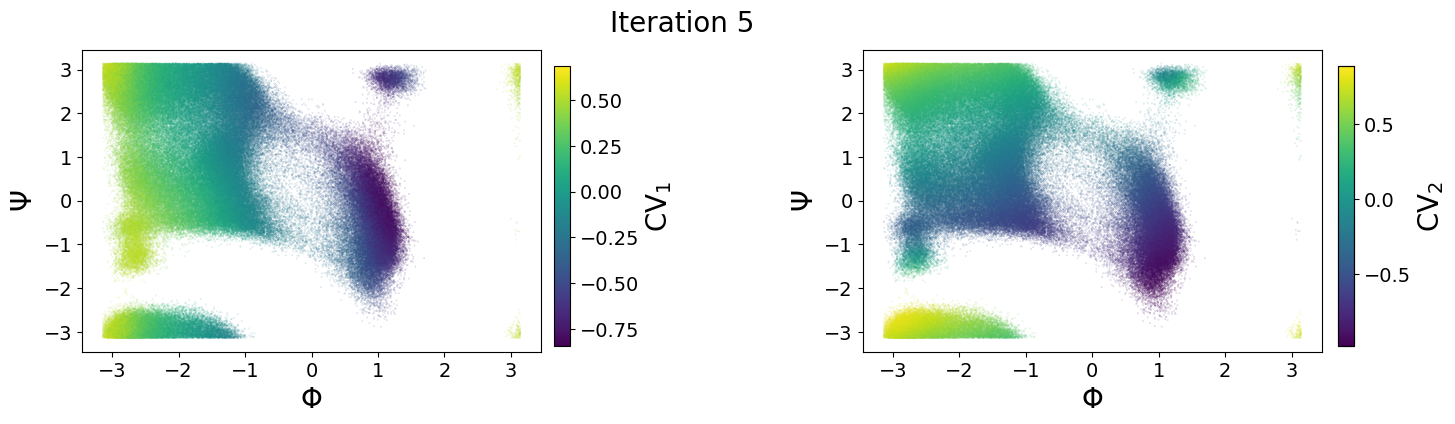}
        \labelnotempty{}
        \vspace{-1em}
    \end{subfigure}
     \begin{subfigure}{0.8\linewidth}
        \centering
        \includegraphics[width=\textwidth]{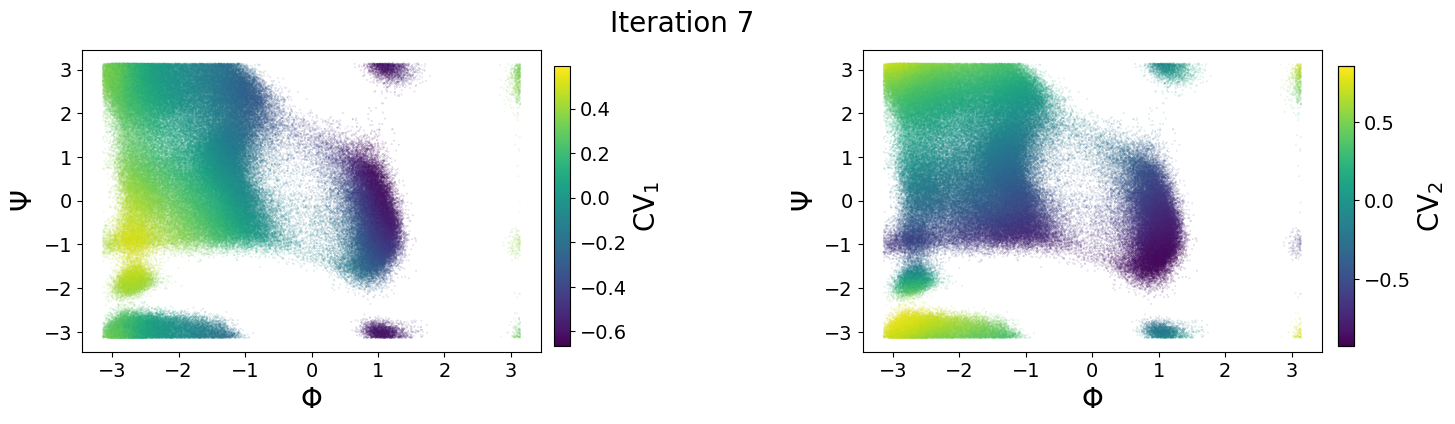}
        \labelnotempty{}
        \vspace{-1.6em}
    \end{subfigure}
   \caption{AE-ABF for 7 iterations (not all are shown here). Ramachandran scatter plots of each trajectory. The first trajectory is unbiased. The coloring corresponds to values of the first component of the CVs (Left) and the values of the second component of the CVs (Right). Visually, the first component of the CV converges approximately to a function of $\Phi$, and the second component to a function of $\Psi$.}
    \labelnotempty{AECVRamach}
\end{figure}

\begin{figure}
\centering  
    \begin{subfigure}{0.49\linewidth}
        \includegraphics[width=1\linewidth]{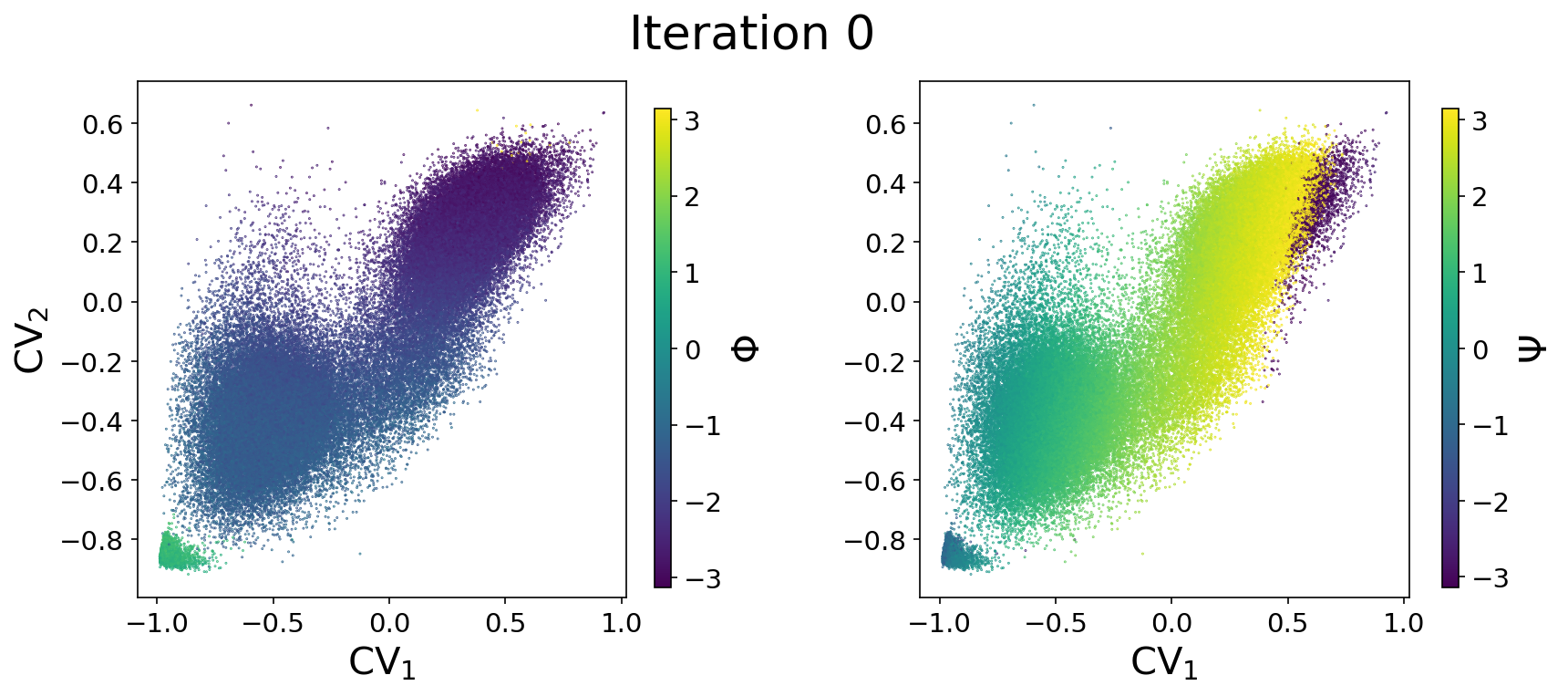} 
        \labelnotempty{}
    \end{subfigure} 
         \begin{subfigure}{0.49\linewidth}
        \centering
        \includegraphics[width=\textwidth]{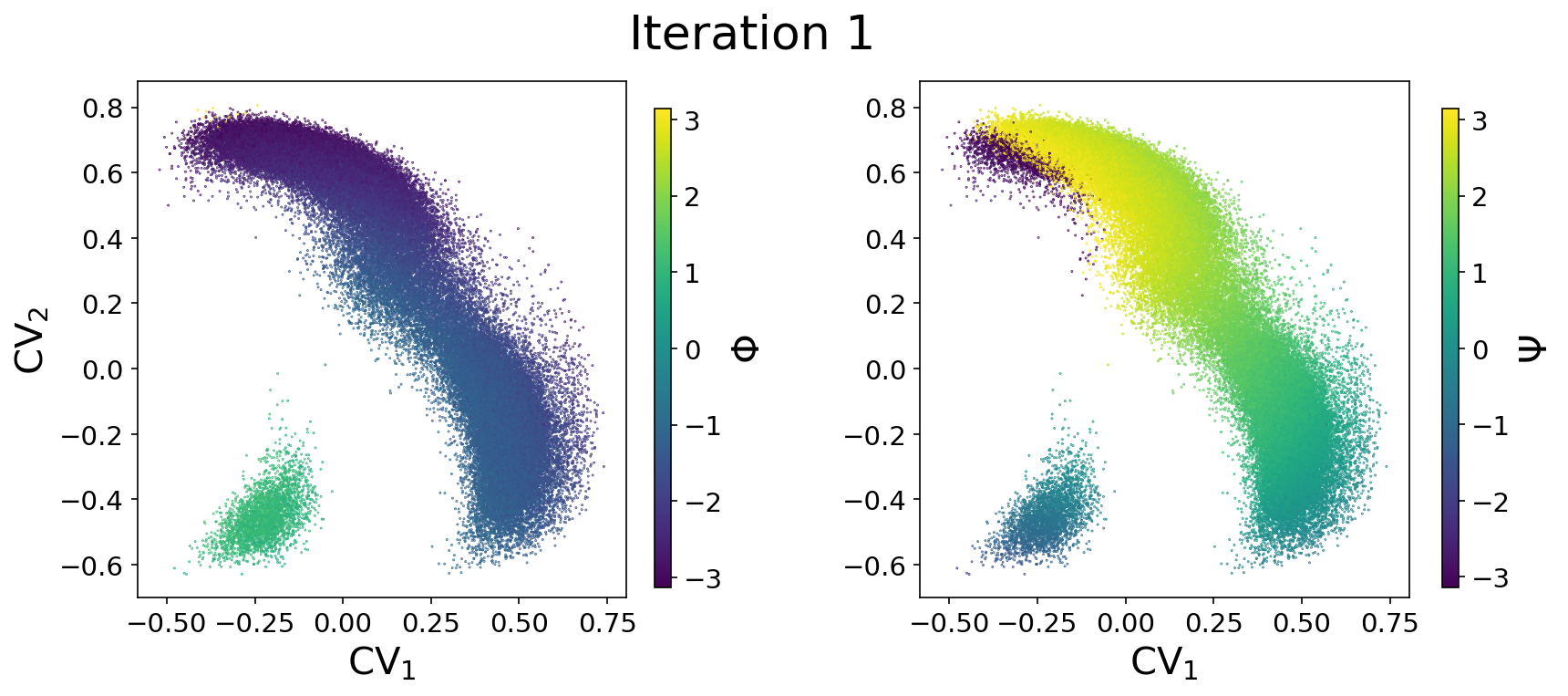}
        \labelnotempty{}
    \end{subfigure}
    \begin{subfigure}{0.49\linewidth}
        \centering
        \includegraphics[width=\textwidth]{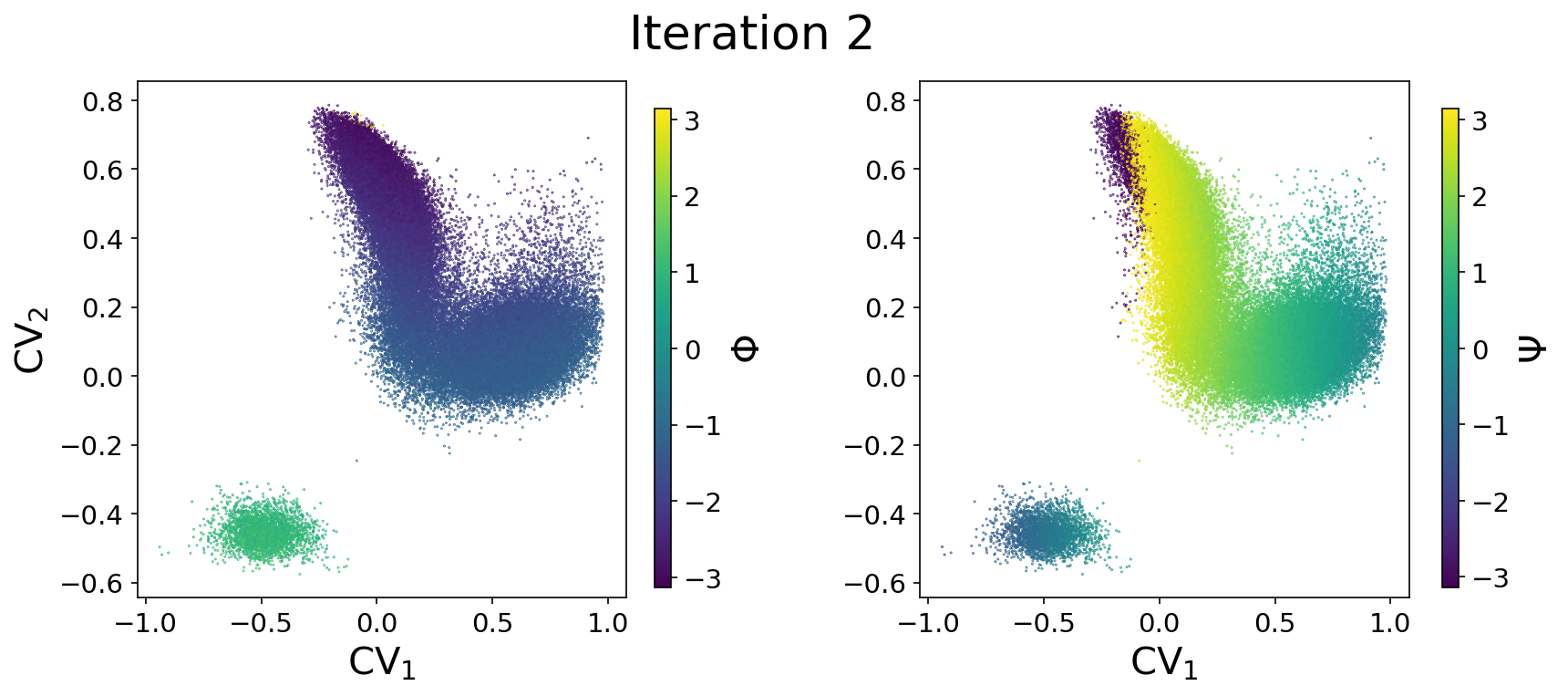}
        \labelnotempty{}
    \end{subfigure}
    \begin{subfigure}{0.49\linewidth}
        \centering
        \includegraphics[width=\textwidth]{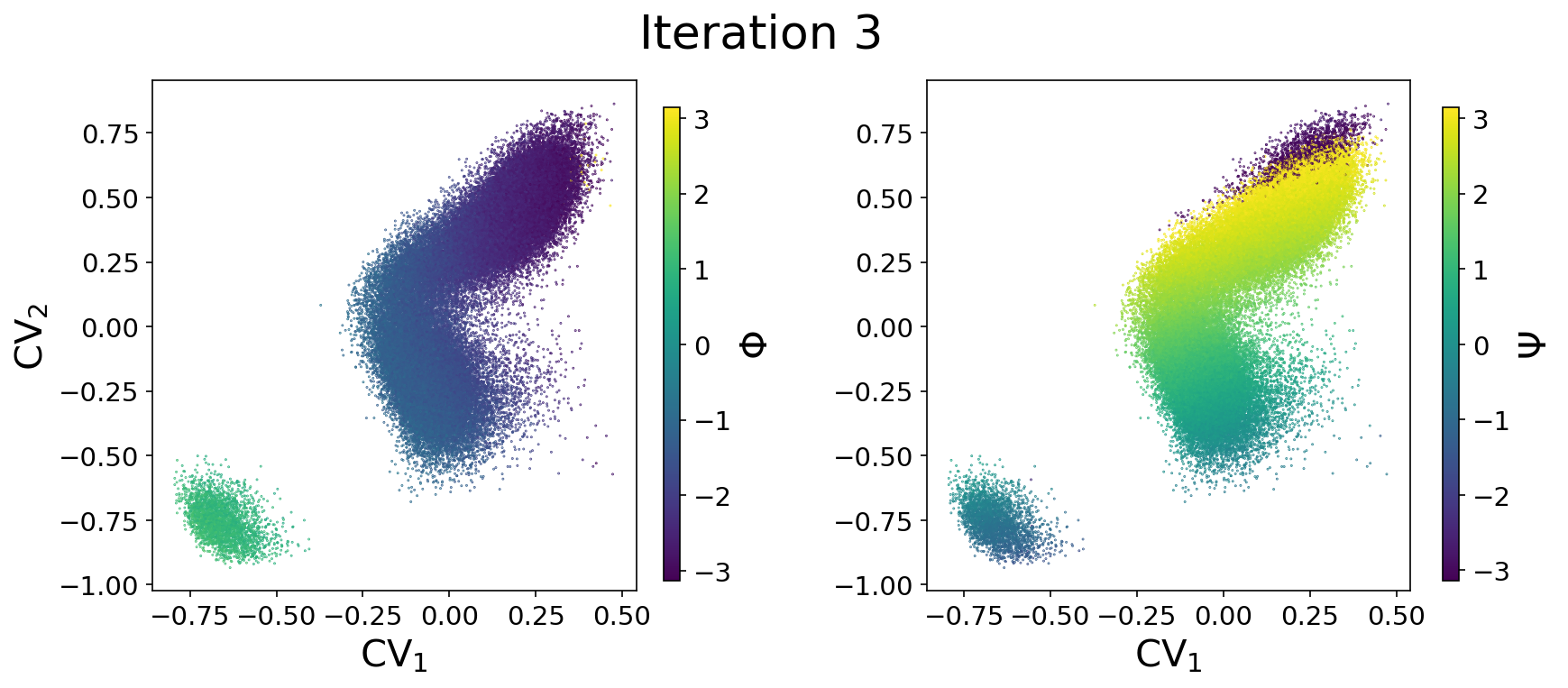}
        \labelnotempty{}
    \end{subfigure}
        \begin{subfigure}{0.49\linewidth}
        \centering
        \includegraphics[width=\textwidth]{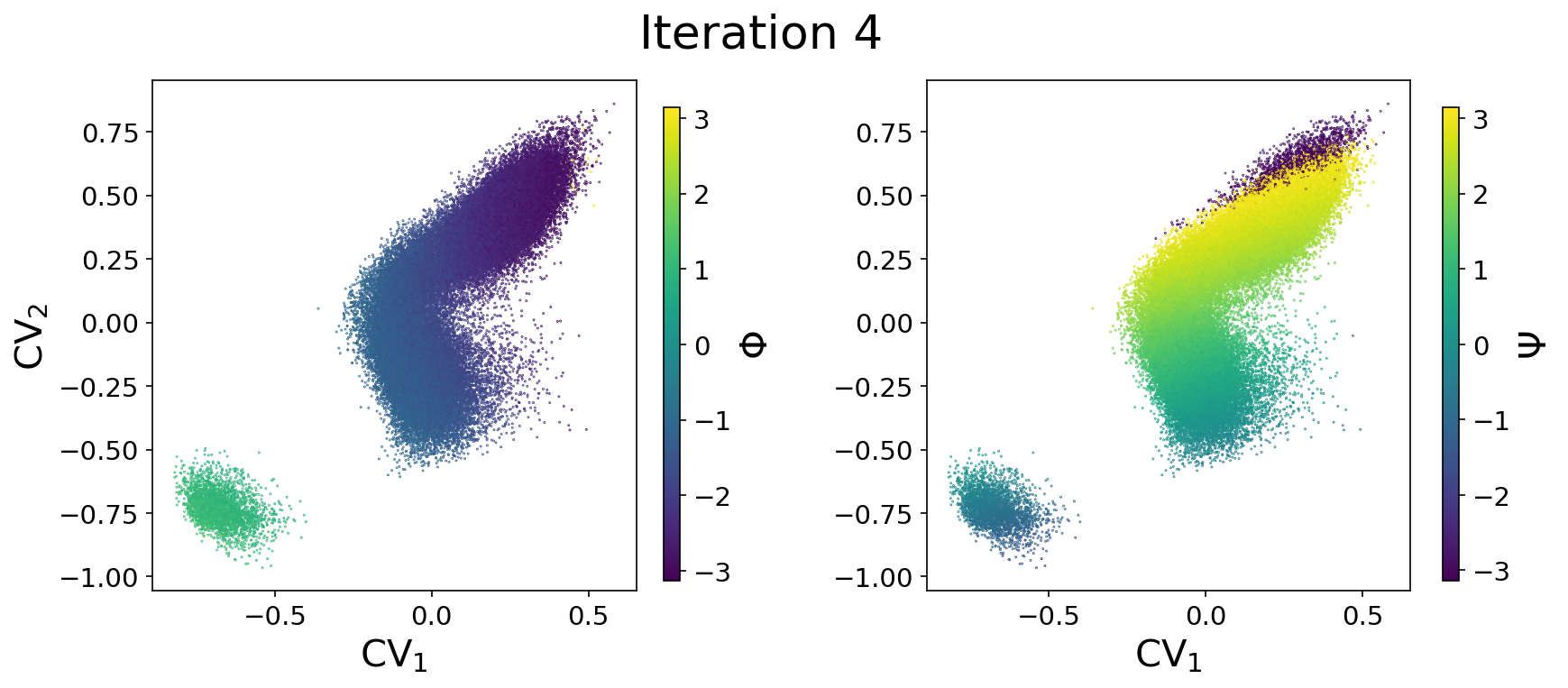}
        \labelnotempty{}
    \end{subfigure}
            \begin{subfigure}{0.49\linewidth}
        \centering
        \includegraphics[width=\textwidth]{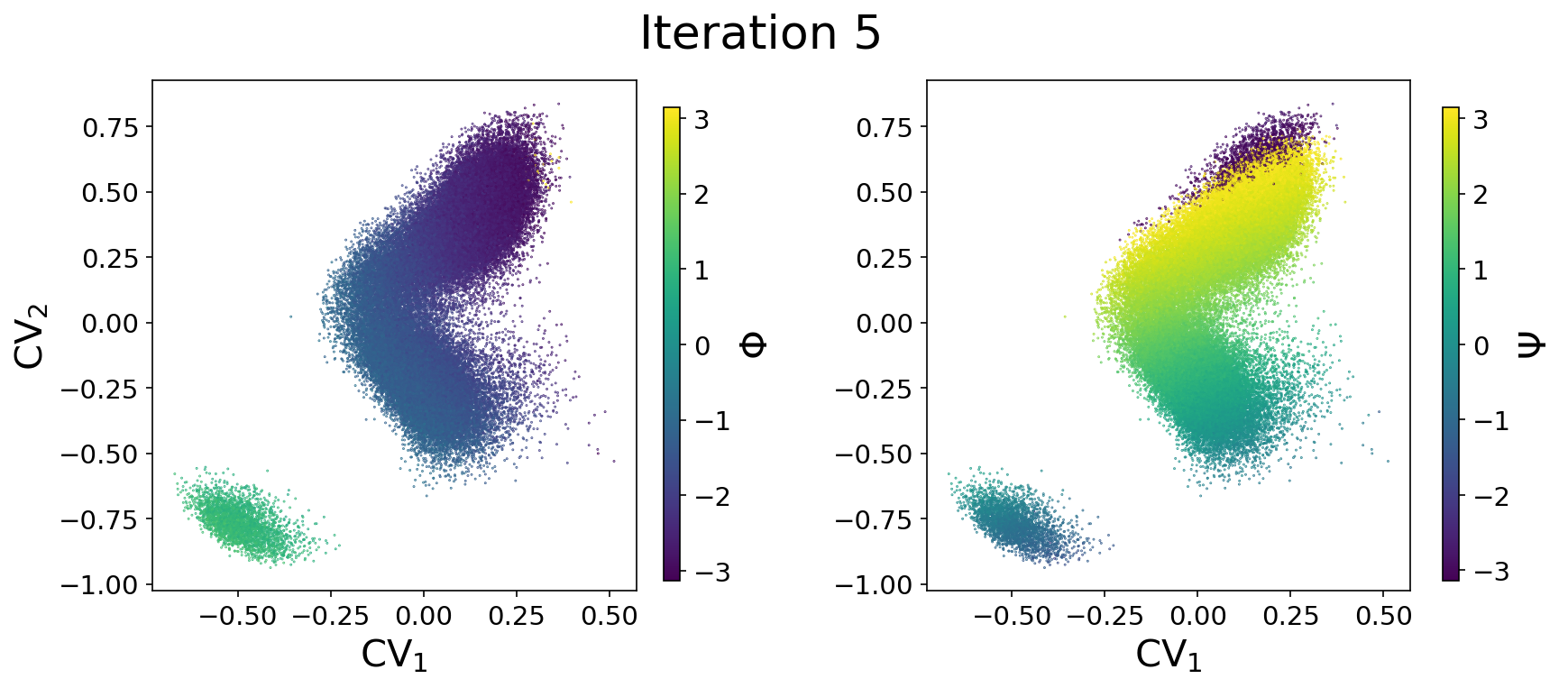}
        \labelnotempty{}
    \end{subfigure}
            \begin{subfigure}{0.49\linewidth}
        \centering
        \includegraphics[width=\textwidth]{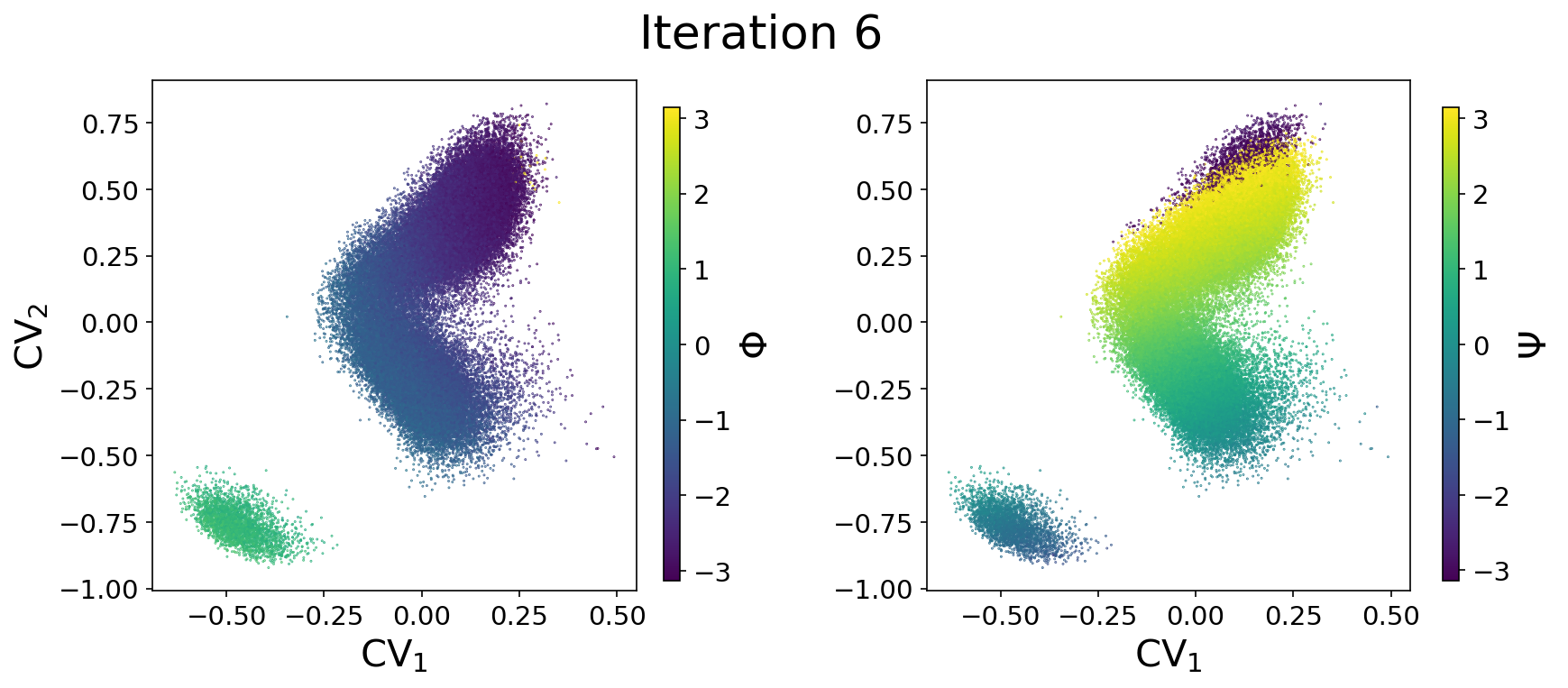}
        \labelnotempty{}
        \vspace{-1em}
    \end{subfigure}
            \begin{subfigure}{0.49\linewidth}
        \centering
        \includegraphics[width=\textwidth]{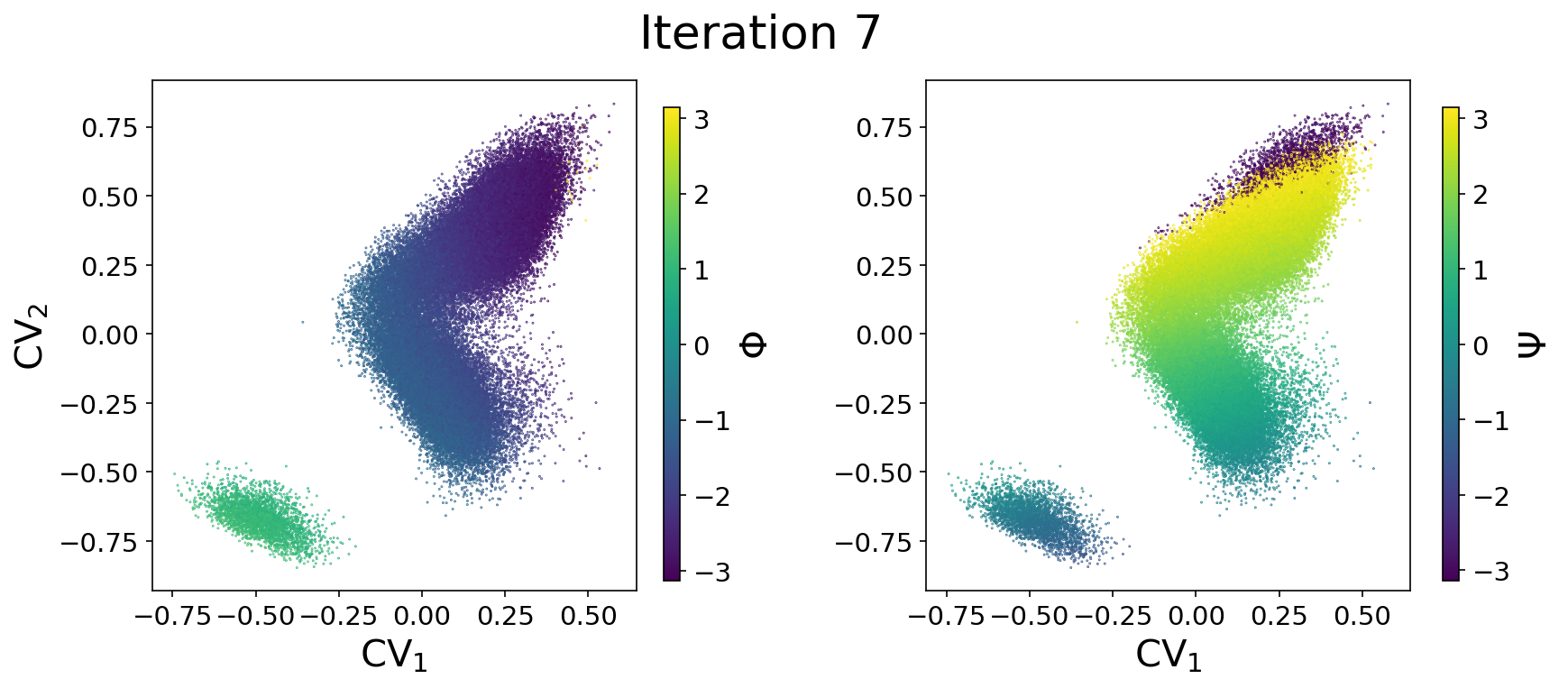}
        \labelnotempty{}
        \vspace{-1em}
    \end{subfigure}

   \caption{Autoencoder CVs through 7 iterations of AE-ABF (in addition to the initial unbiased iteration). The CVs are scatter plotted with $\Phi$ and $\Psi$ colorings. The plots show that the CV obtained with AE-ABF is very similar to the ground truth CV shown in Figure~\ref{GT}. Plots were made using encoder projections over the same unbiased $1.5 \, \mu$s simulation.} 
    \labelnotempty{AEs}
\end{figure}

\begin{table}
\centering
\begin{tabular}{|c|c|c|c|}
\hline
   Iteration & previous CV & Ground Truth CV  & $(\Phi,\Psi)$ \\
 \hline
   $0$ & $-$ & $0.9418$ &  $0.922$\\
 \hline
  $1$ & $ 0.872 $ & $ 0.849$ &  $0.892$\\
   \hline
  $2$ & $ 0.868 $ & $ 0.897$ & $0.853$\\
   \hline
  $3$ & $ 0.922 $ & $ 0.996$ &  $0.973$\\
   \hline
  $4$ & $ 0.999 $ & $ 0.997$ & $0.972$\\
   \hline
  $5$ & $ 0.999 $ & $ 0.996$ &  $0.970$\\
   \hline
  $6$ & $ 0.999 $ & $ 0.996$ &  $0.971$\\
   \hline
  $7$ & $ 0.999 $ & $ 0.995$ &  $0.967$\\
   \hline
  $8$ & $ 0.998 $ & $ 0.993$ &  $0.966$\\
   \hline
  $9$ & $ 0.999 $ & $ 0.994$ &  $0.968$\\
 \hline

\end{tabular}

\caption{Linear regression scores with AE-ABF for 9 iterations. Each line corresponds to an iteration, where the regression score is computed between the  learned CV and: the CV from the previous iteration (2nd column); the ground truth CV of Figure \ref{GT} (third column); and the 2D vector $(\Phi, \Psi)$ (fourth column). CV convergence occurs at iteration $4$, where the regression score is above~$0.996$. The regression score values show that the converged CV is almost perfectly similar to the ground truth CV, and also explains very well $\Phi$ and $\Psi$.  }
\label{tab:scoretable}
\end{table}

As illustrated in Figure~\ref{AECVRamach}, the consecutive eABF simulations all explore the three main metastable states of alanine dipeptide in vacuum. Notably, the transition between the C7ax and C7eq states is more and more sampled through the iterations. In particular, iteration 4 shows an increase of the number of sampled transition state configurations. As shown in Figure~\ref{AEs} and Table~\ref{tab:scoretable}, this coincides with the convergence of the learned CVs to a good approximation of the ground truth CV obtained in Figure~\ref{GT}. 
In conclusion, we have obtained virtually the same CV learned from a $1.5 \, \mu$s trajectory, after only~4 iterations of AE-ABF, equivalent to a total of $40$~ns of biased simulation time, in addition to the first $10$~ns unbiased iteration.

\subsection{Chignolin}

This section provides results of AE-ABF applied to the solvated chignolin mini-protein~\cite{chignolin}. We first present in Section~\ref{chignoprez} the different conformational states of chignolin. Section~\ref{unbchigno} then illustrates the $5$ separate long unbiased trajectories of chignolin that were sampled in order to provide training data for learning a ground truth CV, which is presented in Section~\ref{gtchigno}. The results of AE-ABF applied to solvated chignolin are presented in Section~\ref{aeabfchigno}. Finally, the sampling efficiency of the autoencoder CV is compared to that of more traditional choices of CV (namely well defined interatomic distances).

\subsubsection{Metastable states of chignolin}
\label{chignoprez}
\begin{figure}[ht!]
\centering
\includegraphics[width=0.8\linewidth]{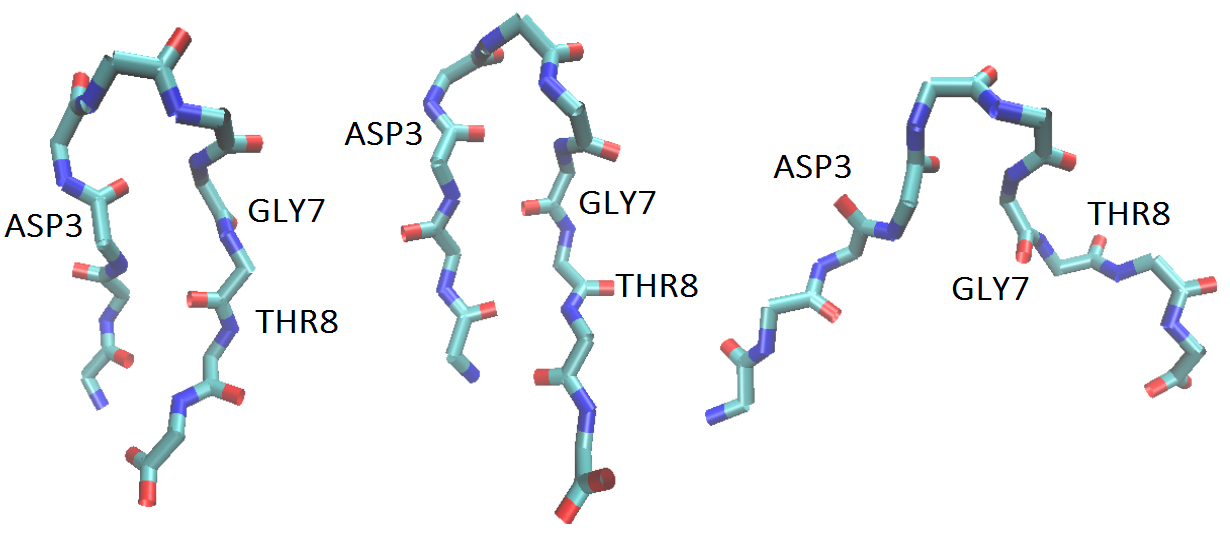} 
\caption{Three main states of chignolin in water. Left to right: Folded, misfolded and unfolded. The difference between the folded and misfolded states can be observed here in the orientation of the oxygen (red) atoms in GLY7 and THR8.}
     \label{chignostates}
\end{figure}

The chignolin miniprotein is a small $10$-residue protein which folds into a $\beta$ hairpin in water. Three main states of the protein can be distinguished: the folded state, the misfolded state and the unfolded state~\cite{chignostate}. These states are illustrated in Figure~\ref{chignostates}. The misfolded state is characterized by the formation of an incorrect hydrogen bond between ASP3N (nitrogen atom of residue 3: ASP) and GLY7O (oxygen atom of residue 7: GLY) instead of the correct hydrogen bond between ASP3N and THR8O (oxygen atom of residue 8: THR) formed in the folded state. In other words, the misfolded state is represented by small values of the distance D(ASP3N-GLY7O) while the folded state corresponds to small values of D(ASP3N-THR8O). The unfolded state shows a more open form of the hairpin and thus represents higher values of both distances.

\subsubsection{Unbiased simulations}
\label{unbchigno}
\begin{figure}[ht!]
\centering
\begin{subfigure}{1\linewidth}
        \centering
        \includegraphics[width=1\linewidth]{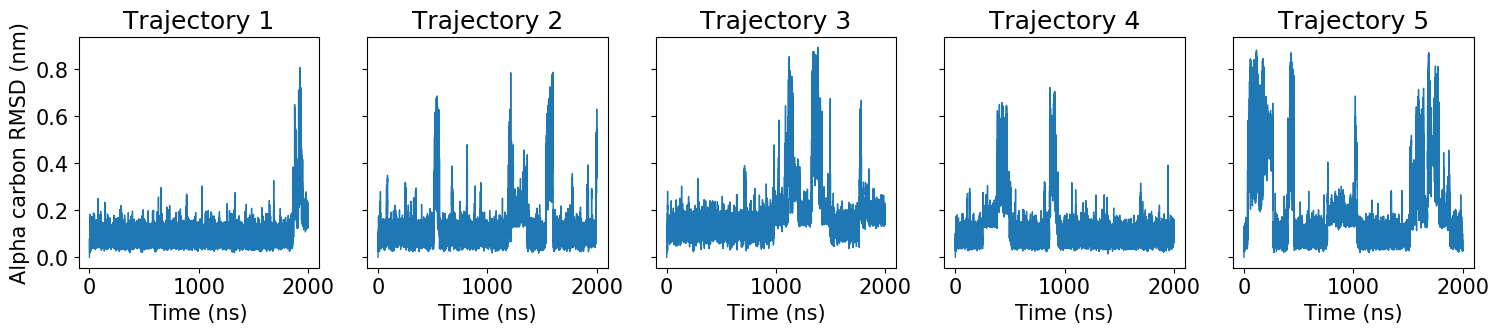} 
    \end{subfigure}
\begin{subfigure}{1\linewidth}
\vspace{2em}
        \centering
        \includegraphics[width=\textwidth]{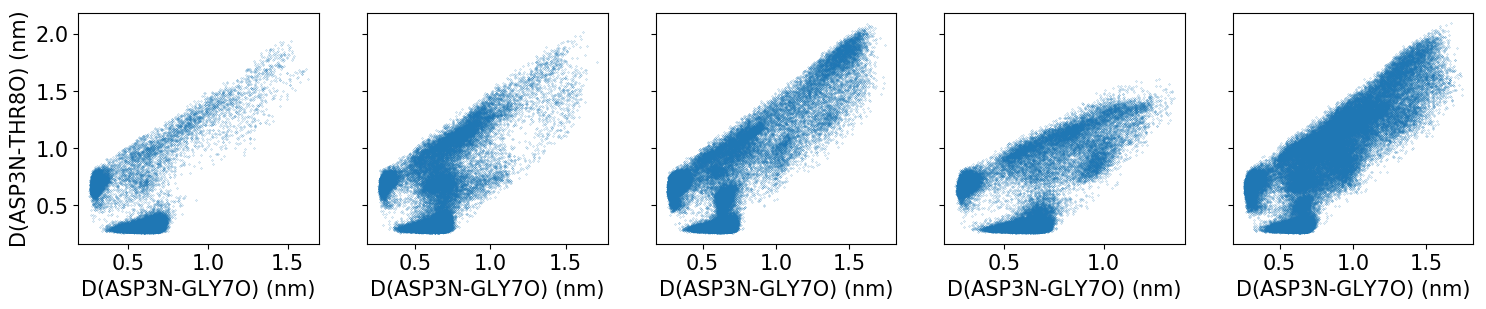}
    \end{subfigure}
     \caption{Top. Alpha carbon RMSD and radius of gyration values over each trajectory. Bottom. Scatter plots of values of the distances D(ASP3N-GLY7O) and D(ASP3N-THR8O) sampled for each trajectory.}
     \label{rmsds}
\end{figure}

Following the protocol described in Section~\ref{setupchigno}, we run $5$ unbiased simulations of $2~\mu$s each. For each simulation, configurations are saved every $5000$ timesteps, i.e $10$ ps. The resulting trajectories thus contain $2\times 10^5$ configurations each. All simulations are started from the same initial configuration in the folded state. Figure~\ref{rmsds} shows the alpha carbon root mean square deviation plotted for each trajectory. These RMSD values are computed with respect to the initial folded conformation. The folded, misfolded and unfolded states correspond approximately to RMSD values in ranges $[0,0.2]$, $[0.2, 0.4]$ are $[0.4, 0.8]$ respectively (these values are expressed in nm). Each simulation accomplishes at least one transition among these three states. Additionally, we include in Figure~\ref{rmsds} the values of the two distances of interest, D(ASP3N-GLY7O) and D(ASP3N-THR8O), observed during the simulations. 

\begin{figure}[ht!]
\centering
\includegraphics[width=0.6\linewidth]{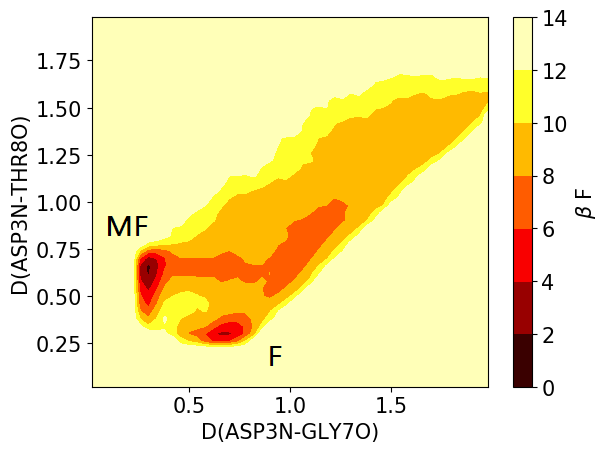} 
\caption{Free energy landscape of the 2D space formed by distances D(ASP3N-GLY7O) and D(ASP3N-THR8O). The folded (F) and misfolded (MF) states correspond to two separate basins. The unfolded state covers a more scattered area.}
     \label{distpmf}
\end{figure}

We also show in Figure~\ref{distpmf} the free energy landscape for the reaction coordinate (D(ASP3N-GLY7O),D(ASP3N-THR8O)) computed using these five simulations using a histogram of $50$ bins in each direction. The folded and misfolded states correspond to local minima, while the unfolded state is shallower and wider.

\subsubsection{The ground truth collective variable}
\label{gtchigno}
\begin{figure}[ht!]
\centering
\includegraphics[width=1\linewidth]{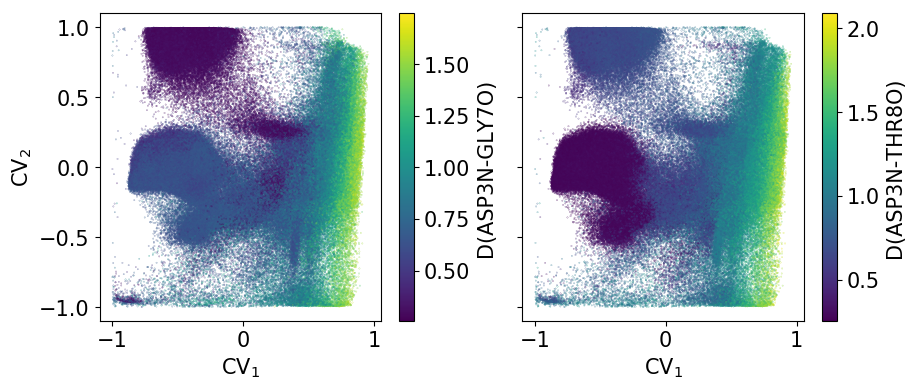} 
\caption{Autoencoder CV projected over the unbiased trajectories and colored according to values of D(ASP3N-GLY7O) (left) and D(ASP3N-THR8O) (right). Two clusters appear: The top cluster (located at CV$_1<0$ and CV$_2>0.5$) corresponds to the misfolded state, while the lower cluster (located at CV$_1<0$ and $-0.5<$ CV$_2<0.5$) represents the folded state. The rest of the 2D space corresponds to the unfolded state.}
     \label{chignocvs}
\end{figure}

The 5 simulations illustrated in the previous section are concatenated to form a dataset of $10^6$ points, which we use to learn a ground truth CV. We train an autoencoder of the structure and parameters described in Section~\ref{setupchigno}. The obtained CV is projected over the dataset in Figure~\ref{chignocvs} and colored according to the values of the distances D(ASP3N-GLY7O) and D(ASP3N-THR8O). Figure~\ref{chignocvs} shows that the obtained CV separates the misfolded and folded states into two clusters, while the unfolded state covers a large part of the remaining space. 

Similarly to alanine dipeptide, we seek to compare the CVs obtained by AE-ABF on chignolin against this ground truth CV.

\subsubsection{Results of AE-ABF}
\label{aeabfchigno}

We apply AE-ABF to chignolin using a fixed number of $7$ iterations. We consider that CV convergence is reached when the regression score is higher than $s_{\min}=0.78$, where $s_{\min}$ is computed using the bootstrapping procedure described in Supp.~Mat.~Section~\ref{appendix:regscore}. The first unbiased simulation had a time horizon of $100$~ns and the following biased simulation a time horizon of $50$~ns. For each biased simulation, atomic positions are saved every $500$ timesteps. 

We compute the values of the carbon alpha RMSD and the two distances at each iteration. These values are plotted in Figure~\ref{chignodists} to illustrate the gradual exploration of the conformational space of chignolin throughout the iterations of AE-ABF. The obtained plots clearly show that the learned CVs are able to accelerate crossing between the folded, misfolded and unfolded states. 

\begin{figure}
\centering  
         \begin{subfigure}{\linewidth}
        \centering
        \includegraphics[width=0.6\textwidth]{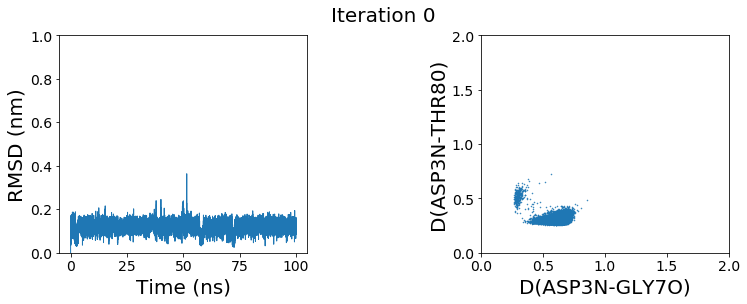}
        \labelnotempty{}
        \vspace{1em}
    \end{subfigure}
    \begin{subfigure}{\linewidth}
        \centering
        \includegraphics[width=0.6\textwidth]{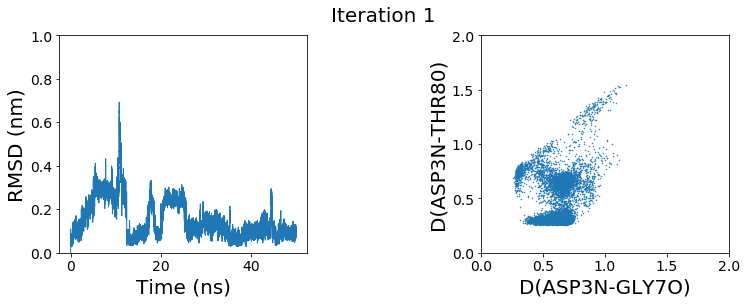}
        \labelnotempty{}
    \end{subfigure}
            \begin{subfigure}{\linewidth}
        \centering
        \includegraphics[width=0.6\textwidth]{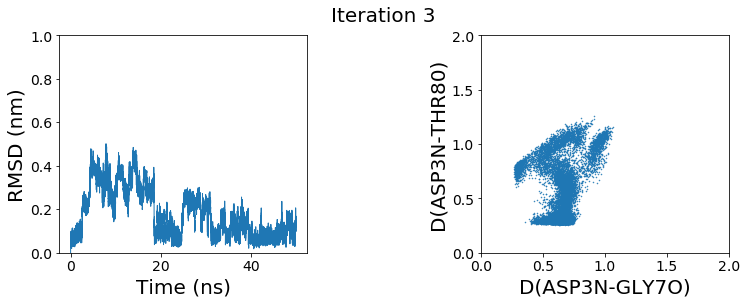}
        \labelnotempty{}
    \end{subfigure}
        \begin{subfigure}{\linewidth}
        \centering
        \includegraphics[width=0.6\textwidth]{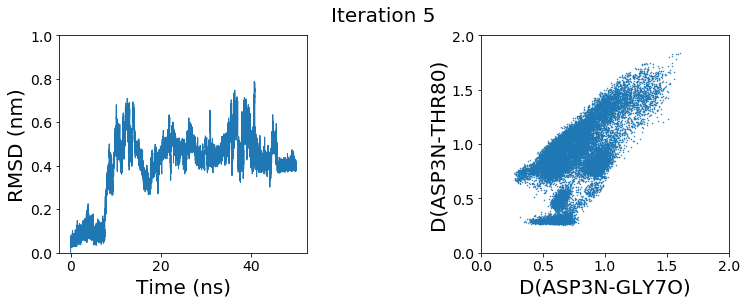}
        \labelnotempty{}        
    \end{subfigure}
                \begin{subfigure}{\linewidth}
        \centering
        \includegraphics[width=0.6\textwidth]{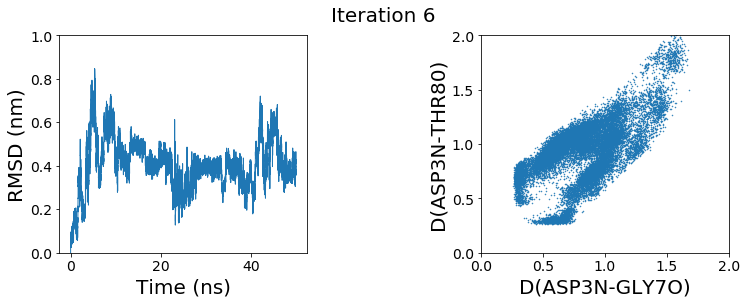}
        \labelnotempty{}
    \end{subfigure}
   \caption{Six iterations of AE-ABF (in addition to the initial unbiased iteration). Left: Evolution of carbon alpha RMSD for each iteration. Right: Sampled values of the two distances. The final rounds of AE-ABF achieve a considerably better exploration of the conformational states. (Not all iterations are shown).} 
    \labelnotempty{chignodists}
\end{figure}

Additionally, we again compute the regression scores between consecutive CVs, as well as between each CV and the ground truth CV. Results are presented in Table~\ref{tab:chignoscoretable}. The obtained regression scores are generally much lower than those obtained for alanine dipeptide, but we still achieve convergence of the CV as defined by the threshold determined in Supp.~Mat.~Section~\ref{appendix:regscore}. 

\begin{table}
\centering
\begin{tabular}{|c|c|c|c|}
\hline
   Iteration & previous CV & Ground Truth CV  & vector of distances \\
 \hline
   $0$ & $-$ & $0.119$ &  $0.099$\\
 \hline
  $1$ & $ 0.344 $ & $ 0.495$ &  $0.363$\\
   \hline
  $2$ & $ 0.349 $ & $ 0.413$ & $0.594$\\
   \hline
  $3$ & $ 0.517 $ & $ 0.772$ &  $0.705$\\
   \hline
  $4$ & $ 0.684 $ & $ 0.765$ & $0.585$\\
   \hline
  $5$ & $ 0.947 $ & $ 0.790$ &  $0.600$\\
   \hline
  $6$ & $ 0.855 $ & $ 0.801$ &  $0.674$\\
   \hline
\end{tabular}

\caption{Linear regression scores with AE-ABF for 6 iterations. Each line corresponds to an iteration, where the regression score is computed between the  learned CV and: the CV from the previous iteration (2nd column); the ground truth CV (third column); and the 2D vector (D(ASP3N,GLY7O), D(ASP3N,THR8O)) (fourth column). CV convergence occurs at iteration $5$ (regression score above~$0.78$). The regression score values show that the converged CV is also well correlated with the ground truth CV. The CVs however do not achieve high regression scores with the distances (D(ASP3N,GLY7O), D(ASP3N,THR8O)).}
\label{tab:chignoscoretable}
\end{table}

\subsubsection{Sampling with the autoencoder CV}
\begin{figure}[ht!]
\centering
\includegraphics[width=0.45\linewidth]{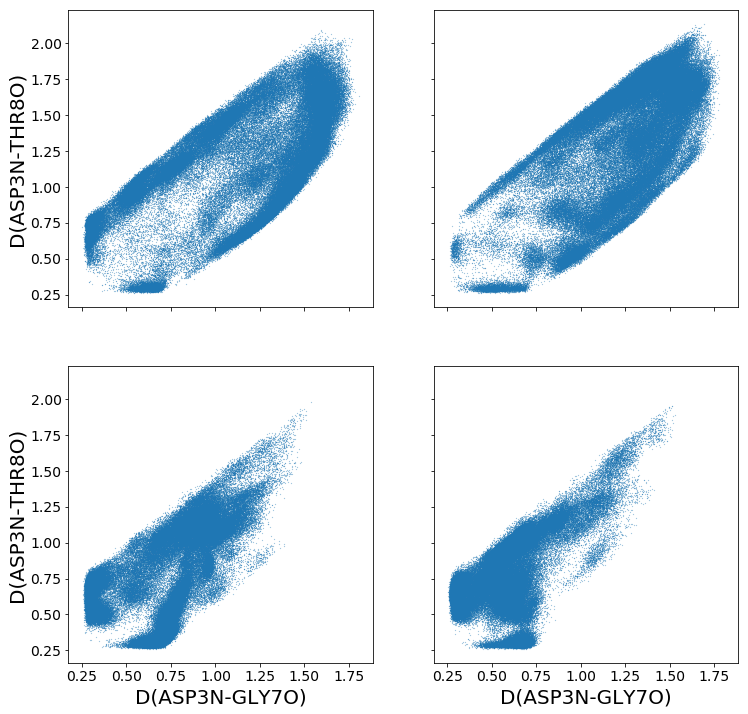}
\caption{Sampled values of the distance space. Top: eABFs using (D(ASP3N,GLY7O), D(ASP3N,THR8O)). Bottom: eABFs using the ground truth CV.}    
     \label{chignobiaseddists}
\end{figure}

\begin{figure}[ht!]
\centering
\includegraphics[width=0.45\linewidth]{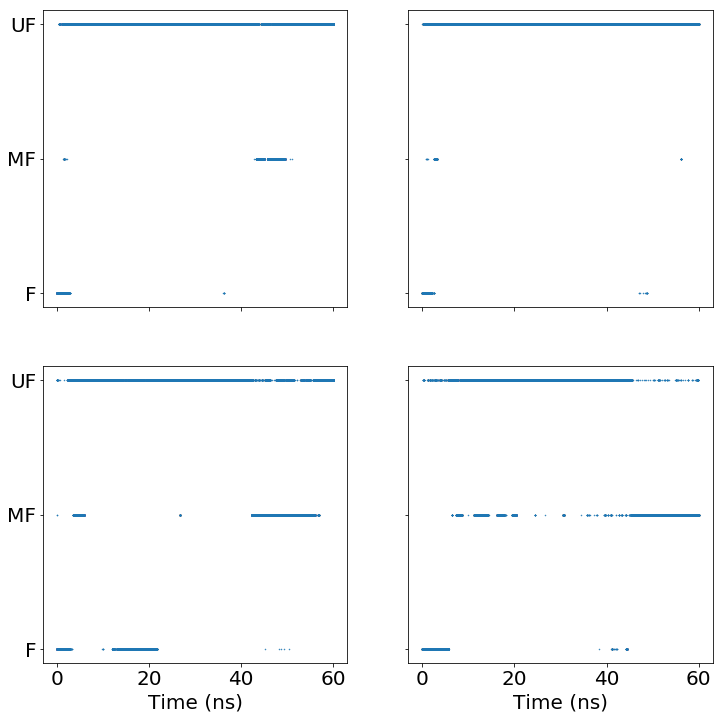}
\caption{State assigned to each sample over $60$ns eABF trajectories. Top: eABFs using (D(ASP3N,GLY7O), D(ASP3N,THR8O)). Bottom: eABFs using the ground truth CV.}    
     \label{chignotransitions}
\end{figure}

The previous section shows that AE-ABF is able to sample the three states of Chignolin in the course of one iteration, i.e. a $50$~ns simulation. Here, we compare sampling efficiency between the autoencoder CV (i.e the ground truth CV introduced in Section~\ref{gtchigno}) and the $2$D CV composed of the distances (D(ASP3N,GLY7O), D(ASP3N,THR8O)). For each CV, we sample $2$ eABF simulations  of $60$ns each. The biasing domain for the autoencoder CV (resp. the distances) is $[-1,1]^2$ (resp. $[0.2,1.75]^2$). We compare the regions of the space visited by the eABF runs. The comparison is done over the distance space, where three regions are distinguished: $[0.4,0.8]\times [0,0.4]$ defines the folded state; $[0,0.4]\times [0.4,0.8]$ defines the misfolded state, while the rest of the space represents the unfolded state. Figure~\ref{chignobiaseddists} shows the sampled values of (D(ASP3N,GLY7O), D(ASP3N,THR8O)), while Figure~\ref{chignotransitions} shows the state assignments for each sample to track transitions. It can first be observed that the trajectories sampled with eABF using (D(ASP3N,GLY7O), D(ASP3N,THR8O)) cover a wider region in the unfolded state. However, these trajectories achieve fewer transitions to the misfolded and folded states, which are thus poorly sampled. This is possibly caused by the small proportion of the space taken up by these states in the distance space (see e.g. Figures~\ref{rmsds} and~~\ref{distpmf})  as opposed to the coordinate space (see Figure~\ref{chignocvs}). Equivalently, this can be viewed as an issue concerning the choice of the biased sampling interval for (D(ASP3N,GLY7O), D(ASP3N,THR8O)). This issue does not arise for the autoencoder CV as it takes values in a bounded domain, making the sampling interval straightforward to determine.

\section{Conclusion and Future Work}
\label{sec:conclusion}
In this paper, we have given a new version of an iterative algorithm for collective variable learning with autoencoders and enhanced sampling. Our method relies on a very important reweighting procedure to ensure the convergence of the CV throughout the iterations. This reweighting procedure was both validated theoretically and in practice on simple toy examples.  
We have then fully described our method and have included some suggestions on several improvements of the basic algorithm based on the transfer of information between consecutive iterations. Finally, we have demonstrated the interest of our method on the alanine dipeptide system in vacuum as well as on the solvated chignolin system. 

Future work involves a more in depth analysis of the information transfer refinements proposed in Section~\ref{transfer}, specifically the free energy initialization procedure. Additionally, a third form of information transfer, where the previously learned model could be used to initialize the new learning model in a transfer learning fashion would also be interesting to investigate. 

The application of our algorithm using other learning models than autoencoders could also be explored. In particular, introducing an iterative algorithm with a transfer operator based method such as VAC is a very attractive possibility. It would also be interesting to explore whether using more sophisticated forms of neural networks e.g. convolutional topologies, could add something to the learned representation. Moreover, the choice of the data input representation is another part of the algorithm that could be optimized. So far, we have only worked with the aligned Cartesian coordinates, but directly using internal coordinates can prove efficient for more complex systems. In addition, more preprocessing steps could be considered to normalize the data, to obtain a set of already independent input variables, etc. For example, whitening transformations could be used to remove underlying correlations.  

Finally, the application of AE-ABF to more complex systems is crucial to test what the method can truly accomplish, and explore any possible limitations. 

\newpage
{\small
\bibliography{Bibliography}
}

\appendix
\counterwithin{figure}{section}
\counterwithin{equation}{section}
\counterwithin{remark}{section}

\section{Theoretical definitions and practical details}

This section highlights some important theoretical and practical points on the implementation of FEBILAE. 

\subsection{Collective variables and free energy biasing}
\label{subsec:FECV}

In MD, a common choice for the biased distribution~$\widetilde{\mu}$ is to consider a free energy biased distribution. In order to make this concept precise, we recall in this section some definitions on collective variables, free energy and free energy based biasing, including general formulas for the Boltzmann--Gibbs measure and the free energy biased measure. 
Here, we consider the positions of a given system $q=(q_1,\dots, q_m)$, where $q_i \in \mathbb{R}^3$ for systems in vacuum or $(L\mathbb{T})^3$ in the case of periodic boundary conditions in a cubic box. We denote by $D=3m$ the dimensionality of the system. 
For this system, the marginal of the canonical measure~$\nu$ with respect to the positions $q$ is defined as:
\begin{equation}
    \nu(dq) = Z_{\nu}^{-1} \mathrm{e}^{-\beta V(q)} dq \text{ ,}
\label{BGPos}
\end{equation}
where $V:\mathbb{R}^D \to \mathbb{R}$ is the potential energy function, $Z_{\nu}=\int_{\mathbb{R}^D} \mathrm{e}^{-\beta V(q)} dq$ is a normalization constant and $\beta=(k_BT)^{-1}$ is proportional to the inverse temperature. Here, $\nu$ corresponds to the original distribution $\mu$ considered in~\eqref{eq:reference_loss}. 

The full sampling of the canonical measure by molecular dynamics simulations is typically infeasible because of metastability. Many methods aim at overcoming this issue. In particular, free energy biasing methods change the original potential $V$ to $V - F \circ \xi$, where $F$ is the free energy associated with the collective variable $\xi: \mathbb{R}^D \rightarrow  \mathbb{R}^d $. The CV $\xi$ is often chosen such that $\xi(q_t) = z_t$ is a slow process given the dynamical evolution equation of $q_t$. The function~$\xi$ typically characterizes the conformational changes between metastable states.
Recall that the free energy associated with a CV $\xi$ and the canonical measure $\nu$ is (up to an irrelevant additive constant):
\begin{equation}
\label{F(z)}
    F(z) = -\frac{1}{\beta} \ln {\left(\int_{\Sigma(z)} \mathrm{e}^{-\beta V(q)} ~ \delta_{\xi(q)-z}(dq)\right)} \text{ ,}
\end{equation}
where $\Sigma(z)=\{ q\in\mathbb{R}^D , \xi(q)=z \}$ and the delta measure $\delta_{\xi(q)-z}(dq)$ is supported on $\Sigma(z)$. The density of the image of $\nu$ by $\xi$ is thus proportional to $\mathrm{e}^{-\beta F}$.

Free energy biasing corresponds to sampling the canonical measure associated with the bias potential $V-F \circ \xi$, that is:
\begin{equation}
    \nu_F(dq) \propto \mathrm{e}^{-\beta(V - F \circ \xi)(q)} dq  \text{ .}
\label{nuF}
\end{equation}
A simple computation shows that the marginal distribution in the $\xi$ variable of this biased measure is uniform. Indeed, the density of the image of $\nu_F$ by $\xi$ is: 
\begin{align}
\begin{split}
\label{eq:1}
       \int_{\Sigma(z)} \mathrm{e}^{-\beta (V(q)-F \circ \xi(q))} ~ \delta_{\xi(q)-z}(dq) 
      = \mathrm{e}^{\beta F(z)} \mathrm{e}^{-\beta F(z)} = 1 \text{ .}
\end{split}
\end{align}
The biased measure $\mathrm{e}^{-\beta(V - F \circ \xi)(q)} dq $ thus has a uniform marginal law along~$\xi$. This implies that $\xi(q_t)$ is less metastable under the free energy biased dynamics. 

The biasing procedure is however only as good as the CV used to perform it: if most of the metastability of the system is captured by the CV~$\xi$ (i.e. when the CV effectively resolves the metastable states), the method effectively renders the dynamics diffusive and annihilates metastability; conversely, if the CV~$\xi$ fails to describe the most important metastable directions, biasing along it will likely be ineffective. This last point motivates the need to find a good choice of collective variable. 

\subsection{Autoencoder optimization step}
The autoencoder is typically trained using a gradient descent optimization algorithm: the parameters $\textbf{p}$ are sequentially modified at each optimization step using a steepest descent algorithm with a stepsize $\eta>0$ called the learning rate. In this paper, we use mini-batching to approximate the gradient of the loss function, meaning that each optimization step only uses a subset (a mini-batch) of the data to modify the parameters $\textbf{p}$. More precisely, we denote by~$b$ the number of points in a mini-batch. The data is first randomly reshuffled, then at each learning step $r$, the loss is computed as the mean loss over datapoints $x^{rb+1},\dots,x^{(r+1)b}$:
$$ 
\mathcal{L}_r (\textbf{p}) = \frac{1}{b} \sum_{i=rb+1}^{(r+1)b} \|x^i-f_{\textbf{p}}(x^i)\|^2 . 
$$
The parameters $\textbf{p}$ are thus updated as follows:
$$
\textbf{p}_{r} = \textbf{p}_{r-1} - \eta \nabla_{\textbf{p}}\mathcal{L}_r(\textbf{p}_{r-1}).
$$

To compute the gradient $\nabla_{\textbf{p}}\mathcal{L}_r(\textbf{p}_{r-1})$, back-propagation is used. The learning proceeds in epochs, where one epoch corresponds to the number of steps $\left \lfloor \frac{N}{b} \right \rfloor$ required to visit the entire dataset (with $\lfloor . \rfloor$ the integer part of a number). A maximum number of epochs is specified, after which training is stopped.  The learning is usually stopped earlier, when the loss no longer decreases. To assess this, part of the data is used as a validation set: This subset is not used in training, and its mean loss is tracked to stop the gradient descent algorithm when it no longer decreases. This procedure is called early stopping. The validation loss is used instead of the total loss to stop the algorithm to avoid overfitting the dataset.

It is very important to note here that during learning and after, the model has access to the gradient of the value of any neuron with respect to the value of any other neuron from a previous layer (including the input). This is useful later, in the context of enhanced sampling, one needs to extract the mapping of the encoder and its gradient.

\subsection{Multiple solutions}
As mentioned in Section~\ref{subsec:learning}, the optimization problem does not necessarily have a unique solution. In addition, the learning loss can have multiple local minima. More precisely, the value $\textbf{p}_{\mu}$ obtained at the end of learning, which corresponds to one local minimum, may depend on the initial value given to $\textbf{p}$ at the beginning of the learning, and on the value of the learning rate $\eta$ used in the gradient descent. As a consequence, when we want to assess the impact of the distribution of the input data on the learned parameters, we make sure that, for a given test case, the numerical setting (initial parameters and topology of the autoencoder, optimization parameters, etc) are kept constant from one experiment to another. 

\subsection{Sample weights normalization}
\label{wnormal}
For all the experiments performed in this paper, the weights used in Equation~\eqref{lossw} are the ones defined in Equation~\eqref{hatw} multiplied by $N$. This is to follow the default weight normalization used by the \texttt{keras} module in \texttt{python}, which we use for all our experiments, and where the weights are normalized to sum to the number of samples: $\displaystyle \sum_{i=1}^N w_i = N$, instead of $1$. Of course, this only adds a multiplicative factor to the loss, and thus does not change anything to the learning from a theoretical viewpoint. This is simply a practical choice that allows us to use default values for most of the parameters of the optimization procedure.

\section{Regression score to assess CV convergence}
\label{appendix:regscore}

The AE-ABF algorithm stops when the CVs learned from the new iteration are sufficiently similar to the ones obtained from the previous iteration. We make precise in this section how this similarity is quantified. 
\subsection*{The $R^2$ score in regression models}
We consider a dataset $\mathcal{Z} = (z^1, \dots , z^N)$, with $z^i \in \mathbb{R}^p$, and corresponding values (called labels) $\mathcal{Y} = (y^1, \dots, y^N)$, with $y^i \in \mathbb{R}^{p'}$. We call an optimized regression model between $\mathcal{Z}$ and $\mathcal{Y}$ a mapping $M$ from the inputs $z^1,\dots,z^N$ to the outputs $y^1,\dots,y^N$ trained so as to minimize the error
$$  \sum_{i=1}^N \|M(z^i)-y^i\|^2 \text{,}   $$
where $\|\cdot \|$ is the Euclidean norm. Here we consider a linear model, which corresponds to $M(z) = Wz+b$, for a given matrix $W \in \mathbb{R}^{p\times p'}$, and bias vector $b \in \mathbb{R}^{p'}$. 

The precision of this regression model, for the data $\mathcal{Z}$ and $\mathcal{Y}$, can be assessed using the $R^2$ score, also called coefficient of determination:
$$R^2(M,\mathcal{Z},\mathcal{Y}) = 1 - \frac{\displaystyle \sum_{i=1}^N \|y^i-M(z^i)\|^2}{\displaystyle \sum_{i=1}^N \|y^i-\bar{y}\|^2} \text{,}$$
where $\bar{y} = \displaystyle \frac{1}{N} \displaystyle \sum_{i=1}^N y^i$. The $R^2$ score is thus simply the fraction of variance explained by the regression model $M$ over the dataset $(\mathcal{Z}, \mathcal{Y})$. Note that this score is equal to $1$ when the model $M$ is able to output exactly $y^i$ for each $z^i$. On the other hand, a baseline model which outputs   $M(z^i) = \bar{y}$ for all $i$ will have a $R^2$ score equal to $0$. 

\subsection*{Using the $R^2$ score for CV comparison}
To compare two CVs $\xi$ and $\xi'$ of dimensions $p$ and $p'$ respectively, we use a test trajectory $ ({q^1, \dots, q^N})$. We then optimize a linear regression model $M$ between inputs $\mathcal{Z} = (z^1, \dots z^N) = (\xi(q^1), \dots,  \xi(q^N))$ and labels $\mathcal{Y} = (y^1, \dots, y^N) = (\xi'(q^1), \dots, \xi'(q^N))$. The CV score between $\xi$ and $\xi'$ is then defined as the coefficient of determination $R^2(M,\mathcal{Z},\mathcal{Y})$. Typically, $p'=p$ and $\xi = \xi_{i-1}$, the CV proposed at the $i$-th iteration of Algorithm~\ref{algo:max}, while $\xi'=\xi_{i}$.

\subsection*{Determining the threshold value $s_{\min}$}
The stopping rule of Algorithm~\ref{algo:max} requires determining a value $s_{\min}$ of the~$R^2$ score above which one can consider that the CVs are sufficiently similar in order to stop the loop. Note that this value  depends on various parameters, one of which is the size $N$ of the datasets that are used to learn the CVs. Let us illustrate one approach to determining $s_{\min}$. In the setting considered in Section~\ref{dialares}, the number of datapoints used for learning a new model at each iteration of AE-ABF is $N=10^5$. To determine a reasonable value for $s_{\min}$ with $N=10^5$, we use the unbiased trajectories obtained as described in Sections~\ref{groundtruth} and~\ref{unbchigno}. The unbiased dataset of $N_{\rm ref}=10^6$ points is randomly split into $10$ subsets $S_1,\dots,S_{10}$. An autoencoder is trained on each subset $S_k$ and a CV, $\xi_k$ is learned. Then the $R^2$ score between pairs $(\xi_k, \xi_{\ell})$ is computed. This provides $\frac{10\times9}{2}=45$ different values of the $R^2$~score when $N_{\rm ref}/N=10$. This operation can be repeated $r$ times to obtain $45 r$ realisations of the $R^2$~score. This procedure allows to approximate the distribution of this score for $N=10^5$. We choose to define $s_{\min}$ as the $5$\% percentile of this distribution, that is to say, $s_{\min}$ is such that $95\%$ of the $45 r$ values are larger than $s_{\min}$. Figure~\ref{histr2} shows the regression scores computed for alanine dipeptide and chignolin. For alanine dipeptide, using $r=30$, and thus $1350$ realisations of~$R^2$, we find $s_{\min}\approx 0.996$. This motivates the choice of the value $s_{\min}=0.996$ considered in Section~\ref{groundtruth}. For chignolin, using $r=15$, and thus $675$ realisations of~$R^2$, we find a much lower value: $s_{\min} \approx 0.78$. The difference between the two values obtained for two different systems shows the necessity of determining the threshold $s_{\min}$ for each new system. 

Note that this method relies on having a large  dataset (here $N_{\rm ref} = 10N = 10^6$ samples) to sample $N$-sized subsets. In practice, this dataset may not be available at the beginning of the simulation. However, a dataset of size $N$ is sampled at each iteration of AE-ABF. After a number $N_{\rm it}$ of iterations of the algorithm is achieved, the $N_{\rm it}$ datasets which have been obtained can be compiled and used as a larger dataset of size $N_{\rm ref} = N_{\rm it} N$ to determine~$s_{\rm min}$. This means that $N_{\rm it}$ iterations of Algorithm~\ref{algo:max} are performed without checking the convergence of the CV, after which a value of $s_{\rm min}$ is determined and Algorithm~\ref{algo:max} is continued as usual. 
\begin{figure}[ht!]
\centering
\begin{subfigure}{0.45\linewidth}
\centering
     \includegraphics[width=\linewidth]{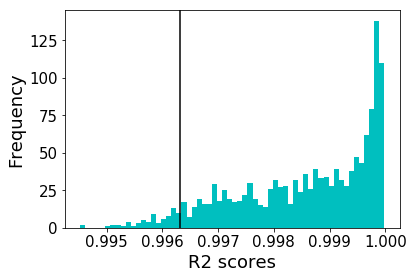}
\end{subfigure}
\begin{subfigure}{0.45\linewidth}
\centering
     \includegraphics[width=\linewidth]{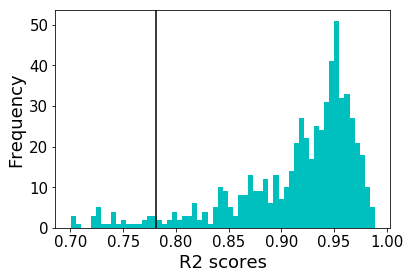}
\end{subfigure}
    \caption{Histogram of the $R^2$ scores obtained using subsets of $N=10^5$ points out of $10^6$ points. The vertical black line indicates the $5$th percentile. Right. Alanine dipeptide. Left. Chignolin.}
    \labelnotempty{histr2}
\end{figure}

\section{Transferring information between iterations}
\subsection{Using trajectories from previous iterations}
\label{appendix:prevtraj}
This section illustrates results of AE-ABF applied with a sliding window of $n_T=1$, $2$ or $3$ training trajectories: At iteration $i \geq 1$, the autoencoder is trained on the last $\min(n_T,i)$  sampled trajectories.

Recall that the results for alanine dipeptide given in Section~\ref{sec:results} correspond to an implementation of the algorithm which includes a sliding window of $n_T=2$ trajectories. Here, we use the 2D system of Section~\ref{subsec:2D} as an illustrative example. We use  a reduced trajectory horizon of only $2\times 10^5$ timesteps at each iteration with $\Delta t = 10^{-3}$ and save samples every $5$ timesteps. We therefore sample relatively small trajectories that do not always visit the three states of the configurational space, and simulations that are not sufficiently long the free energy profile to converge. We demonstrate here that for this case, using the two or three last trajectories for training ($n_T=2$ or $3$ respectively) enables convergence, contrarily to using only the new trajectory ($n_T=1$). 

We run three instances of AE-ABF for 6 iterations each, using respectively a sliding window of $n_T=1$, $2$ and $3$ trajectories. The corresponding CVs are plotted using an unbiased test trajectory and shown in Figure~\ref{slides}. These plots show that learning on only one trajectory is not enough to achieve convergence to the CV $\xi(x,y) = x$, because the free energy does not converge, and the trajectories do not always explore the three metastable states. However, combining the trajectories enables convergence, as it allows for the resulting dataset to be complete (even if each trajectory only visits one of the two deep wells).

\begin{figure}[h!]
    
    \begin{subfigure}{\linewidth}
        \centering
        \includegraphics[width=\textwidth]{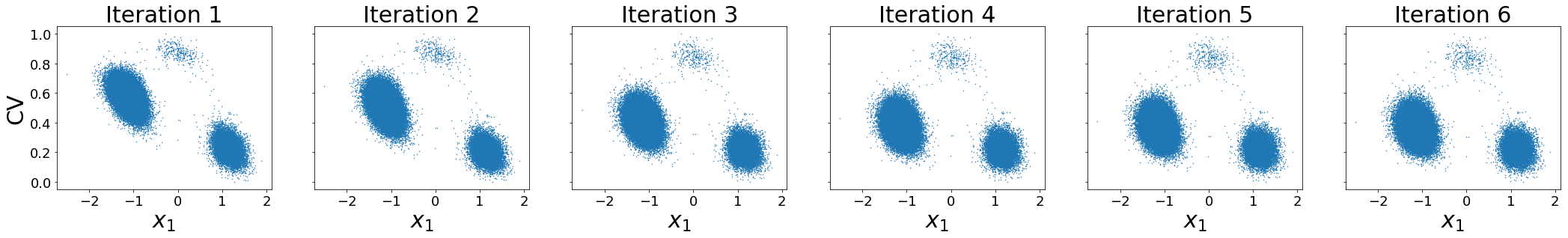}
        \caption{$n_T=1$}
        \labelnotempty{slide1}
    \end{subfigure}
    \begin{subfigure}{\linewidth}
        
            \includegraphics[width=\textwidth]{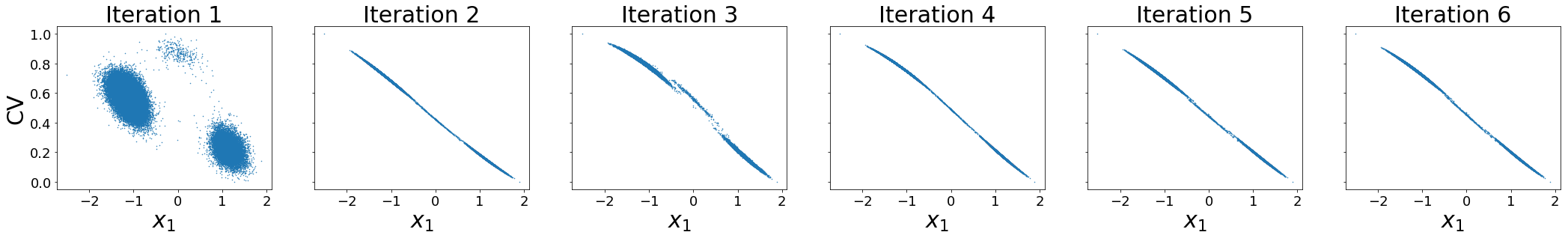}
            \caption{$n_T=2$}
        \labelnotempty{slide2}
    \end{subfigure}
        \begin{subfigure}{\linewidth}
        
            \includegraphics[width=\textwidth]{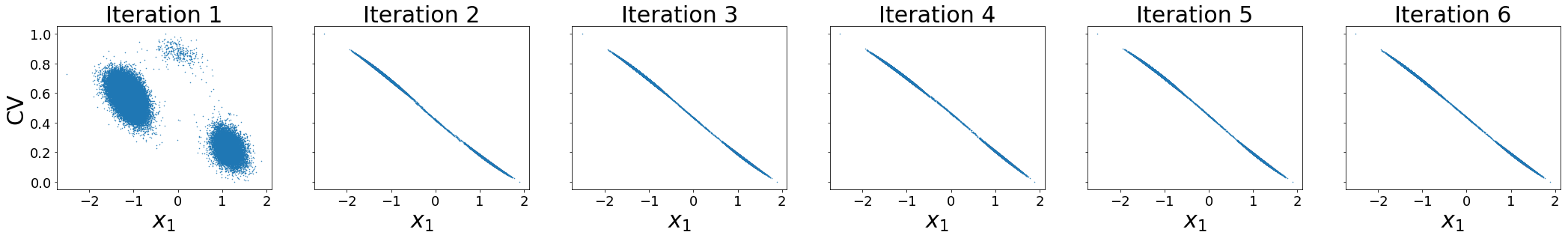}
            \caption{$n_T=3$}
        \labelnotempty{slide3}
    \end{subfigure}
   \caption{CVs obtained at consecutive iterations of AE-ABF using different numbers $n_T$ of previous trajectories as training data.}
    \labelnotempty{slides}
\end{figure}

\subsection{Results with free energy initialization on the 2D toy example}
\label{appendix:feinit}
At each iteration of AE-ABF, an eABF run can be divided into two periods over each region of the configurational space. The first period only ends when a good estimate of the free energy has been computed. During this part of the simulation, the estimated bias is either not applied (when not enough samples have been gathered in the corresponding bin), or is not an accurate enough estimate to allow for a full exploration of the configuration space and thus to obtain a correct free energy. The second period represents the actual biasing of the system using the converged estimate. The free energy initialization scheme introduced as a refinement of our algorithm in Section~\ref{sec:ae-abf} allows to reduce the time required for the first period, leaving more simulation time for the more important second period.

In this section, the free energy initialization procedure is implemented for the 2D three well potential system, to observe in practice the accuracy of this initialization and its impact on the convergence of the algorithm.  This procedure is equivalent to running the same eABF while changing the biasing CV at each iteration of AE-ABF. To keep this ongoing eABF run consistent, at each new iteration, the simulation is started from the last sampled point of the previous iteration. 

Note that the mean force initialization established in Equation~\eqref{cvmappreg} assumes a strong correlation between the consecutive CVs. Consequently, the initialization scheme is only applied when the regression score between the consecutive CVs is high enough. We chose to apply it when it is above $R^2_{\min}=0.9$. 

Here, we reduce the trajectory time horizon to only $5\times 10^4$ timesteps at each iteration with $\Delta t = 10^{-3}$, saving samples every $5$ timesteps. Training is done over a sliding window of $n_T=4$ trajectories. Without using free energy initialization, this time horizon is not enough to obtain CV convergence (even using $n_T=4$ trajectories for learning), because the estimated free energy is far from stabilized.

Figure~\ref{prevtraj} shows the progressive convergence of the CV learned at each iteration to a monotonic function of the coordinate $x_1$, and the matching values of the free energy initialization. It is directly observed that, as the learned CV stabilizes, the free energy initialization is more and more accurate. 

\begin{figure}[h!]
    
    \begin{subfigure}{\linewidth}
        \centering
        \includegraphics[width=\textwidth]{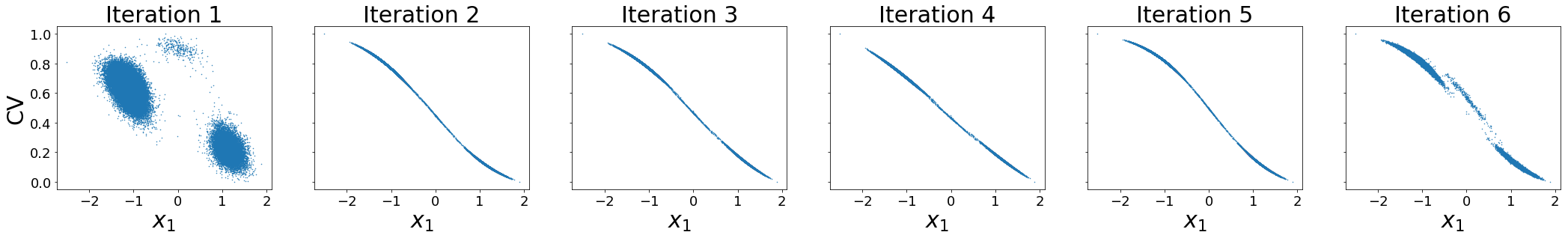}
        \caption{Learned CVs}
        \labelnotempty{cvs}
    \end{subfigure}
    \begin{subfigure}{1.1\linewidth}
       \hspace{-1em}
       \includegraphics[width=\textwidth]{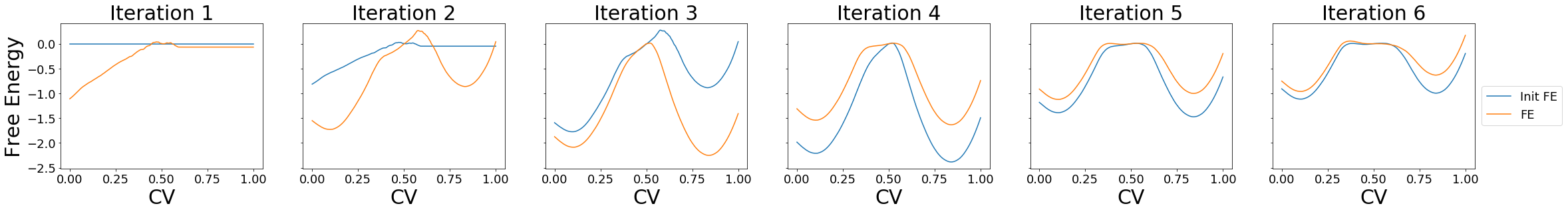}
            \caption{Free Energies: initial guesses and final estimates}
        \labelnotempty{initmfs}
    \end{subfigure}

   \caption{The initial guess of the free energy at iteration~$i$ (blue lines) is increasingly more similar to the free energy estimate at the end of iteration~$i$ (orange lines), as the CV converges (top scatter plots).}
    \labelnotempty{prevtraj}
\end{figure}

\section{Additional details and results on alanine dipeptide}
\label{appendix:diala}

\subsection{States and collective variables }

Alanine dipeptide is a small molecule of 22 atoms, including 8 backbone atoms. Alanine dipeptide in vacuum  has been extensively studied and used in previous works, which makes it a good choice for bench marking. Figure~\ref{dialamol} shows the structure of the molecule and highlights the three metastable states of alanine dipeptide in vacuum.

\begin{figure}[ht!]
        \centering
    \begin{subfigure}{0.4\linewidth}
        \centering
            \includegraphics[width=\textwidth]{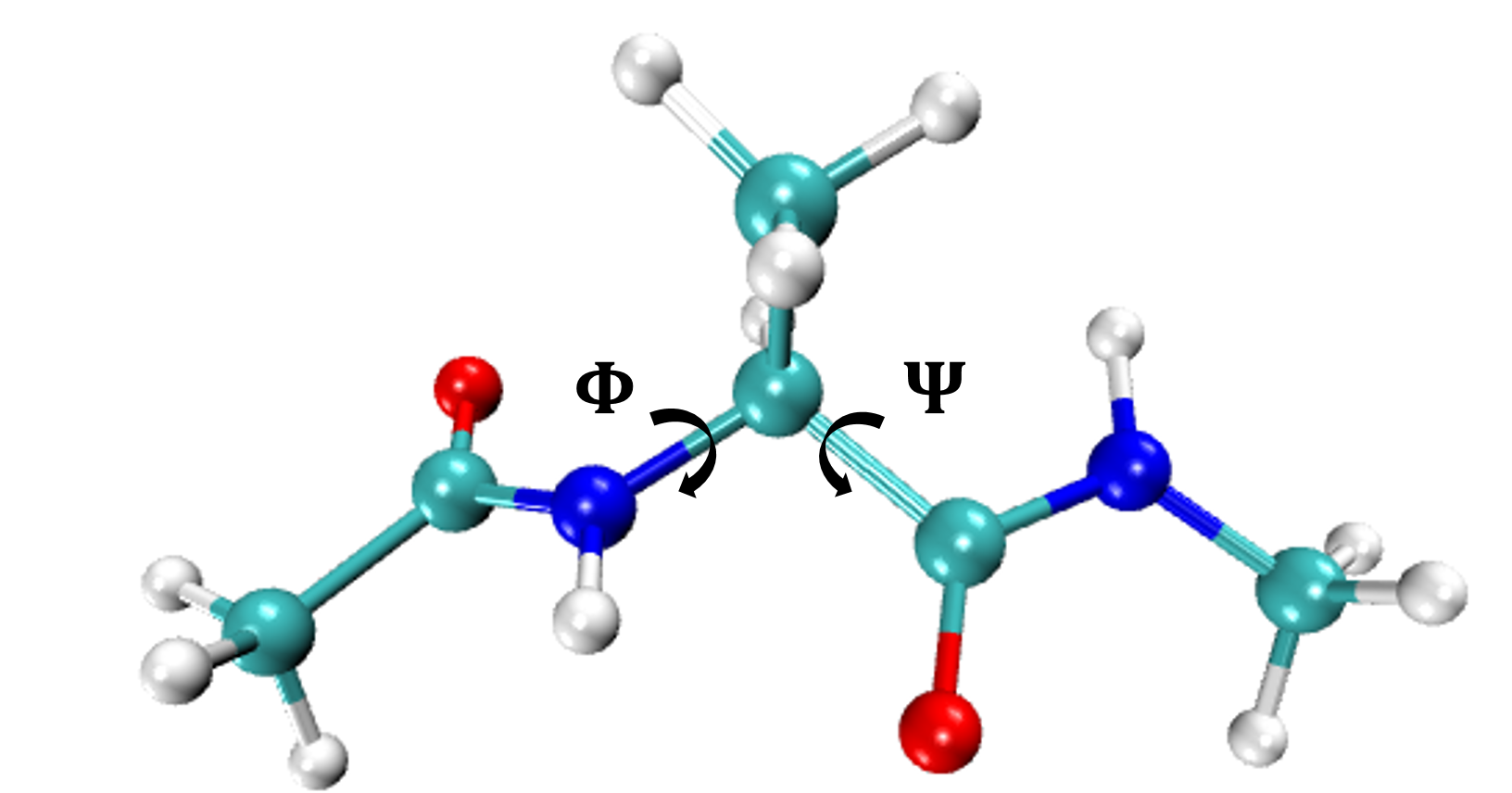}
        \labelnotempty{dialastruct}
    \end{subfigure}
    \begin{subfigure}{0.4\linewidth}
        \centering
            \includegraphics[width=\textwidth]{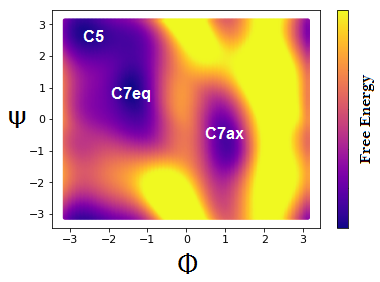}
        \labelnotempty{dialapmf}
    \end{subfigure}
        \caption{Alanine dipeptide. Left. Molecular structure. Right. Metastable states of the molecule in vacuum represented on the Ramachandran space using the free energy profile under a temperature of $300$K (using $180$bins in each direction). Transition times between C$5$ and C$7$eq states are fairly short, unlike the transition times between these states and the C$7$ax state. }
        \labelnotempty{dialamol}
\end{figure} 

\subsection{The FVE plateau method}
Using the $1.5~\mu$s unbiased trajectory obtained as described in Section~\ref{groundtruth}, we train $10$ different autoencoders with respective bottleneck dimensions $1,\dots, 10$. We then compute as explained in Remark~\ref{cvdimknee} the FVE value for each model, and plot these values to find a visual plateau in the FVE curve which allows to determine the optimal bottleneck dimensionality. The obtained FVE curve is shown in Figure~\ref{dialafve}. A knee is observed at bottleneck dimension~$d=2$.

\begin{figure}[ht!]
\centering
     \includegraphics[width=0.6\linewidth]{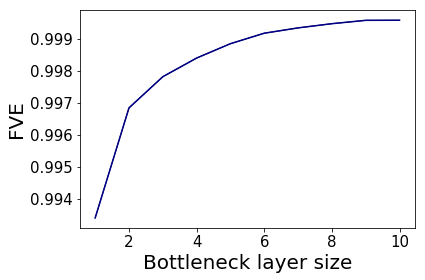}
    \caption{FVE curve obtained using the $1.5 \, \mu$s unbiased trajectory as training data. The plateau occurs at dimension~$d=2$.}
    \labelnotempty{dialafve}
\end{figure}
\subsection{Real time regression scores of AE-ABF using the sampled trajectories}

Table~\ref{tab:scoretable} above shows the $R^2$ scores between consecutive CVs of AE-ABF, using the $1.5 \, \mu$s unbiased trajectory. As mentioned in the latter section, Table~\ref{tab:scoretable} only serves to analyze our results post run. In order to use the regression scores to establish a stopping rule, the long unbiased trajectory can of course not be used. Instead, the last two sampled \emph{biased} trajectories are used at each iteration. Note that these trajectories are again reweighted to obtain the unbiased regression score. Table~\ref{tab:realscorestable} shows the obtained scores, compared to the ones obtained by comparing to the reference CV constructed from the unbiased trajectory. The scores are quite similar. In particular, the convergence of the CV happens at the same iteration of AE-ABF for the score obtained from the long unbiased trajectory, and the ones obtained from the AE-ABF trajectories. 

\begin{table}
\centering
\begin{tabular}{|c|c|c|}
\hline
   Iteration & over last $2$ sampled trajectories & over reference trajectory \\
 \hline
   $0$ & $-$ & $-$ \\ 
 \hline
  $1$ & $ 0.399$ & $0.872$ \\ 
   \hline
  $2$ & $ 0.927 $ & $0.868$\\
   \hline
  $3$ & $ 0.904 $ & $0.922$\\
   \hline
  $4$ & $ 0.998 $ & $0.999$ \\
   \hline
  $5$ & $ 0.999 $ & $0.999$\\
   \hline
  $6$ & $ 0.999 $ & $0.999$\\
   \hline
  $7$ & $ 0.998 $ & $0.999$\\
   \hline
  $8$ & $ 0.996 $ & $0.998$\\
   \hline
  $9$ & $ 0.998 $ & $0.999$\\
 \hline
\end{tabular}
\caption{Linear regression scores, AE-ABF for 9 iterations. Regression scores between consecutive CVs computed during the AE-ABF run using the sampled biased trajectories (first column), and using the long unbiased trajectory (second column).}
\label{tab:realscorestable}
\end{table}

\subsection{Additional results: reweighted versus unweighted CVs}
\label{biasvsnonbias}

One of the goals of our algorithm is to find the same autoencoder CV that would have been obtained using long unbiased simulations. The results in Section~\ref{dialares} demonstrate that this goal is attained: The converged AE-ABF CV achieves a high regression score with respect to the ground truth CV. 
We note however that the ground truth CV by construction contains little information on transition states compared to metastable states, as it is trained on unbiased simulations. Yet incorporating some of this information could potentially improve the CV's ability to sample these transition states. Consequently, a CV that is trained on biased data may be more efficient in biasing the dynamics to enhance sampling.

In this section, we use the alanine dipeptide system to discuss this point. First, we compare in Section~\ref{sec:gt_CV_vs_ub_CV} the sampling properties of the ground truth CV of Figure~\ref{GT} and a CV obtained from unweighted biased data. Second, we perform in Section~\ref{sec:AE_ABF_without_reweighting} an AE-ABF run where the reweighting step is omitted. The sampled trajectories on this unweighted AE-ABF are visually compared to those of a regular AE-ABF. 

\subsubsection{Comparison between the ground truth CV and a CV obtained from unweighted biased trajectories}
\label{sec:gt_CV_vs_ub_CV}

\begin{figure}[ht!]
\centering
\includegraphics[width=1\linewidth]{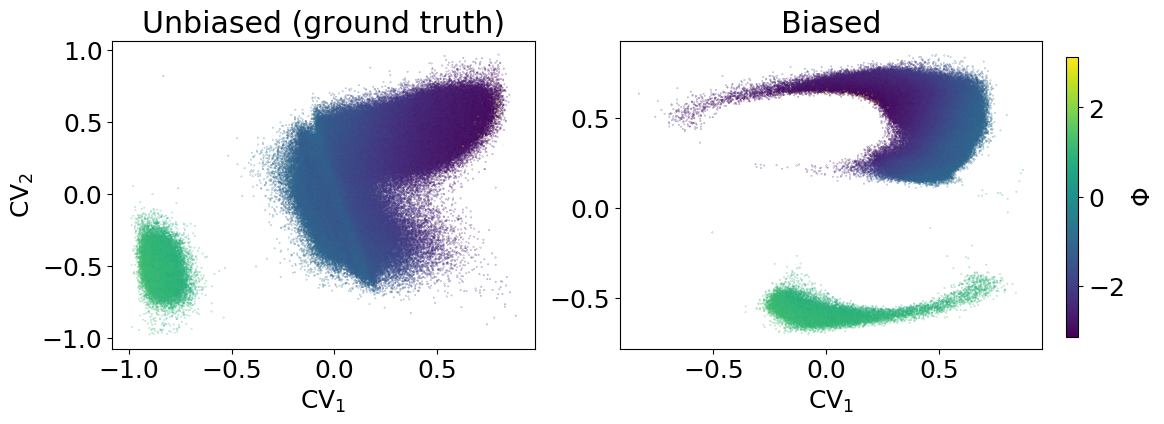} 
\caption{Projections of CV$_{\rm{g}}$ (left) and CV$_{\rm{u}}$ (right) over the $1.5~\mu$s unbiased data, colored according to $\Phi$ values for clarity.}
     \label{gtvsbs}
\end{figure}

We compare here the ground truth CV shown in Figure~\ref{GT} with a CV obtained from biased data. The latter CV is constructed by training an autoencoder on a simulation of $300$~ns biased using the free energy associated with~$(\Phi,\Psi)$, and applying no reweighting to the data points. The biased trajectory visits the quasi-totality of the $(\Phi,\Psi)$ space and sample multiple transitions. In this section, we denote by CV$_{\rm{g}}$ and CV$_{\rm{u}}$ the ground truth CV and the CV obtained from biased data, respectively (the subscript 'u' standing for 'unweighted'). The projections of CV$_{\rm{u}}$ and CV$_{\rm{g}}$ on the unbiased $1.5~\mu$s dataset are shown in Figure~\ref{gtvsbs}.

To compare the two CVs, we use the following methods:
\begin{itemize}
\item{\textbf{Regression scores:}} we measure the CVs' correlation to the dihedrals $\Phi$ and $\Psi$ using the value of the regression score between each CV and the $(\Phi,\Psi)$ vector. CV$_{\rm{g}}$ achieves a regression score of $0.967$, whereas the regression score of CV$_{\rm{u}}$ is $0.880$. However, these scores are computed over configuration from the $1.5~\mu$s unbiased trajectory, which contains only few transition states. We therefore also compute the regression scores using this time the $300$ ns $(\Phi,\Psi)$-biased trajectory which served as training data for learning CV$_{\rm{u}}$. This biased trajectory achieves a large number of transitions (over $5000$ transitions between the C5-C7eq and the C7ax states). Over this trajectory, the computed regression scores between the CVs and $(\Phi,\Psi)$ are in general significantly lower. However, CV$_{\rm{g}}$ still achieves a higher regression score ($0.781$) than CV$_{\rm{u}}$ ($0.410$). When measuring these regression scores over the transition states alone, both scores are lower than $0.5$, but this time CV$_{\rm{u}}$ outperforms CV$_{\rm{g}}$ ($0.45$ to $0.39$). Here, transition state regions were determined approximately using values of $\Phi$, with transition states corresponding to $\Phi \in [-0.5,0.5]\cup [1.8,2.6]$.

These results indicate that in the case of alanine dipeptide, the unweighted CV does not seem to gain valuable information over the ground truth CV from the transition states in the dataset, and even loses $(\Phi,\Psi)$-related information overall. One possible explanation can be the presence of "unrealistic" conformations in the biased data that are not canceled out by reweighting. 

\item{\textbf{Convergence of the free energy.}}  We run two eABF simulations, using CV$_{\rm{g}}$ and  CV$_{\rm{u}}$.  As done in Section~\ref{dialares} for the final free (see in particular~\eqref{eq:free_energy_estimation_histogram}), at any time $t$ of the obtained trajectory, the intermediate estimates $G_t$ of the CV's free energy can be used to compute an intermediate approximation $\tilde{F_t}$ of the $(\Phi,\Psi)$ free energy $F$ by reweighting histograms. 

The error between the current estimate of the mean force $\tilde{F}_t$ and the final estimate $\tilde{F}_{\text{final}}$ obtained at the end of the simulation is computed at each timestep $t$ as a weighted $\ell_2$-distance between these two quantities:
\begin{equation}
\displaystyle  \Delta F_t = \sqrt{ \displaystyle \frac{\displaystyle \sum_{j,l = 1}^{k} \exp(-\beta F^{j,l})~ (\tilde{F_t}^{j,l} - \tilde{F}_{\text{final}}^{j,l})^2}{\displaystyle \sum_{j,l = 1}^k \exp(-\beta F^{j,l})}  }  \text{,} 
\label{deltaFt}
\end{equation}
where the sum is over all bins $(j,l)$ of $(\Phi,\Psi)$ and $F^{j,l}$ is the value of the reference free energy in the bin ${j,l}$. 

\begin{remark}
The free energies are defined up to additive constants. In this paper, two values (e.g. estimates) $F_1$ and $F_2$ of the free energy are always optimally aligned by shifting them so as to minimize the $\ell_2$-distance defined in~\eqref{deltaFt}, where $(\tilde{F_t}^{j,l} - \tilde{F}_{\text{final}}^{j,l})$ is replaced by $(F_1^{j,l} - F_2^{j,l})$.
\label{l2distance}
\end{remark}

In order to assess the convergence speed of the free energy for each eABF run, we plot the time dependent value $\Delta F_t$. As a point of comparison, the same protocol is used to compute approximations of $F$ and their convergence speed using eABFs with the dihedrals $(\Phi,\Psi)$. In all three cases, the biasing CV (i.e. the ground truth CV, the $(\Phi,\Psi)$ dihedrals, or the CV obtained from unweighted biased data) is rescaled to have a range in $[-1,1]^2$, and the number of bins in each direction is $50$. Sampled points are saved every $10$~timesteps. The simulations are $250$ns long, which allows for free energy convergence. Figure~\ref{dft} shows the value of $\beta\Delta F_t$ as a function of the eABF simulation time for the three CVs.

\begin{figure}[ht!]
\centering
\includegraphics[width=0.7\linewidth]{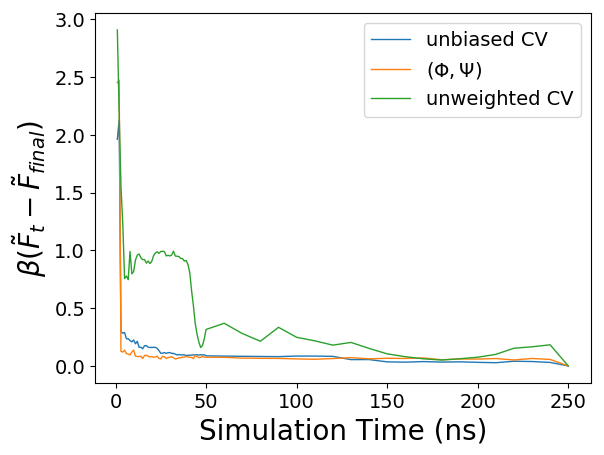} 
\caption{Evolution of the distance $\beta \Delta F_t$ between the final and current estimates of the free energy profiles, as a function of time.}
     \label{dft}
\end{figure}

It can be observed that the ground truth CV free energy profile stabilizes within a small simulation time comparable to that of the $(\Phi,\Psi)$ CV.  This indicates that the conformational space is relatively well explored in a small amount of time when biasing with the ground truth CV. The convergence of the free energy profile for the CV obtained from unweighted biased data is slower, and even seems to not completely be reached at $250$~ns.


\item{\textbf{Sampling transition states:}} Finally, we compare CV$_{\rm{g}}$ and CV$_{\rm{u}}$ by estimating their ability to sample transitions between the C$5$/C$7$eq and the C$7$ax states. For this, we sample $2$ eABF trajectories for each of CV$_{\rm{g}}$, CV$_{\rm{u}}$, and of $(\Phi,\Psi)$ for comparison. We divide the $\Phi$ space into $4$ regions: C$5$/C$7$eq, TS$1$, C$7$ax, and TS$2$, corresponding respectively to: $[-\pi,-0.5]\cup[2.6,\pi]$, $[-0.5,0.5]$, $[0.5,1.7] $, and $[1.8,2.6]$. We then plot in Figure~\ref{transitions} the regions sampled along the $6$ eABF simulations using the sampled values of $\Phi$. Figure~\ref{transitions} shows that CV$_{\rm{g}}$ enables a more efficient sampling of regions C$5$/C$7$eq, TS$1$ and C$7$ax than CV$_{\rm{u}}$. The CV$_{\rm{u}}$ eABF trajectories seem to occasionally stay trapped in the C$5$/C$7$eq basin. Neither CV enables very high sampling of TS$2$ (compared to the other three regions, and especially compared to eABFs with $(\Phi,\Psi)$). However CV$_{\rm{u}}$ eABF trajectories visit this region more often than CV$_{\rm{g}}$ eABF trajectories do. 

\begin{figure}[ht!]
\centering
\includegraphics[width=0.8\linewidth]{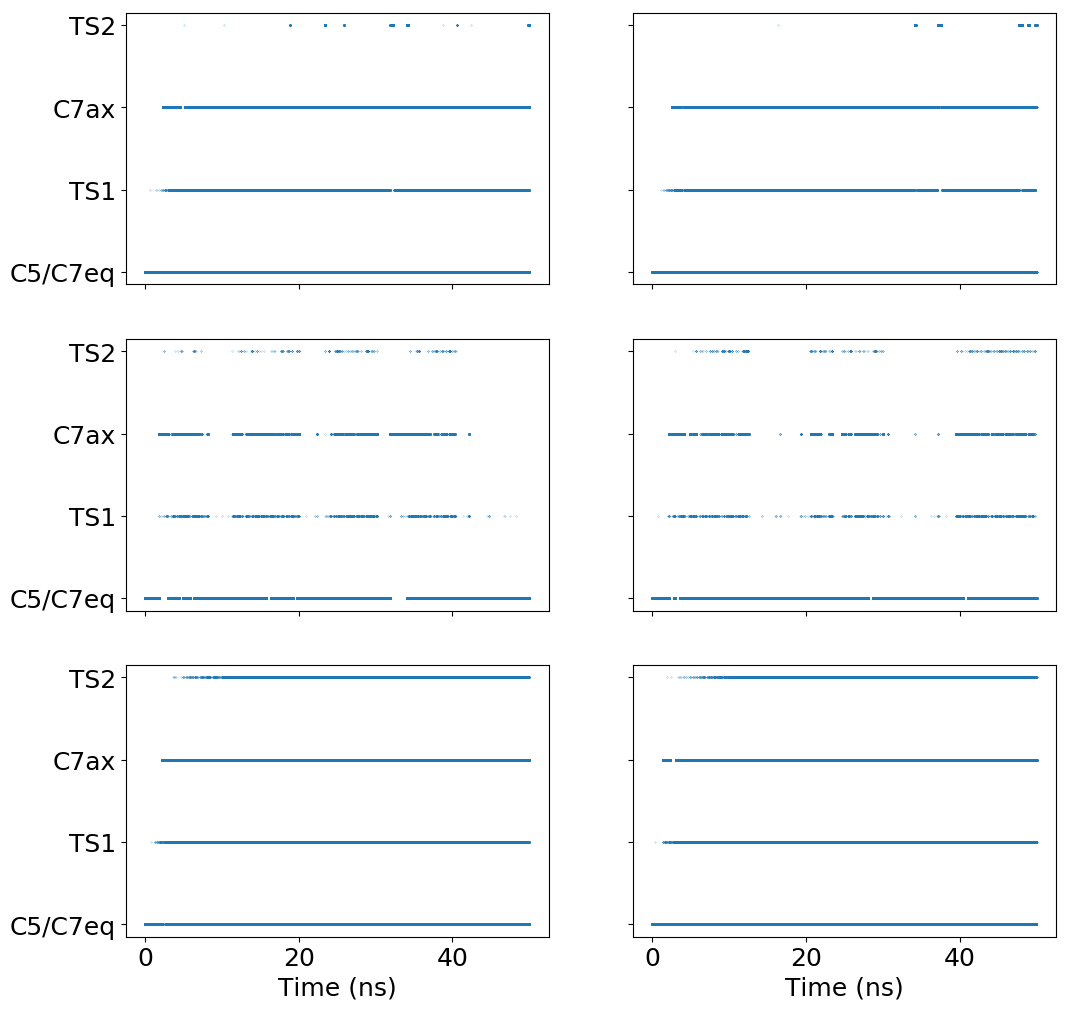} 
\caption{State assigned to each sample over $50$ns eABF trajectories. Top: eABFs using CV$_{\rm{g}}$. Middle: eABFs using CV$_{\rm{u}}$. Bottom: eABFs using $(\Phi,\Psi)$.}    
     \label{transitions}
\end{figure}
\end{itemize}

In general, the obtained results suggest that CV$_{\rm{g}}$ is more efficient at sampling than CV$_{\rm{u}}$. The latter even seems to be a rather poor choice of CV. However, it is important to note that our choice of a unweighted biased data does not constitute the only way of learning a CV from biased data, nor of incorporating transition state information into the CV. One could for instance only apply some partial reweighting to the data points, in order to remove the most unlikely configurations under the Boltzmann--Gibbs measure, while still keeping transition states associated with sufficiently low free energy barriers. 

\subsubsection{AE-ABF without reweighting}
\label{sec:AE_ABF_without_reweighting}

This section  presents the results of performing AE-ABF without reweighting. The unweighted AE-ABF is run for $9$ iterations, using the same parameters as in Section~\ref{setup}, but of course without the use of reweighting to unbias the training. Just as was done for the reweighted results, we compute, for each iteration, the regression score of the learned CV with respect to the previous CV and with respect to $(\Phi, \Psi)$. Table~\ref{tab:biasedtable} shows the obtained values. The regression scores between consecutive CVs are consistently lower than those obtained with the reweighted AE-ABF. More importantly, these regression scores are lower than the threshold $s_{\min}=0.996$ determined in Supp.~Mat.~Section~\ref{appendix:regscore}, indication that we cannot consider the CV to be converged. It is also important to note that the regression scores between the CVs and $(\Phi, \Psi)$ is lower than with reweighted AE-ABF. These results are in accordance with the results obtained in the comparison between the unbiased ground truth and the biased CV in the previous section.

\begin{table}
\centering
\begin{tabular}{|c|c|c|}
\hline
   Iteration & previous CV &  $(\Phi,\Psi)$ \\
 \hline
   $0$ & $-$ & $0.935$\\
 \hline
  $1$ & $ 0.850 $ & $ 0.896$\\
   \hline
  $2$ & $ 0.867 $ & $ 0.963$\\
   \hline
  $3$ & $ 0.915 $ & $ 0.930$\\
   \hline
  $4$ & $ 0.987 $ & $ 0.924$\\
   \hline
  $5$ & $ 0.911 $ & $ 0.910$\\
   \hline
  $6$ & $ 0.935 $ & $ 0.902$\\
   \hline
  $7$ & $ 0.882 $ & $ 0.935$\\
   \hline
  $8$ & $ 0.850 $ & $ 0.879$\\
  \hline
  $9$ & $ 0.901 $ &  $ 0.890$\\
  \hline
\end{tabular}

\caption{Linear regression scores with unweighted AE-ABF for 9 iterations. Each line corresponds to an iteration. The regression score is computed between the  learned CV and: the CV from the previous iteration (second column); the 2D vector $(\Phi, \Psi)$ (third column). CV convergence does not occur. The regression score values with respect to $(\Phi, \Psi)$ are generally lower than those obtained with reweighted AE-ABF. }
\label{tab:biasedtable}
\end{table}

In addition, we plot in Figure~\ref{bAECVRamach} the sampled regions of the Ramachandran space at each iteration. It can be observed that the sampled trajectories are similar to those obtained with the regular weighted AE-ABF. Figure~\ref{bAEs} shows the learned CV at each iteration projected on the unbiased $1.5~\mu$s trajectory. It can be seen that, as implied by the values of the regression scores reported in Table~\ref{tab:biasedtable}, the learned CV does not converge.

\begin{figure}
    \centering
          \begin{subfigure}{0.8\linewidth}
        \centering
        \includegraphics[width=\textwidth]{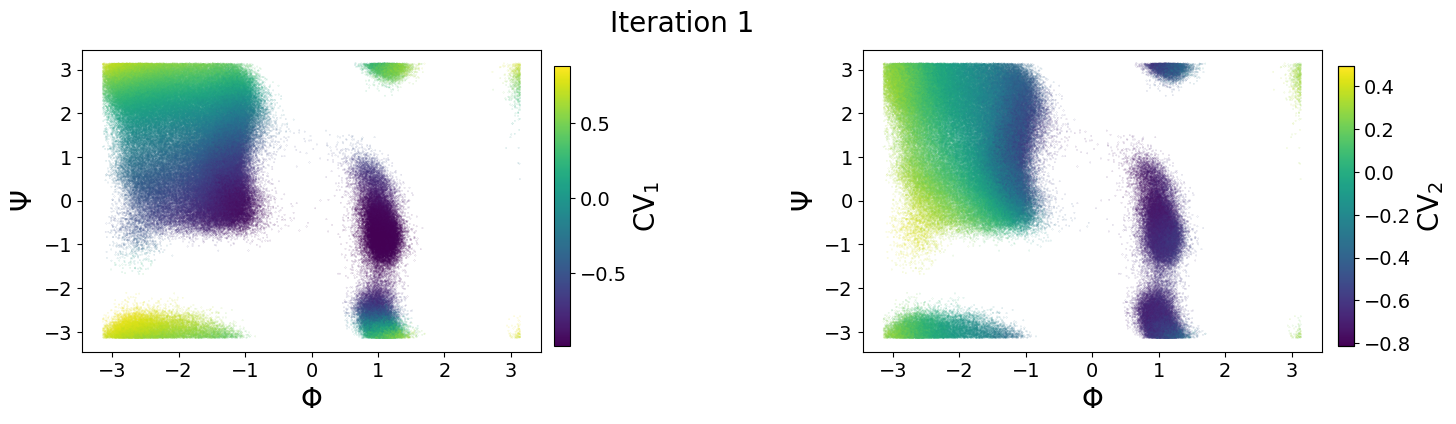}
        \labelnotempty{}
        \vspace{-1em}
    \end{subfigure}            \begin{subfigure}{0.8\linewidth}
        \centering
        \includegraphics[width=\textwidth]{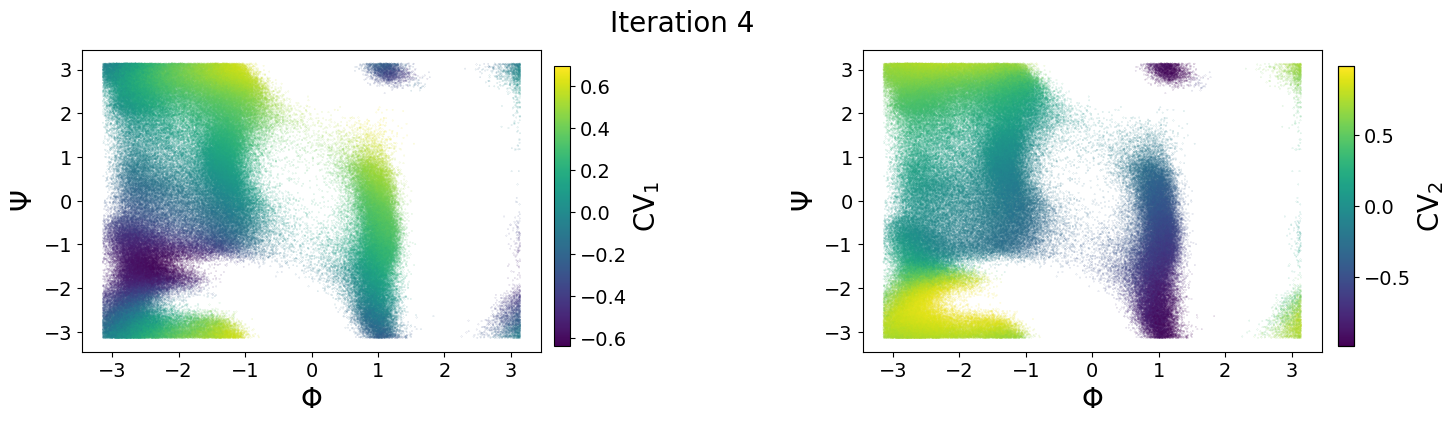}
        \labelnotempty{}
        \vspace{-1em}
    \end{subfigure}            \begin{subfigure}{0.8\linewidth}
        \centering
        \includegraphics[width=\textwidth]{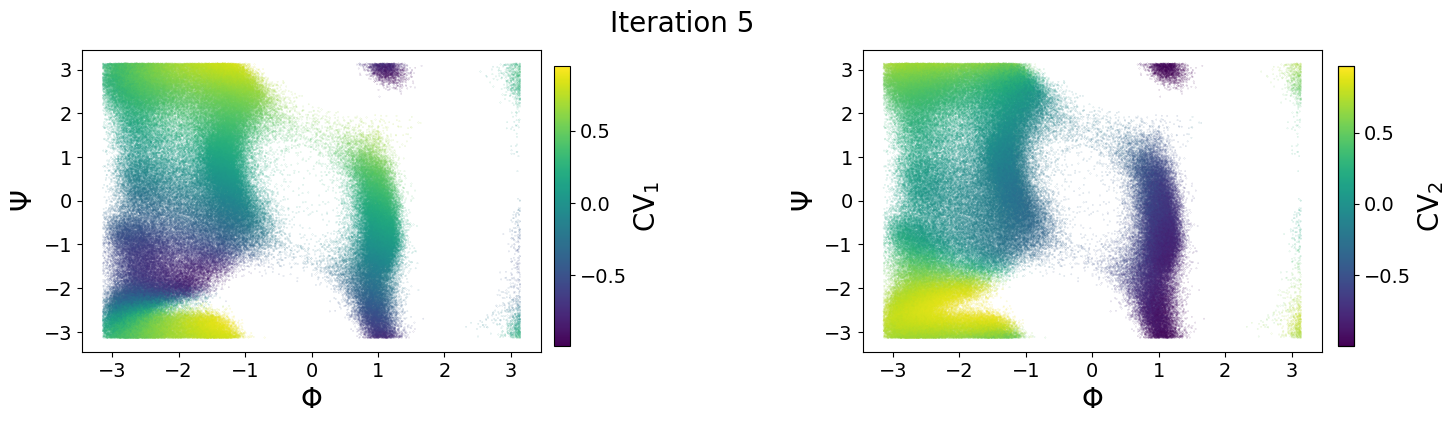}
        \labelnotempty{}
        \vspace{-1em}
    \end{subfigure}
     \begin{subfigure}{0.8\linewidth}
        \centering
        \includegraphics[width=\textwidth]{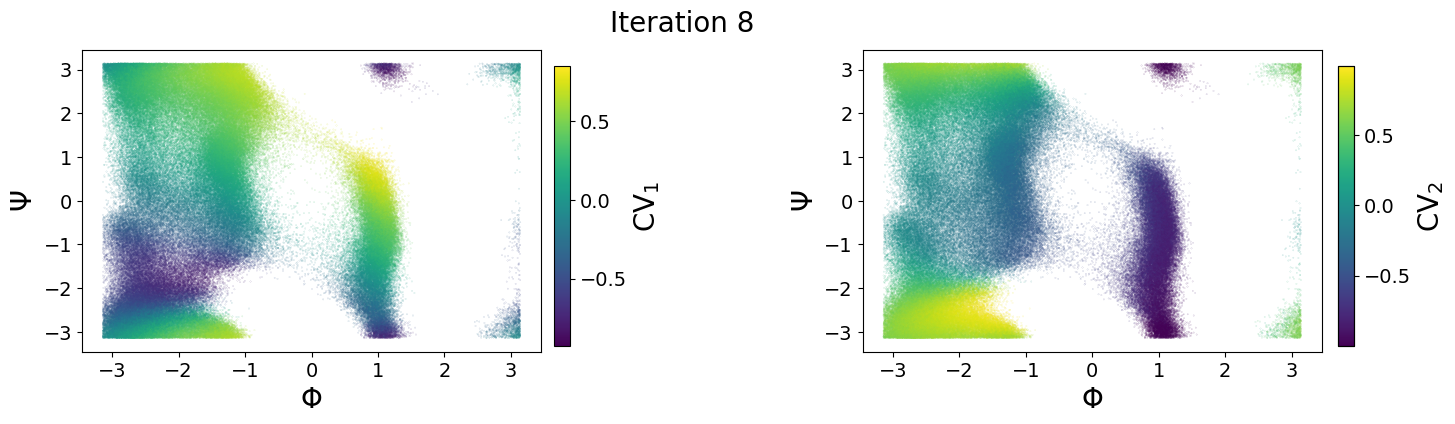}
        \labelnotempty{}
        \vspace{-1.6em}
    \end{subfigure}
   \caption{AE-ABF for 8 iterations without reweighting (not all are shown here). Ramachandran scatter plots of each trajectory. The coloring corresponds to values of the first component of the CVs (Left) and the values of the second component of the CVs (Right). The sampled $(\Phi,\Psi)$ regions are generally the same as those sampled during reweighted AE-ABF.}
    \labelnotempty{bAECVRamach}
\end{figure}

\begin{figure}
\centering  
    \begin{subfigure}{0.49\linewidth}
        \includegraphics[width=1\linewidth]{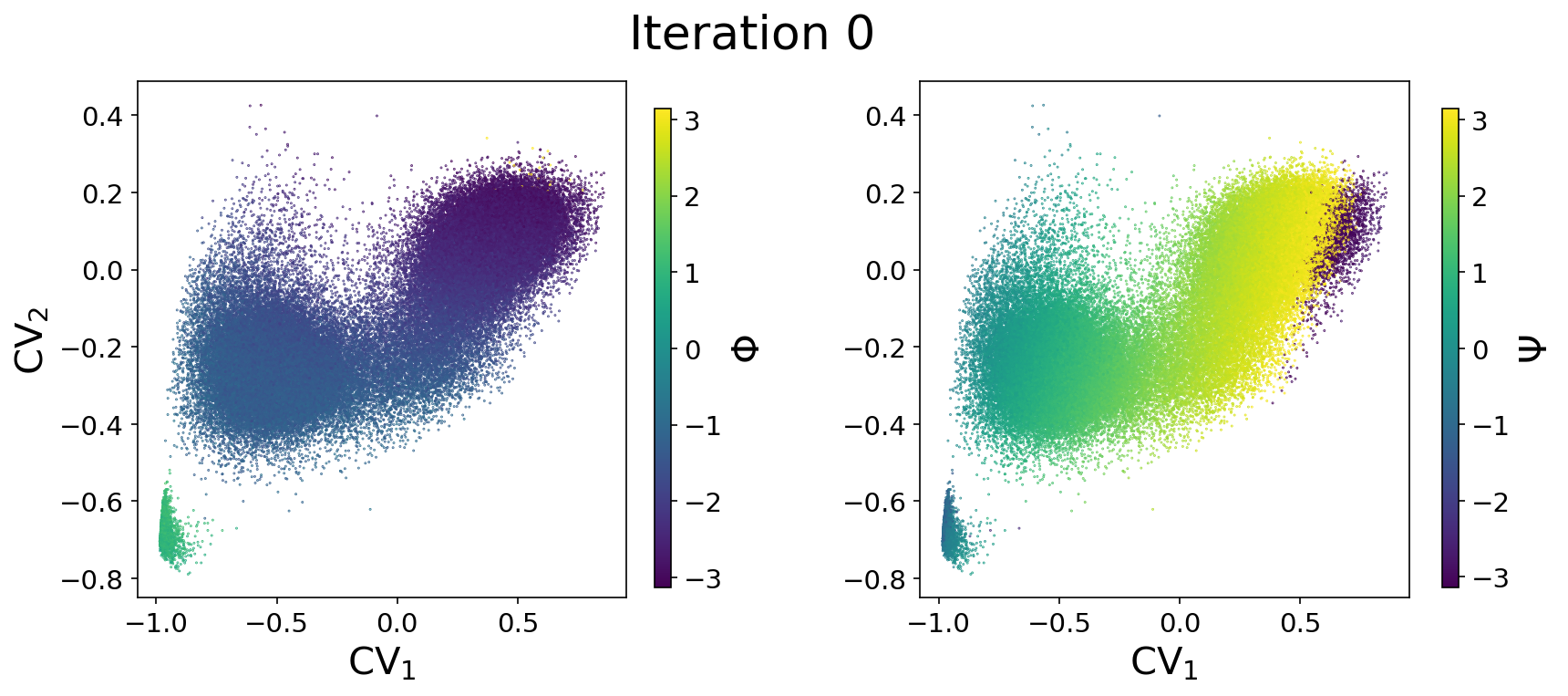} 
        \labelnotempty{}
    \end{subfigure} 
         \begin{subfigure}{0.49\linewidth}
        \centering
        \includegraphics[width=\textwidth]{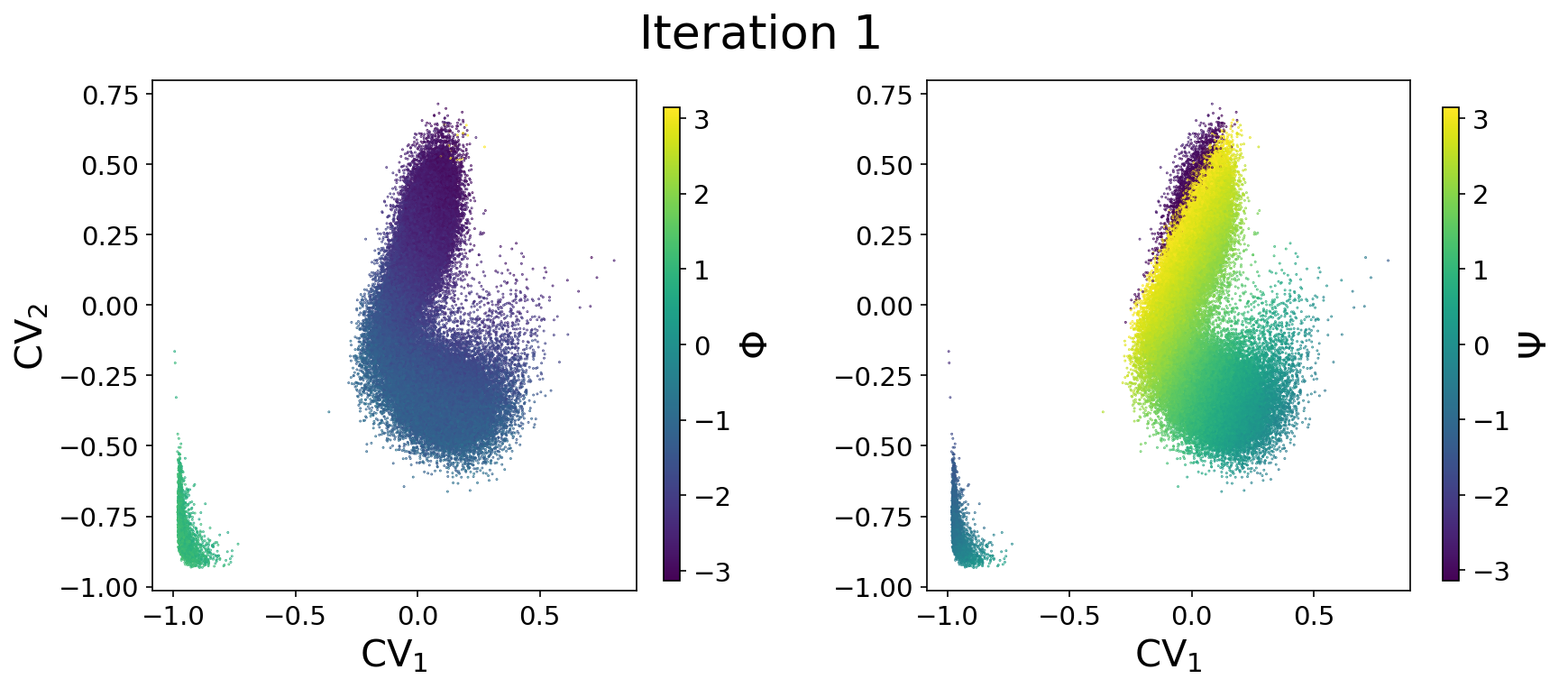}
        \labelnotempty{}
    \end{subfigure}
    \begin{subfigure}{0.49\linewidth}
        \centering
        \includegraphics[width=\textwidth]{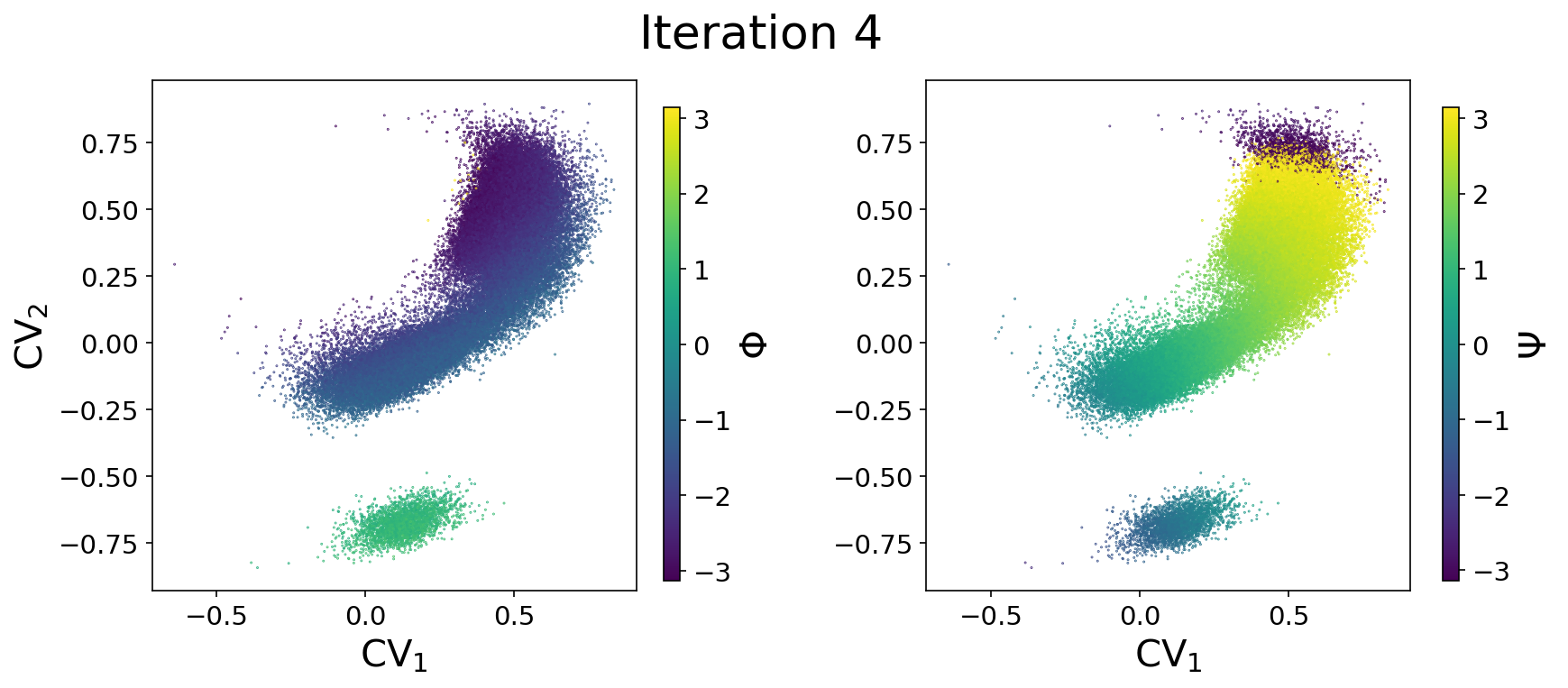}
        \labelnotempty{}
    \end{subfigure}
    \begin{subfigure}{0.49\linewidth}
        \centering
        \includegraphics[width=\textwidth]{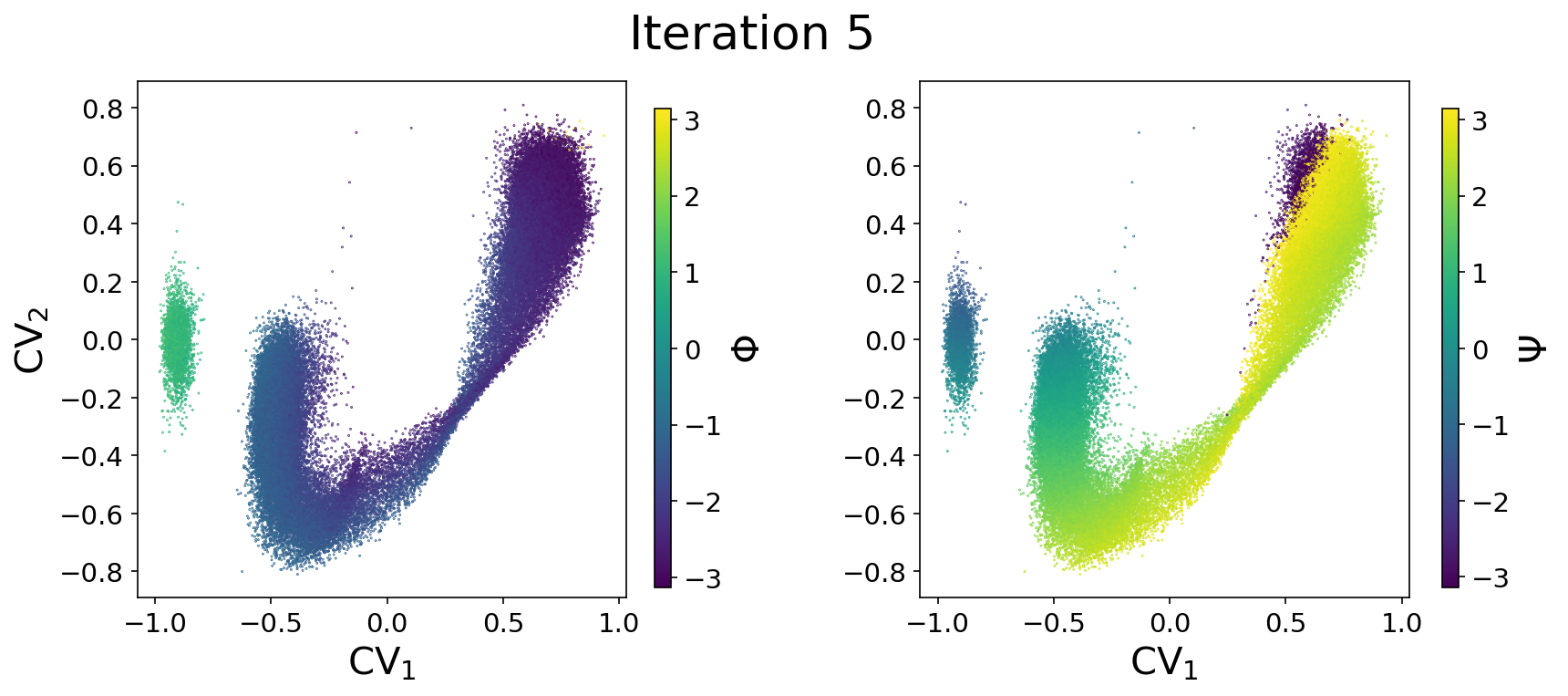}
        \labelnotempty{}
    \end{subfigure}
        \begin{subfigure}{0.49\linewidth}
        \centering
        \includegraphics[width=\textwidth]{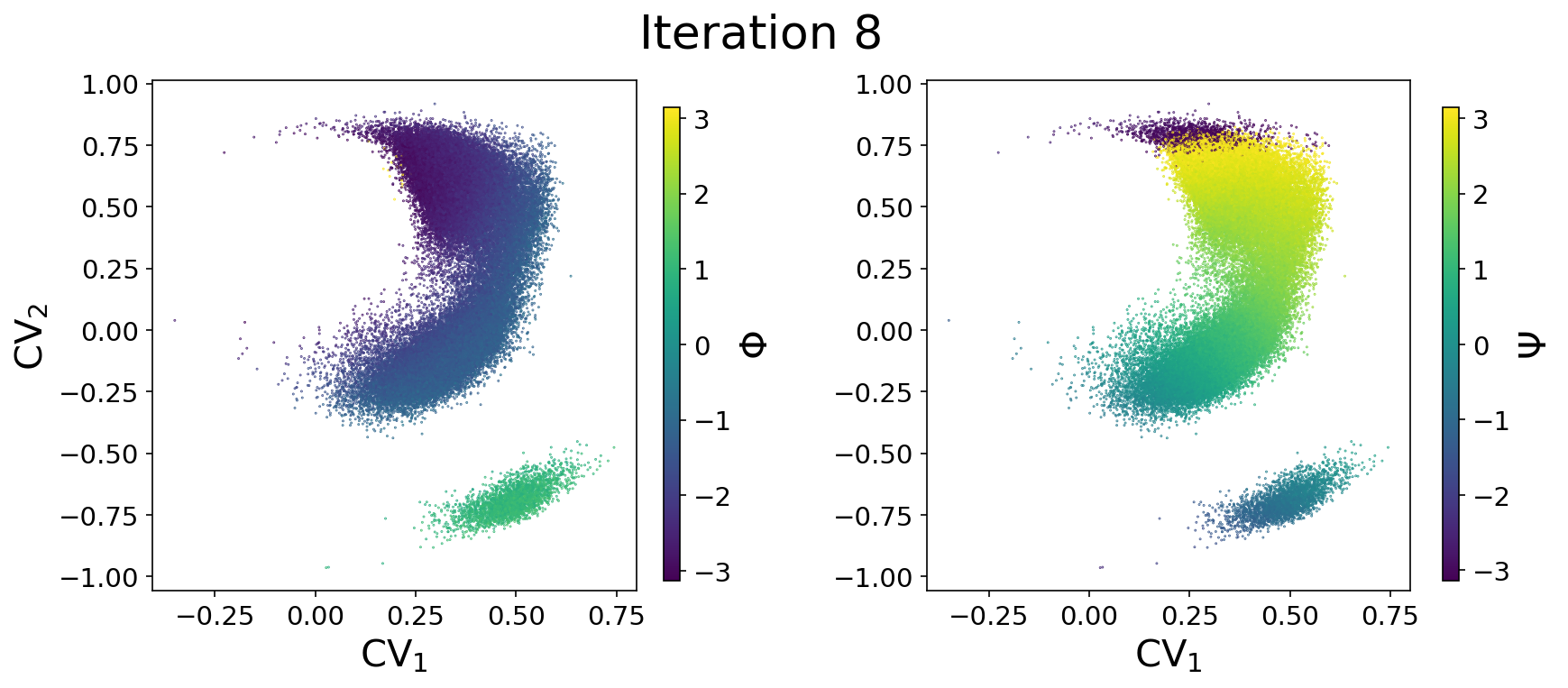}
        \labelnotempty{}
    \end{subfigure}
            \begin{subfigure}{0.49\linewidth}
        \centering
        \includegraphics[width=\textwidth]{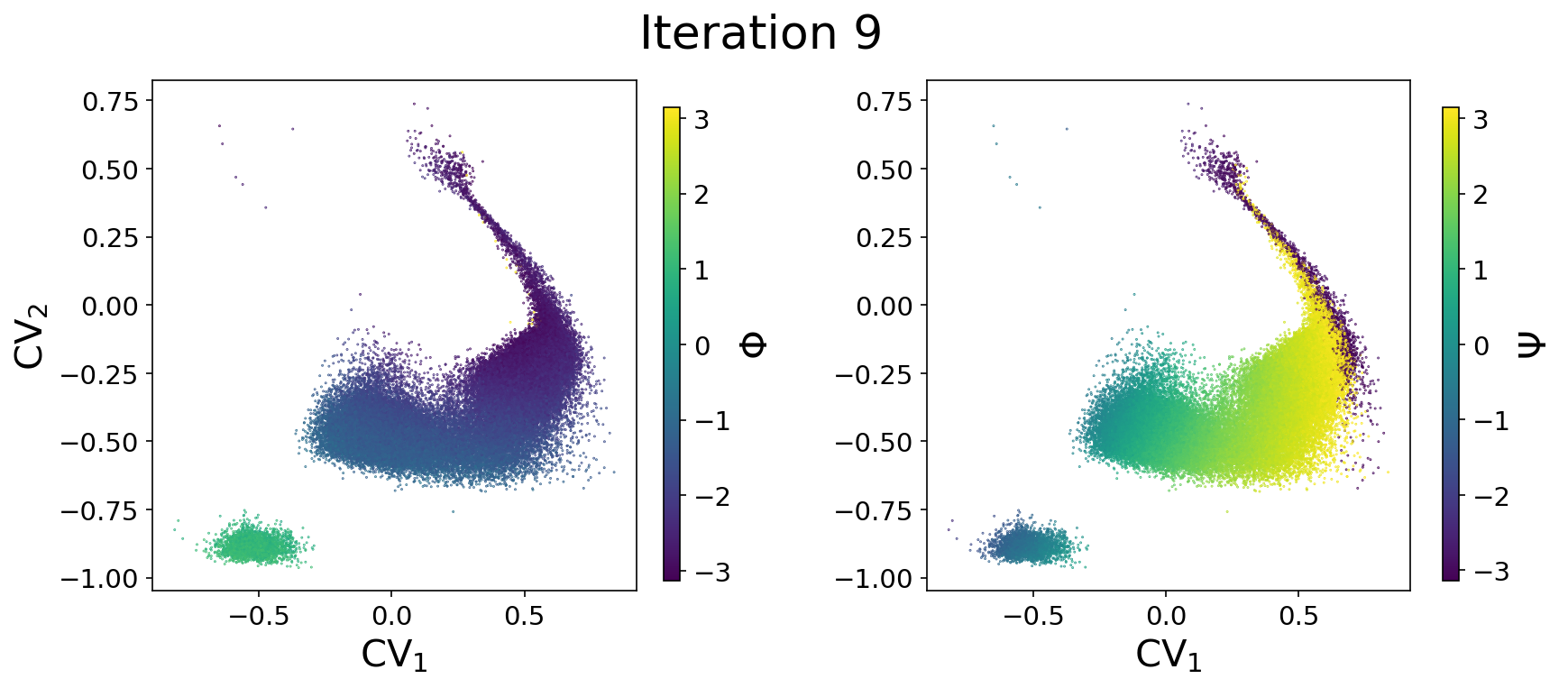}
        \labelnotempty{}
    \end{subfigure}
   \caption{Autoencoder CVs through 9 iterations of AE-ABF without reweighting. The CVs are scatter plotted with $\Phi$ and $\Psi$ colorings. The plots show that the CVs obtained at each iteration are not visually similar to the ground truth CV shown in Figure~\ref{GT}, or to each other.} 
    \labelnotempty{bAEs}
\end{figure}

While CVs trained on biased data do not seem to outperform CVs trained on unbiased (or reweighted) data in this particular test case, we cannot generalize this statement to every system. In cases where biased CVs would provide a sampling advantage, it would be interesting to test a hybrid version of our iterative algorithm where two CVs are computed at each iteration: one with reweighting to check the convergence of the algorithm, and one without reweighting to perform the next round of enhanced sampling.

\end{document}